\documentclass[letterpaper,twocolumn,10pt]{article}

\usepackage{zhanggroup}
\usepackage{tikz}
\usepackage{amsfonts}
\usepackage{xspace} 
\usepackage{amsmath}
\usepackage{mathrsfs}
\usepackage{amssymb}
\usepackage{amsmath}
\usepackage{amsthm}
\usepackage{subcaption}
\usepackage{booktabs}
\hypersetup{
  colorlinks,
  linkcolor={blue!70!green},
  citecolor={green!70!blue},
  urlcolor={orange!70!red}
}
\usepackage[absolute]{textpos}
\usepackage{graphicx}
\usepackage{textcomp}
\usepackage{eucal}
\usepackage{epsfig,endnotes}
\usepackage{pifont}
\usepackage[normalem]{ulem}
\usepackage{gensymb}
\usepackage{multirow}
\usepackage{subcaption}
\usepackage{caption}
\usepackage{mathtools}
\usepackage{bbm}
\usepackage{booktabs}
\usepackage{makecell}
\usepackage{enumitem}
\setlist[itemize]{leftmargin=*}
\usepackage[ruled,linesnumbered,noend]{algorithm2e}
\usepackage[export]{adjustbox}
\usepackage{cellspace}
\usepackage{colortbl}
\usepackage{etoolbox}
\newcommand{\mypara}[1]{\noindent\textbf{#1.}}

\newcommand{\tuple}[1]{\ensuremath{\langle \mathsf{#1} \rangle}}
\DeclareSymbolFont{cmsymbols}{OMS}{cmsy}{m}{n}
\SetSymbolFont{cmsymbols}{bold}{OMS}{cmsy}{b}{n}
\DeclareSymbolFontAlphabet{\mathcal}{cmsymbols}
\begin{document}

\date{}

\title{\Large \bf On the Privacy Risks of Cell-Based NAS Architectures}

\author{
{\rm Hai Huang\textsuperscript{1}}\ \ \
{\rm Zhikun Zhang\textsuperscript{1}}\ \ \
{\rm Yun Shen\textsuperscript{2}}\ \ \
{\rm Michael Backes\textsuperscript{1}}\ \ \
{\rm Qi Li\textsuperscript{3}}\ \ \
{\rm Yang Zhang\textsuperscript{1}}
\\
\\
\textsuperscript{1}\textit{CISPA Helmholtz Center for Information Security}\ \ \
\textsuperscript{2}\textit{NetApp}\ \ \\\
\textsuperscript{3}\textit{Tsinghua University \& Zhongguancun Lab}
}

\maketitle

\begin{textblock}{15}(1.9,1)
To Appear in 2022 ACM SIGSAC Conference on Computer and Communications Security, November 2022
\end{textblock}

% ----------------------------------------------------
\begin{abstract}
% ----------------------------------------------------

Existing studies on neural architecture search (NAS) mainly focus on efficiently and effectively searching for network architectures with better performance.  
Little progress has been made to systematically understand if the NAS-searched architectures are robust to privacy attacks while abundant work has already shown that human-designed architectures are prone to privacy attacks.
In this paper, we fill this gap and systematically measure the privacy risks of NAS architectures. 
Leveraging the insights from our measurement study, we further explore the cell patterns of cell-based NAS architectures and evaluate how the cell patterns affect the privacy risks of NAS-searched architectures. 
Through extensive experiments, we shed light on how to design robust NAS architectures against privacy attacks, and also offer a general methodology to understand the hidden correlation between the NAS-searched architectures and other privacy risks.\footnote{The source code of our experiments can be found at \url{https://github.com/MiracleHH/nas_privacy}}

% ----------------------------------------------------
\end{abstract}
% ----------------------------------------------------

% ----------------------------------------------------
\section{Introduction}
% ----------------------------------------------------

Deep neural networks (DNNs) have enjoyed a remarkable boom in recent decades and achieved superior performance in many real-world tasks (e.g., image classification~\cite{SZ15,KSH12,HZCKWWAA17}, object detection~\cite{RHGS15,LAESRFB16}, text classification~\cite{K14,YML19}, etc.). 
With significant advances in computing power, DNNs have become more complicated.
They are both deeper~\cite{SS17,L20} and wider~\cite{LPWHW17,L20} to further improve model performance (e.g., GPT-1/2/3~\cite{RNSS18,RWCLAS19,BMRSKDNSSAAHKHCRZWWHCSLGCCBMRSA20}, DALL-E~\cite{RPGGVRCS21}, etc.).  
Despite their success, manually designing those complex networks in a trial-and-error way remains a tedious task, requiring both architectural engineering skills and domain expertise. 

\begin{figure}[!t]
\centering
\includegraphics[width=0.4\textwidth]{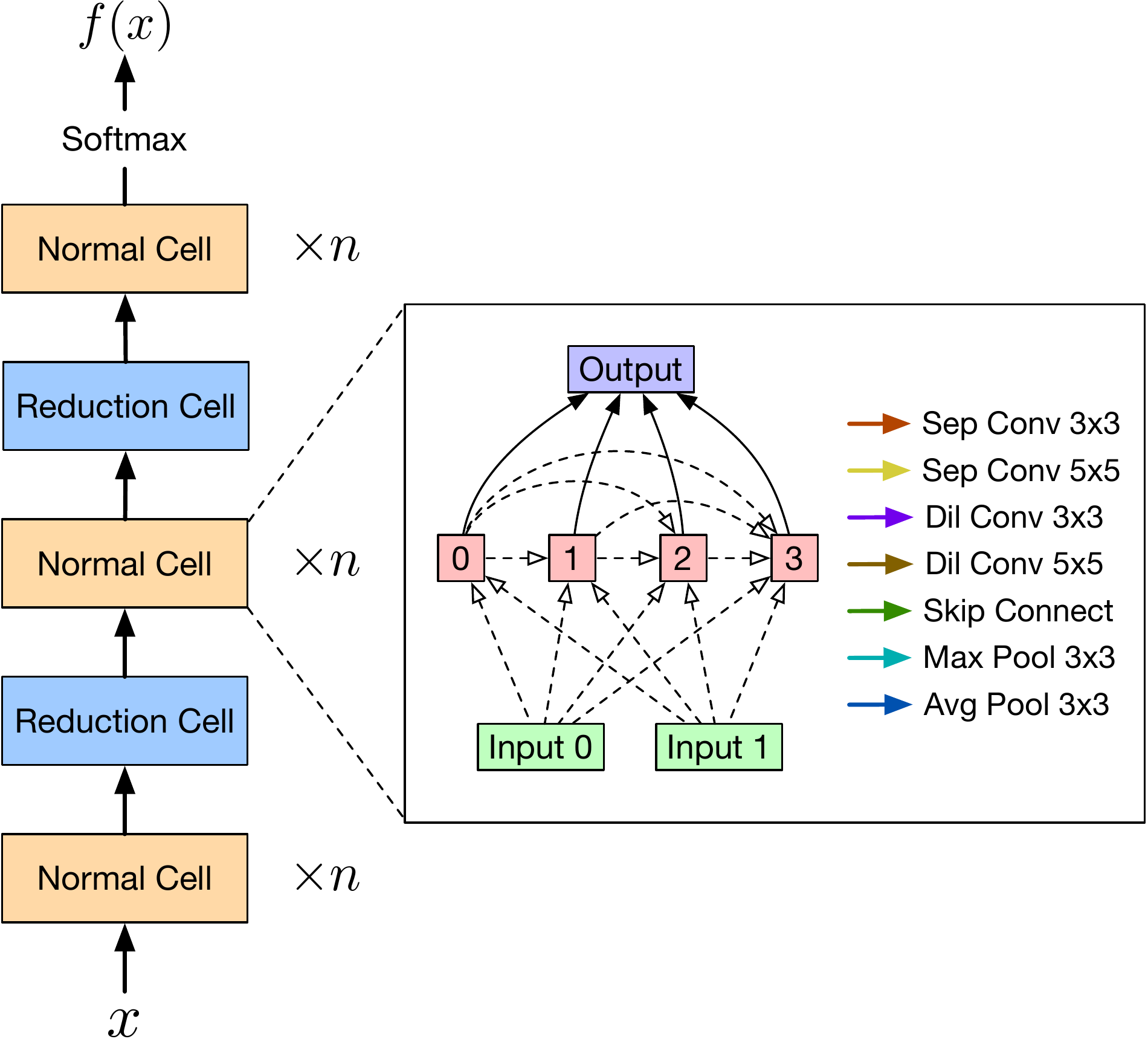}
\caption{Illustration of the cell-based NAS architecture. 
The normal and reduction cells are repeated multiple times and connected together to constitute the whole network architecture, and the cell structures for normal and reduction cells will be searched in the search process of NAS. 
In the box on the right, there is a general DARTS~\cite{LSY19} search space prototype where the arrows with dashed lines represent possible edges and each selected edge will be assigned with one operation indicated by one colored arrow on the right.
}
\label{fig:cell_nas}
\end{figure}

\textit{Neural architecture search} (NAS)~\cite{ZL16,BGNR17} is a natural step towards resolving the above challenge. 
It aims at automating the search process for the most suitable deep neural network architectures for specific tasks.
The core idea of NAS is using a \textit{search strategy} to select an architecture from a predefined \textit{search space}, and leverage a performance estimation strategy to guide the search process.
Directly searching all eligible architectures is computationally expensive; thus the mainstream cell-based NAS methods treat the whole network architecture as a combination of specific modules to reduce the search space (\autoref{fig:cell_nas} illustrates a typical cell-based NAS architecture).
Previous work has shown that the architectures selected by cell-based NAS techniques can attain comparable performance or even outperform those traditional human-designed architectures on tasks such as image classification~\cite{ZVSL18,GWLYWLC21,LHH21}, object detection~\cite{WGCWTSZ20, GLL19}, etc.
Existing NAS research mainly focuses on two directions --- identify efficient and effective search strategies to obtain the best architecture candidates~\cite{CGW21,AMDL21,WCCTH21} and improve the robustness of NAS-searched architectures against adversarial examples~\cite{DAMGB21,GYXLL20,LYWX21}.
On the other hand, although abundant work has already shown that human-designed architectures are prone to privacy attacks, little progress has been made to systematically understand if the NAS-searched architectures are robust to privacy attacks given their complex network topology~\cite{PXJLW22}.
In privacy attacks, an adversary's goal is to gain knowledge of the target model's training data, which is not intended to be shared.
The most popular attack in this domain is membership inference attack (MIA)~\cite{SSSS17} which enables an attacker to infer whether a sample is used to train a target model.
So far, little attention has been paid on the membership privacy of NAS-searched architectures.

\mypara{Our Work}
In this paper, we focus on the aforementioned membership inference attack against NAS-searched architectures due to its pervasiveness and the privacy impact in the real world.
Concretely, we address two research questions in this study --- \emph{whether NAS-searched architectures are robust to membership inference attacks} and \emph{how architectural information affects such robustness.}
Note that the first research question has been discussed by some preliminary work~\cite{PXJLW22} in a limited scenario, i.e., label-only membership inference~\cite{CTCP21,LZ21}.

To this end, we first conduct a comprehensive measurement study to systematically evaluate the privacy risks of NAS-searched architectures identified by cell-based NAS algorithms, the most popular and influential NAS methods~\cite{WREL22}. 
We find that NAS-searched architectures are generally more robust against MIAs.
For instance, 9 out of 10 NAS-searched architectures are more robust to membership inference attacks than all other 10 human-designed architectures for the \tuple{Black\mbox{-}Box, Shadow} attack setting (see \autoref{sec:measurement} for more details) on the CIFAR10 dataset~\cite{CIFAR} in our study.
Interestingly, our findings show that the robustness against MIAs varies from architecture to architecture.
As we see in \autoref{figure:blackbox_shadow_mia_eval}, it is evident that certain NAS-searched architectures are more vulnerable to MIAs than others.
For example, architectures searched by TENAS~\cite{CGW21} tend to be more prone to membership inference attacks.

Following the new insights from our measurement study, we move on to understand how different internal structures impact the privacy robustness of NAS-searched architectures.
To this end, we introduce a general framework to extract cell patterns related to MIA performance (see \autoref{sec:cell_pattern}). 
Concretely, we first evaluate the MIA effectiveness on the well-performed architectures, and then extract the cell patterns for the relatively vulnerable and robust architectures separately under the assistance of a regression model.
Here, this regression model is trained with the former evaluation results to give direct MIA performance predictions on cell architectures.
Finally, we use the extracted cell patterns to modify the internal structures of the target cell architectures and compare the MIA performance before and after the modifications to estimate the effectiveness of the cell patterns.
This framework is generic and can be applied to any other analysis work similar to our objective.
Following this framework, we evaluate the MIA performance of 2,678 NAS-searched architectures from the NAS-Bench-301 dataset~\cite{SZZLKH20} and show that we can identify certain cell patterns that promote or demote MIA performance. 

We further apply our analysis framework to mitigating MIA on NAS-searched architectures.
The evaluation results show that our cell patterns can successfully promote or demote the MIA performance on the target NAS architectures in the majority of cases.
In addition, our cell patterns, though extracted from NAS-Bench-301 dataset, can be transferred to different datasets, attacks, and search spaces.
Finally, we show that our work is complementary to existing MIA defense mechanisms when applying the MIA demotion cell patterns to the target NAS architectures, and can further enhance the effectiveness of existing defense strategies.
We hope that our findings will inspire future work on designing more robust NAS architectures against membership leakage. 

\mypara{Contributions}
In summary, we make the following contributions.
\begin{itemize}
    \item 
    We systematically evaluate the privacy risks of NAS-searched architectures. 
    Through extensive experiments on four datasets, we show that NAS-searched architectures are usually more robust than human-designed architectures. 
    Our finding is contrary to the results from Pang et al.~\cite{PXJLW22}.
    However, the privacy risks of the NAS-searched architectures must be individually evaluated.
    \item 
    We introduce a general framework to analyze the relationship between the NAS-searched architectures and corresponding privacy attacks. 
    We successfully instantiate it to understand, analyze, and identify the hidden cell patterns that impact MIA performance.
    \item 
    We explore the correlation between the NAS-searched architectures and the robustness against MIAs and identify certain cell patterns that impact MIA performance. 
    The experimental results show that our cell patterns can successfully demote or promote MIA performance on target NAS architectures in most cases. 
    Furthermore, our cell patterns can transfer to new attack settings and are complementary to existing defense work.
\end{itemize}

% ----------------------------------------------------
\section{Preliminaries}
% ----------------------------------------------------

% ----------------------------------------------------
\subsection{Cell-based Neural Architecture Search}
\label{subsec:nas}
% ----------------------------------------------------

\mypara{Overview}
The \textit{search space} of NAS can be extremely large if we directly search for and enumerate every single candidate structure of the whole network.
To cope with this challenge, \textit{cell-based} NAS algorithms treat the whole network architecture as a combination of specific small modules, which is referred to as \textit{cells}.
As such, we only need to search for the structures of the basic cells, which significantly reduces the search space and speeds up the search process.
Due to the outstanding performance and high flexibility, cell-based NAS algorithms have dominated the NAS research~\cite{WREL22}. 
\autoref{fig:cell_nas} illustrates a typical cell-based NAS architecture.

There are usually two types of cells in such architectures, \textit{normal cell} and \textit{reduction cell}.
The normal cell preserves the dimension of the input, while the reduction cell reduces the spatial dimension of the input.
The reduction cells are usually placed at the 1/3 and 2/3 positions of the total number of cells, and the rest are normal cells~\cite{LSY19}.
Under this setting, a cell-based NAS architecture usually has only two reduction cells and tends to have more normal cells especially when the network is deep.
Both normal and reduction cells are composed of topological combinations of candidate operations (e.g., separable convolution and skip connection). 

The discrete candidate operations can be represented as continuous architectural parameters such that the whole architecture can be differentially optimized regarding both the model weights and architectural parameters in the search process.
We can gradually train a super network that contains all possible edges in a cell to search for suitable structures of both normal and reduction cells.
When we evaluate the performance of candidate architectures, only a limited number (e.g., 2) of input edges with the highest architectural parameters for each intermediate node in the cell will be retained as real corresponding operations. 
The model weights will be inherited from the super network by the current candidate architecture to evaluate the performance on a validation dataset.
In this way, the search space and computation cost are significantly reduced compared to previous NAS methods.

\mypara{DARTS}
We use DARTS~\cite{LSY19}, the most typical cell-based NAS algorithm, to demonstrate the general concepts and workflow of the cell-based NAS methods.
The right box of \autoref{fig:cell_nas} illustrates the cell search space of the DARTS algorithm, which is represented by a directed acyclic graph (DAG) containing 7 nodes, i.e., 2 input nodes, 4 intermediate nodes, and 1 output node, and multiple edges.
The nodes represent the state of the data, and the edges represent the operations on the data.
The operation $o^{(i,j)}$ between node $i$ and $j$ is selected from a predefined \textit{operation set} containing $K=7$ different operations. 
For an intermediate node $j$, it obtains the intermediate data $x^{(j)}$ by aggregating all of its predecessors, i.e., $x^{(j)}=\sum_{i<j}o^{(i,j)}x^{(i)}$.
The outputs of all intermediate nodes will be concatenated to the output node in the end.
In the final cell architecture, each intermediate node is only allowed to have 2 input nodes. 
Therefore, though there are 14 possible edges to connect intermediate nodes in \autoref{fig:cell_nas}, we can only choose ($4\times2=8$) edges from them and fill these edges with the most likely operations. 
For simplicity, we refer to the search space defined in the DARTS algorithm as \textit{DARTS search space}.
Note that, there are also some other kinds of search spaces, the most representative one among them is NAS-Bench-201~\cite{DY20}, a much simpler and smaller search space. 
More details can be found in \autoref{app:nb201}.

% ----------------------------------------------------
\subsection{Membership Inference Attack}
\label{subsec:membership_inference}
% ----------------------------------------------------

Membership inference attacks (MIAs) against machine learning models aim to infer whether a target sample $x$ is used to train a target model $f_\text{target}$.
As such, MIA directly leads to a privacy breach, allowing the adversaries to learn sensitive information about the training data.
For example, in the real world, $x$ can be a clinical record or an individual.
MIA enables the attackers can infer whether this clinical record or individual has been used to train a model associated with a certain disease.
This is evidently a privacy and confidentiality violation.

Consider the most common attack setting where the adversary has black-box access to the target model~\cite{SSSS17}, to launch such an attack, the attackers first train a shadow model $f_\text{shadow}$ using a shadow dataset $\mathcal{D}_\text{shadow}^\text{train}$, which performs the same task as $f_\text{target}$ (e.g., classification). 
The attackers then query $f_\text{shadow}$ using both  $\mathcal{D}_\text{shadow}^\text{train}$ (member data) and $\mathcal{D}_\text{shadow}^\text{test}$ (non-member data) and obtain the query responses $R = R_\text{member} \cup R_\text{non-member}$.
They can build an attack model $f_\text{attack}: R \rightarrow \{0,1\}$, where the responses of member data are labeled as 1 and those of non-member are labeled as 0.
At the attack time, the attackers query $f_\text{target}$ using the data instance $x$ and use $f_\text{attack}$ to infer whether $x \in \mathcal{D}_\text{target}^\text{train}$ or not using the response from the target model $f_\text{target}$.

% ----------------------------------------------------
\section{Privacy Measurement of NAS-searched Architectures}
\label{sec:measurement}
% ----------------------------------------------------

% ----------------------------------------------------
\subsection{Motivation}
\label{subsec:measurement_goal}
% ----------------------------------------------------

Many papers point out that overfitting of the target ML models (i.e., the target model performs much better on its training data than test data) is the main factor contributing to the success of MIAs~\cite{LWHSZBCFZ22,YGFJ18,SZHBFB19,SSSS17}.
Shu et al.~\cite{SWC20} show that the cell architectures searched by cell-based NAS tend to be shallow and wide to make the loss value converge stably and fast during the search process, which usually leads to competitive generalization performance though it is not guaranteed to be the best.
In other words, the architectures searched by these NAS algorithms usually have low overfitting levels. Therefore, a natural hypothesis is that NAS-searched architectures are more robust against MIAs than traditional human-designed ones.
Yet, recent work from Pang et al.~\cite{PXJLW22}, using empirical results, demonstrates that the NAS architectures are more vulnerable to MIAs than those human-designed architectures, even though the former have better normal model performance than the latter. 
Note that the conclusion of Pang et al.~\cite{PXJLW22} is based on the label-only scenario for MIA in a black-box setting. 
In light of the conflicting results, in this section, we comprehensively evaluate the performance of MIAs on NAS-searched architectures and human-designed architectures in both black-box and white-box settings with different levels of knowledge.
Our goal is to eliminate the potential experimental bias introduced in the previous research and compare the MIA performance on both categories in a wider spectrum of attack scenarios.

% ----------------------------------------------------
\subsection{Measurement Setting}
\label{subsec:measurement_setting}
% ----------------------------------------------------

\mypara{Datasets} 
We use 4 diverse benchmark datasets to conduct the measurement experiments, including CIFAR10~\cite{CIFAR}, CIFAR100~\cite{CIFAR}, STL10~\cite{CNL11}, and CelebA~\cite{LLWT15}. 
We refer the readers to \autoref{app:details_measurement} for the details of these datasets.

\mypara{NAS Algorithms} 
\label{sec:nas_algos}
We use 10 representative NAS algorithms to conduct the experiments: (1) DARTS-V1~\cite{LSY19}; (2) DARTS-V2~\cite{LSY19}; (3) ENAS~\cite{PGZLD18}; (4) GDAS~\cite{DY19}; (5) SETN~\cite{DY192}; (6) Random~\cite{CYZHY18}; (7) TENAS~\cite{CGW21}; (8) DrNAS~\cite{CWCTH21}; (9) PC-DARTS~\cite{XXZCQTX20}; (10) SDARTS~\cite{CH20}. 
We defer the detailed description of them to \autoref{app:details_measurement}.

\mypara{Manual Architectures} 
For comparison, we also select 10 representative human-designed architectures to conduct the experiments: (1) ResNet~\cite{HZRS16}; (2) ResNext~\cite{XGDTH17}; (3) WideResNet~\cite{ZK16}; (4) VGG~\cite{SZ15}; (5) DenseNet~\cite{HLMW17}; (6) EfficientNet~\cite{TL19}; (7) RegNet~\cite{RKGHD20}; (8) CSPNet~\cite{WLYWCH20}; (9) BiT~\cite{KBZPYGH20}; (10) DLA~\cite{YWSD18}. 
We refer the readers to \autoref{app:details_measurement} for the details of these architectures.

\mypara{Data Configuration} 
To facilitate the fair comparison of different MIA methods, we split the original dataset $\mathcal{D}$ into 4 disjoint parts with the same size, i.e., $\mathcal{D}_{\mathrm{target}}^{\mathrm{train}}$, $\mathcal{D}_{\mathrm{target}}^{\mathrm{test}}$, $\mathcal{D}_{\mathrm{shadow}}^{\mathrm{train}}$ and $\mathcal{D}_{\mathrm{shadow}}^{\mathrm{test}}$.
$\mathcal{D}_{\mathrm{target}}^{\mathrm{train}}$ and $\mathcal{D}_{\mathrm{target}}^{\mathrm{test}}$ serve as the training and testing dataset of the target model, while $\mathcal{D}_{\mathrm{shadow}}^{\mathrm{train}}$ and $\mathcal{D}_{\mathrm{shadow}}^{\mathrm{test}}$ are utilized as the training and testing dataset of the shadow model. 
As for the NAS algorithm, we further split $\mathcal{D}_{\mathrm{target}}^{\mathrm{train}}$ into two disjoint parts with the same size to obtain $\mathcal{D}_{\mathrm{target}}^{{\mathrm{train}}^{\mathrm{t}}}$ and $\mathcal{D}_{\mathrm{target}}^{{\mathrm{train}}^{\mathrm{v}}}$, where $\mathcal{D}_{\mathrm{target}}^{{\mathrm{train}}^{\mathrm{t}}}$ and $\mathcal{D}_{\mathrm{target}}^{{\mathrm{train}}^{\mathrm{v}}}$ are used as the training and validation datasets respectively of the NAS algorithm in the search process. 
After the final NAS architecture is generated, we train it from scratch using $\mathcal{D}_{\mathrm{target}}^{\mathrm{train}}$ and test its performance on $\mathcal{D}_{\mathrm{target}}^{\mathrm{test}}$.
Note that our NAS training/validation data split is in line with the latest research by Oymak et al.~\cite{OLS21}, which states that the train-validation accuracy gap decreases rapidly when the validation data is mildly large.
Take CIFAR10 dataset for example, the size of $\mathcal{D}_{\mathrm{target}}^{{\mathrm{train}}^{\mathrm{v}}}$ is 7,500, which is fairly sizeable and matching the recommendation by Oymak et al.~\cite{OLS21}.
In turn, we are more likely to make our NAS-searched architectures achieve better generalization performance. 

\mypara{Attacker's Knowledge} 
According to the different knowledge levels the attacker has about the target model, we classify the existing MIAs into 5 types and use them to evaluate the above 10 NAS algorithms and 10 human-designed models. 
In this way, we can conduct a thorough and objective evaluation of both NAS-searched and human-designed architectures in all known MIA attach scenarios~\cite{LWHSZBCFZ22, HSSDYZ21}.

\begin{itemize}
    \item 
    \tuple{Black\mbox{-}Box, Shadow}. 
    In this scenario, the attacker only has black-box access to the target model with a local shadow dataset and does not know the training dataset of the target model.
    \item
    \tuple{Black\mbox{-}Box, Partial}. 
    The attacker has black-box access to the target model and has partial knowledge of the training dataset of the target model.
    \item
    \tuple{White\mbox{-}Box, Shadow}. 
    The attacker has white-box access to the target model with only a local shadow dataset and does not know the training dataset of the target model.
    \item
    \tuple{White\mbox{-}Box, Partial}. 
    The attacker has white-box access to the target model and has partial knowledge of the training dataset of the target model. 
    This is the strongest attack scenario.
    \item
    \tuple{Label\mbox{-}Only}. 
    The attacker only has black-box access to the target model and can only infer information from the output labels of the target model, which is the weakest attack scenario.
\end{itemize}

\noindent Note that the attacker obtains all confidence scores returned by the target model except for the \tuple{Label\mbox{-}Only} MIA scenario. 
We defer the training details of the target models and attack models to \autoref{app:details_measurement}.

% ----------------------------------------------------
\subsection{Measurement Results}
\label{subsec:measurement_results}
% ----------------------------------------------------

\begin{table}[!t]
\centering
  \caption{Normal test accuracy of NAS-searched and human-designed architectures on four different datasets.
  The numbers in the parentheses stand for the corresponding overfitting levels.}
  \label{tab:measure_test_acc}
  \setlength{\tabcolsep}{5pt}
  \scalebox{0.62}{
    \begin{tabular}{c|c|c|c|c|c}
\toprule
\multicolumn{2}{c|}{\multirow{2}{*}{Architecture}}&\multicolumn{4}{c}{Dataset}\cr\cline{3-6}
    \multicolumn{2}{c|}{}&CIFAR10&CIFAR100&STL10&CelebA\cr
\midrule    
    \multirow{10}{*}{\rotatebox{90}{Human-designed}}
    &       ResNet & 0.6587 (0.3413) & 0.2993 (0.7005) & 0.4948 (0.5052) & 0.7438 (0.2562)\cr
    &      ResNext & 0.6410 (0.3590) & 0.3026 (0.6973) & 0.4711 (0.5289) & 0.7468 (0.2532)\cr
    &   WideResNet & 0.6431 (0.3569) & 0.3028 (0.6971) & 0.4662 (0.5338) & 0.7516 (0.2484)\cr
    &          VGG & 0.7796 (0.2204) & 0.4395 (0.5603) & 0.6102 (0.3895) & 0.7627 (0.2373)\cr
    &     DenseNet & 0.7509 (0.2491) & 0.4128 (0.5871) & 0.5815 (0.4185) & 0.7506 (0.2494)\cr
    & EfficientNet & 0.5637 (0.4360) & 0.2401 (0.7597) & 0.3862 (0.6132) & 0.7240 (0.2687)\cr
    &       RegNet & 0.5360 (0.4640) & 0.2252 (0.7744) & 0.4206 (0.5794) & 0.7366 (0.2632)\cr
    &       CSPNet & 0.6745 (0.3255) & 0.3151 (0.6848) & 0.5169 (0.4831) & 0.7434 (0.2566)\cr
    &          BiT & 0.6165 (0.3835) & 0.2417 (0.7581) & 0.4274 (0.5726) & 0.7401 (0.2599)\cr
    &          DLA & 0.6245 (0.3755) & 0.3049 (0.6948) & 0.4517 (0.5483) & 0.7411 (0.2589)\cr
\midrule    
    \multirow{10}{*}{\rotatebox{90}{NAS-searched}}
    &   DARTS-V1 & 0.7043 (0.0979) & 0.4071 (0.5929) & 0.5917 (0.0911) & 0.7663 (0.0272)\cr
    &   DARTS-V2 & 0.7028 (0.0962) & 0.3895 (0.6105) & 0.6123 (0.3877) & 0.7704 (0.0777)\cr
    &       ENAS & 0.6343 (0.0270) & 0.3895 (0.6057) & 0.3726 (0.0302) & 0.7664 (0.2331)\cr
    &       GDAS & 0.6621 (0.0657) & 0.4111 (0.1956) & 0.6357 (0.3154) & 0.7639 (0.2311)\cr
    &       SETN & 0.7108 (0.0765) & 0.4107 (0.1991) & 0.5206 (0.0412) & 0.7672 (0.2278)\cr
    &     Random & 0.8112 (0.1886) & 0.4313 (0.3671) & 0.6452 (0.3517) & 0.7659 (0.2032)\cr
    &      TENAS & 0.7937 (0.2063) & 0.4353 (0.5645) & 0.5754 (0.4246) & 0.7559 (0.2403)\cr
    &      DrNAS & 0.7768 (0.2232) & 0.4224 (0.5774) & 0.6025 (0.3975) & 0.7625 (0.2013)\cr
    &   PC-DARTS & 0.7647 (0.2353) & 0.4273 (0.5726) & 0.6040 (0.3960) & 0.7612 (0.1726)\cr
    &     SDARTS & 0.7641 (0.2359) & 0.4410 (0.5485) & 0.6400 (0.3600) & 0.7727 (0.0790)\cr
\bottomrule    
    \end{tabular}
    }
\end{table}

\mypara{Model Performance}
The model performance and overfitting of both NAS-search and human-designed models on 4 benchmark datasets are shown in \autoref{tab:measure_test_acc}. 
We can observe that the performance of the architectures searched by NAS is comparable to or even better than that of the human-designed architectures. 
Our model performance results are consistent with the previous research from both the ML community~\cite{ZVSL18, CH20} and Pang et al.~\cite{PXJLW22}.
The overfitting levels on the CIFAR100 and STL10 datasets are relatively high due to the fact that the test accuracy on these two datasets is much lower than that on the CIFAR10 and CelebA datasets, and the training accuracy on all four datasets can reach round 1.0 for most architectures under the same training settings.
We can see that, most architectures comply with the general rule that a higher test accuracy tends to lead to a lower overfitting level.

Moreover, we observe that the NAS-searched architectures usually have lower overfitting levels than the human-designed architectures on all four datasets, which leads us to expect the robustness of the NAS-searched architectures against MIAs should be better than that of the human-designed ones, since previous work~\cite{LWHSZBCFZ22} have shown that a higher overfitting level of the target model usually leads to a better MIA performance.
However, this conjecture conflicts with the observations of Pang et al.'s work~\cite{PXJLW22}, i.e., the NAS-searched architectures are more vulnerable to MIAs than the human-designed architectures, which motivates us to further explore the reason behind the conflicts and estimate the real privacy threats of various MIAs on the NAS-searched architectures.
In fact, Pang et al. ~\cite{PXJLW22} only consider the label-only attack scenario in which the attacker has only access to the output labels of the target model, and their data sampling method for testing the MIA performance in their released code implementation\footnote{\url{https://github.com/ain-soph/autovul/blob/main/projects/membership.py}} is actually biased to some data samples with extremely high confidence scores.
To test our conjecture and comprehensively study the MIA threats on NAS-searched architectures with reliable and unbiased experimental results, we further evaluate the MIA performance with the aforementioned 5 attack settings on both the NAS-searched and the human-designed architectures.

\begin{figure*}[!t]
\centering
\begin{subfigure}{0.6\columnwidth}
\includegraphics[width=\columnwidth]{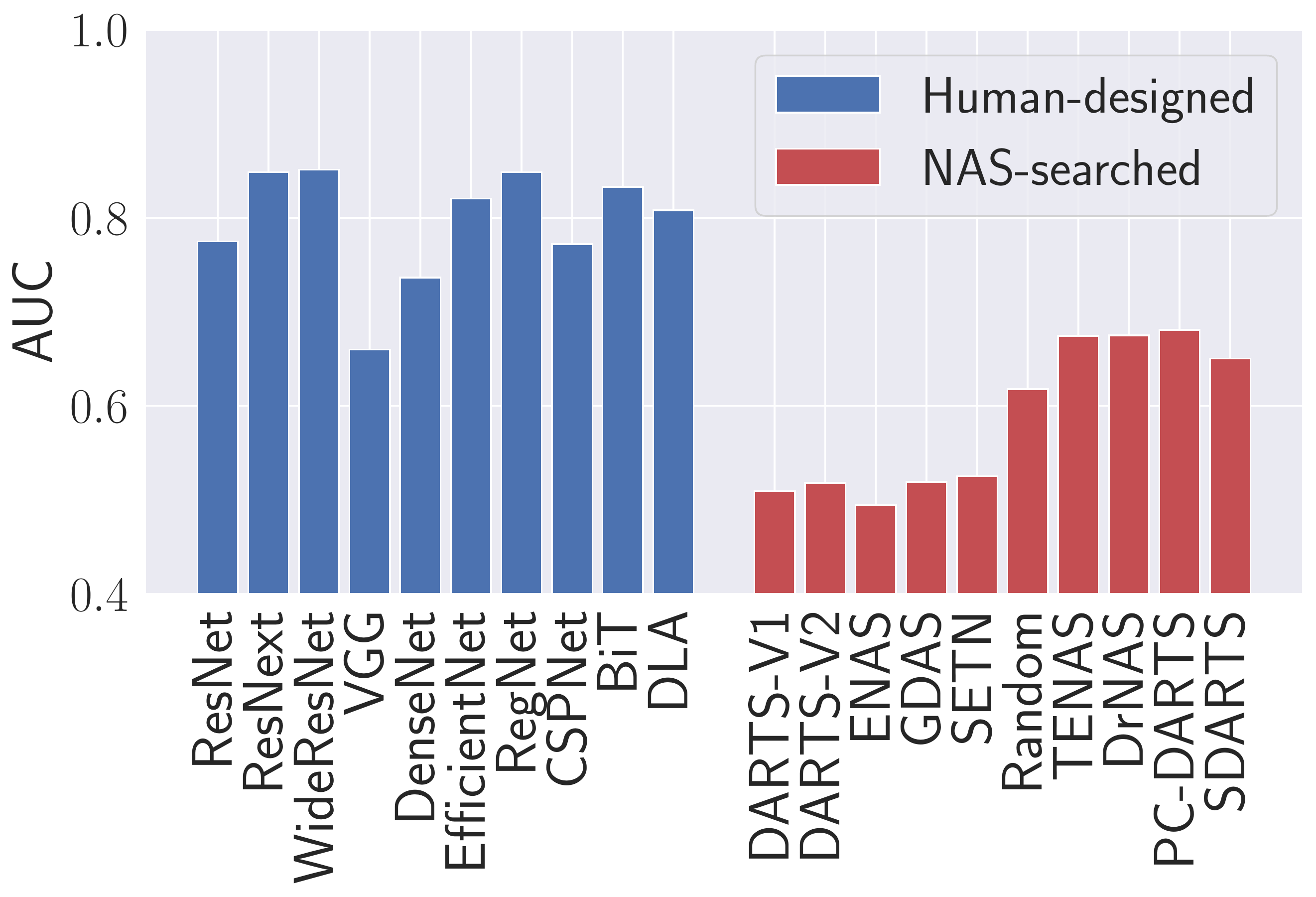}
\caption{CIFAR10}
\label{figure:label_only_cifar10}
\end{subfigure}
\begin{subfigure}{0.6\columnwidth}
\includegraphics[width=\columnwidth]{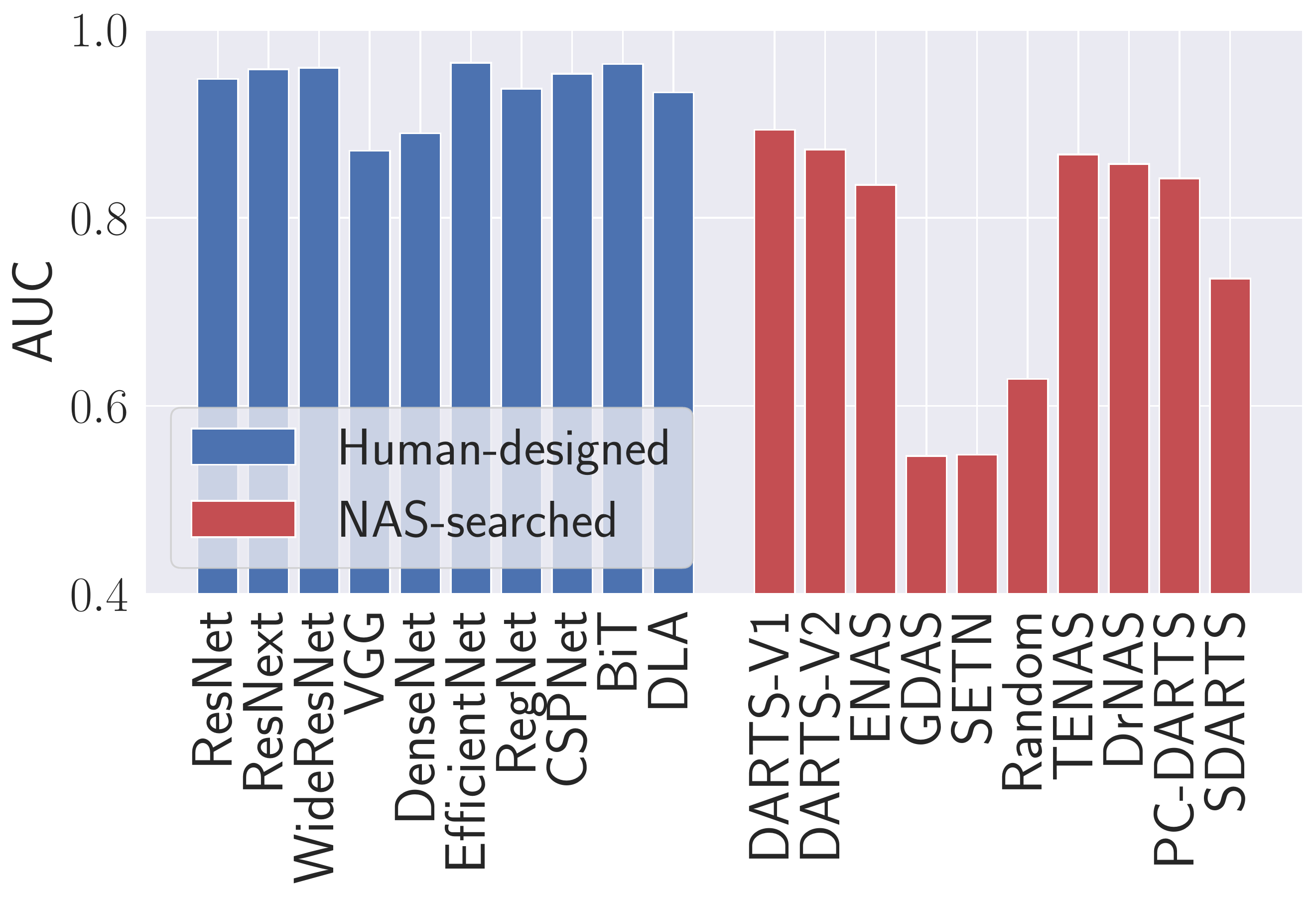}
\caption{CIFAR100}
\label{figure:label_only_cifar100}
\end{subfigure}
\begin{subfigure}{0.6\columnwidth}
\includegraphics[width=\columnwidth]{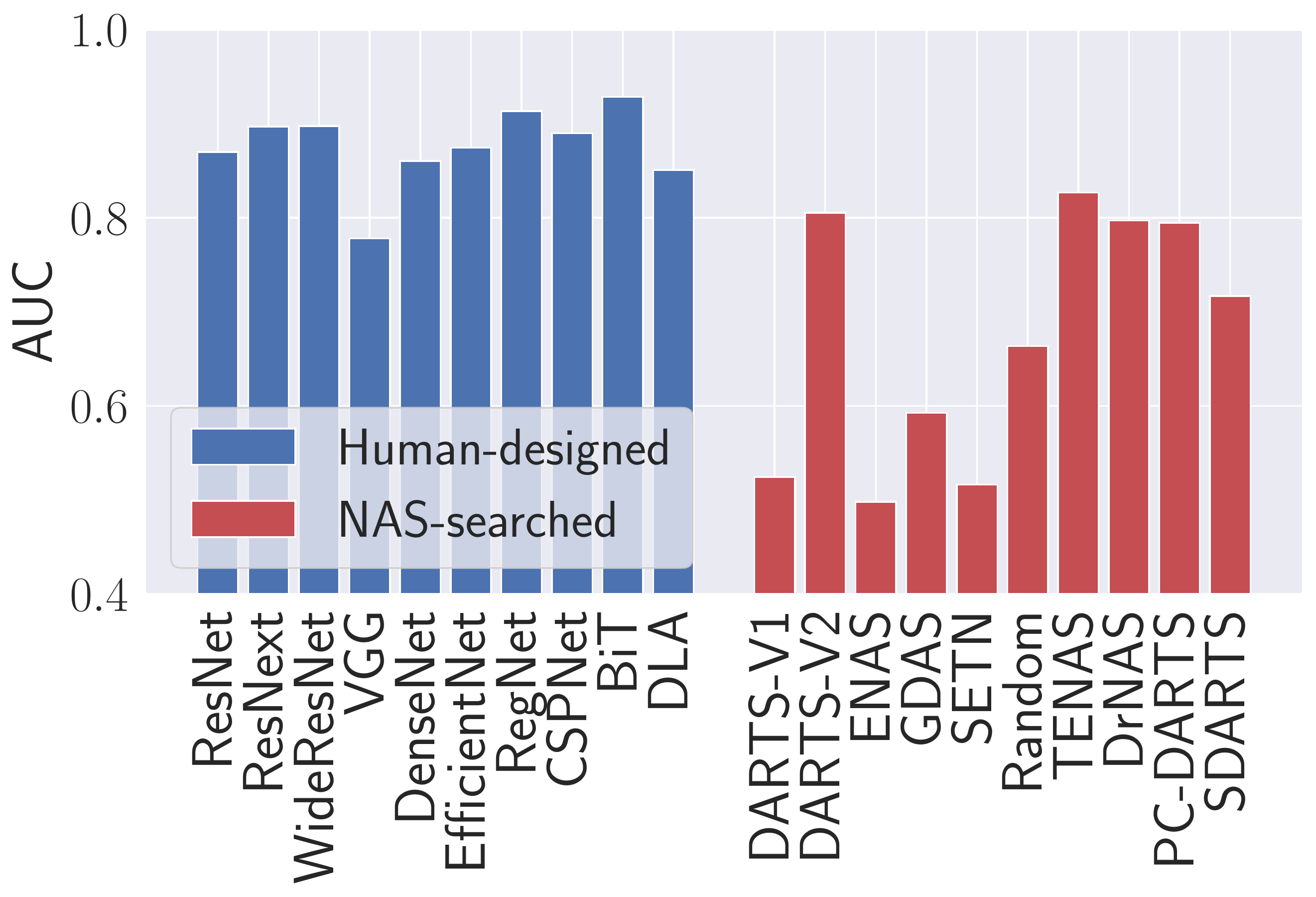}
\caption{STL10}
\label{figure:label_only_stl10}
\end{subfigure}
\caption{The performance of MIAs with the \tuple{Label\mbox{-}Only} setting on different datasets.}
\label{figure:label_only_mia_eval}
\end{figure*}

\begin{figure*}[!t]
\centering
\begin{subfigure}{0.6\columnwidth}
\includegraphics[width=\columnwidth]{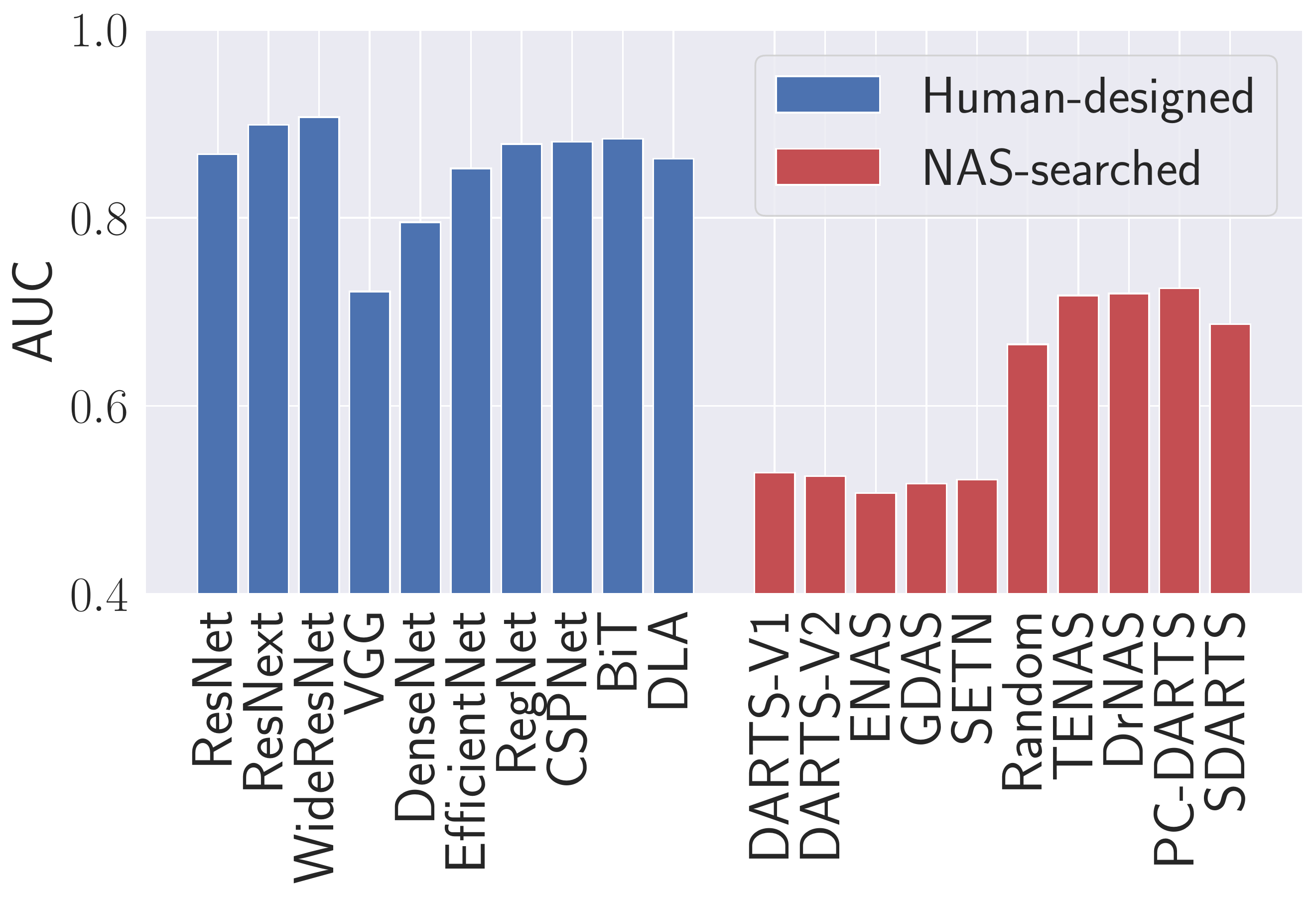}
\caption{CIFAR10}
\label{figure:blackbox_shadow_cifar10}
\end{subfigure}
\begin{subfigure}{0.6\columnwidth}
\includegraphics[width=\columnwidth]{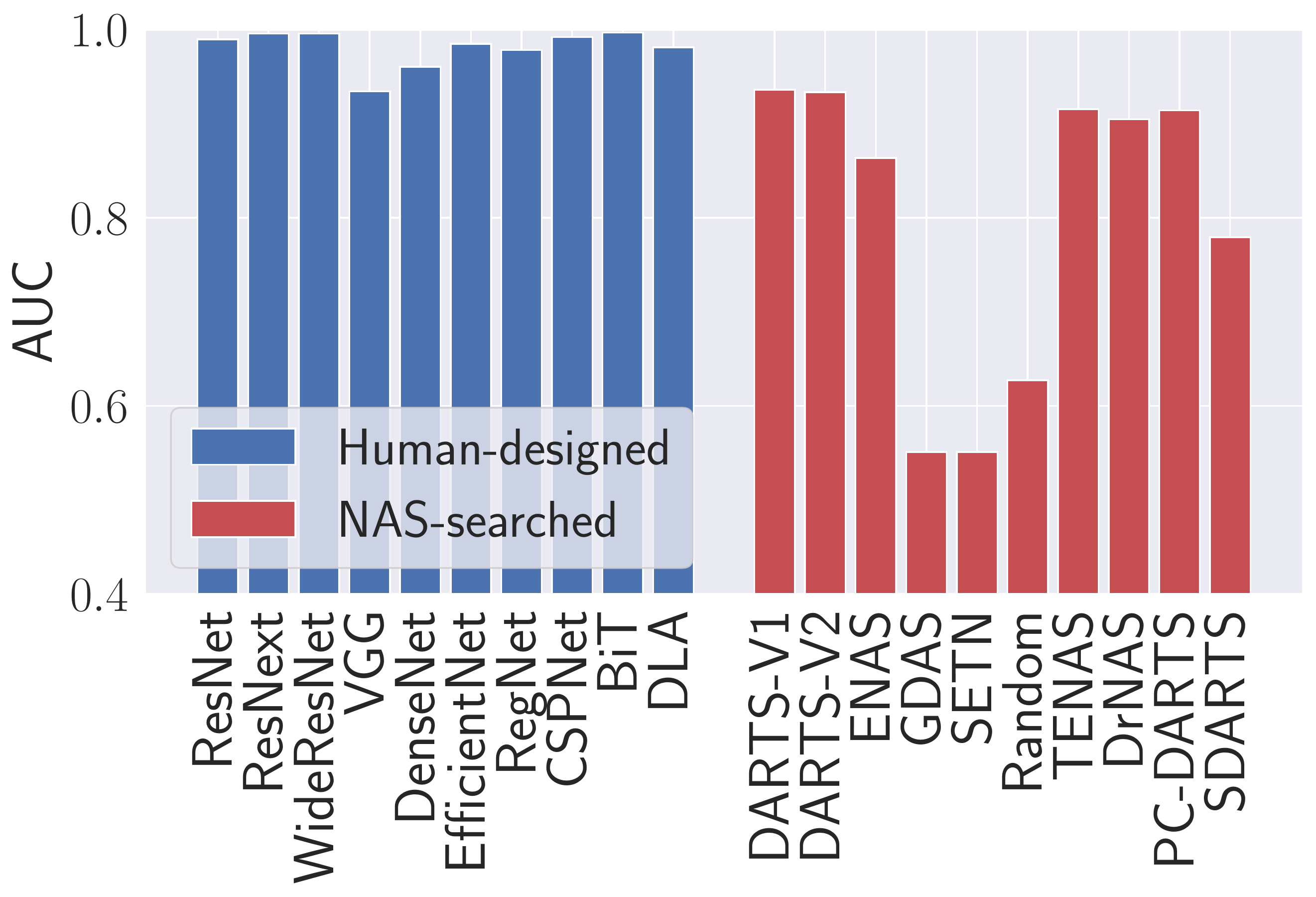}
\caption{CIFAR100}
\label{figure:blackbox_shadow_cifar100}
\end{subfigure}
\begin{subfigure}{0.6\columnwidth}
\includegraphics[width=\columnwidth]{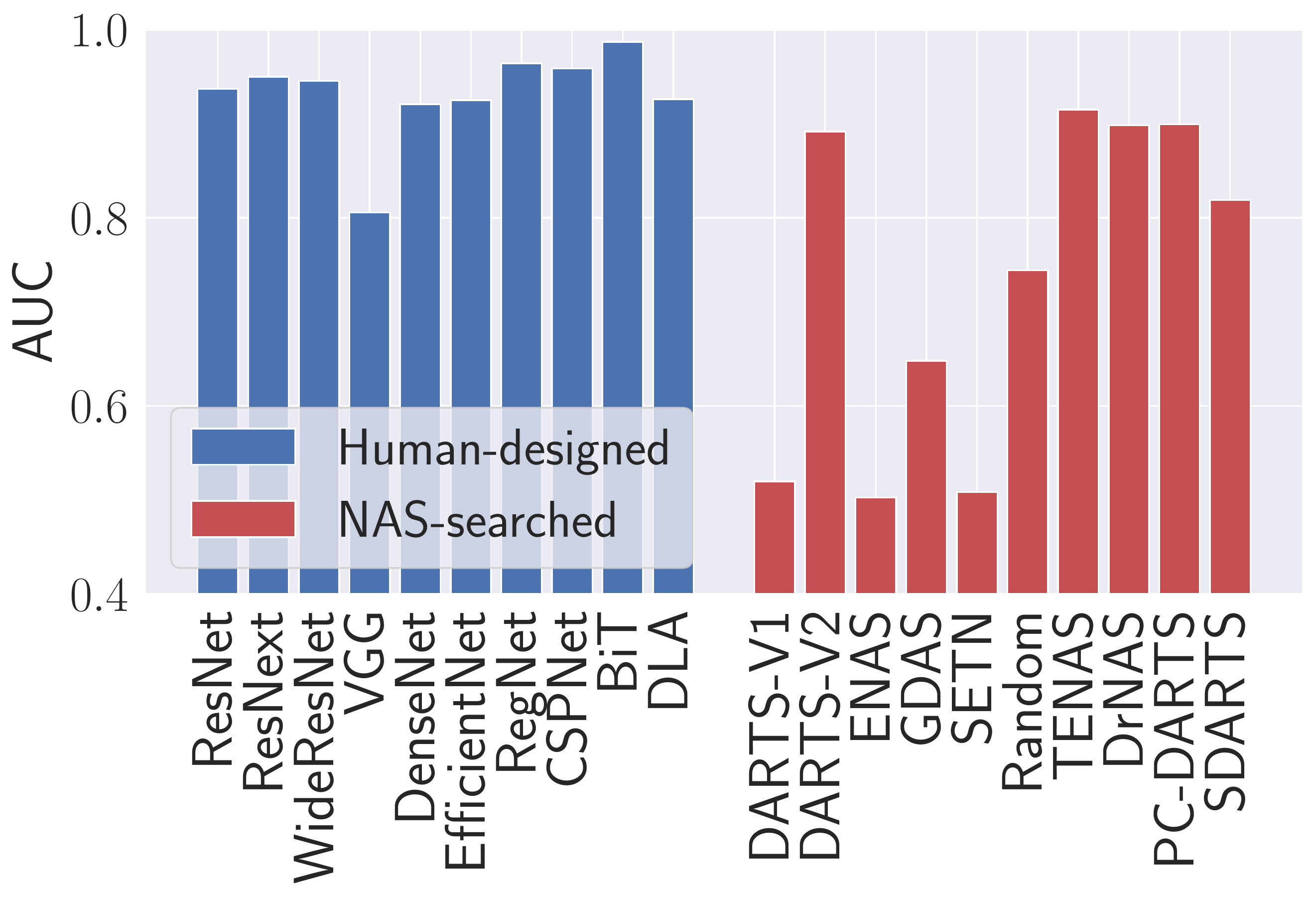}
\caption{STL10}
\label{figure:blackbox_shadow_stl10}
\end{subfigure}
\caption{The performance of MIAs with the \tuple{Black\mbox{-}Box, Shadow} setting on different datasets.}
\label{figure:blackbox_shadow_mia_eval}
\end{figure*}

\begin{figure*}[!t]
\centering
\begin{subfigure}{0.6\columnwidth}
\includegraphics[width=\columnwidth]{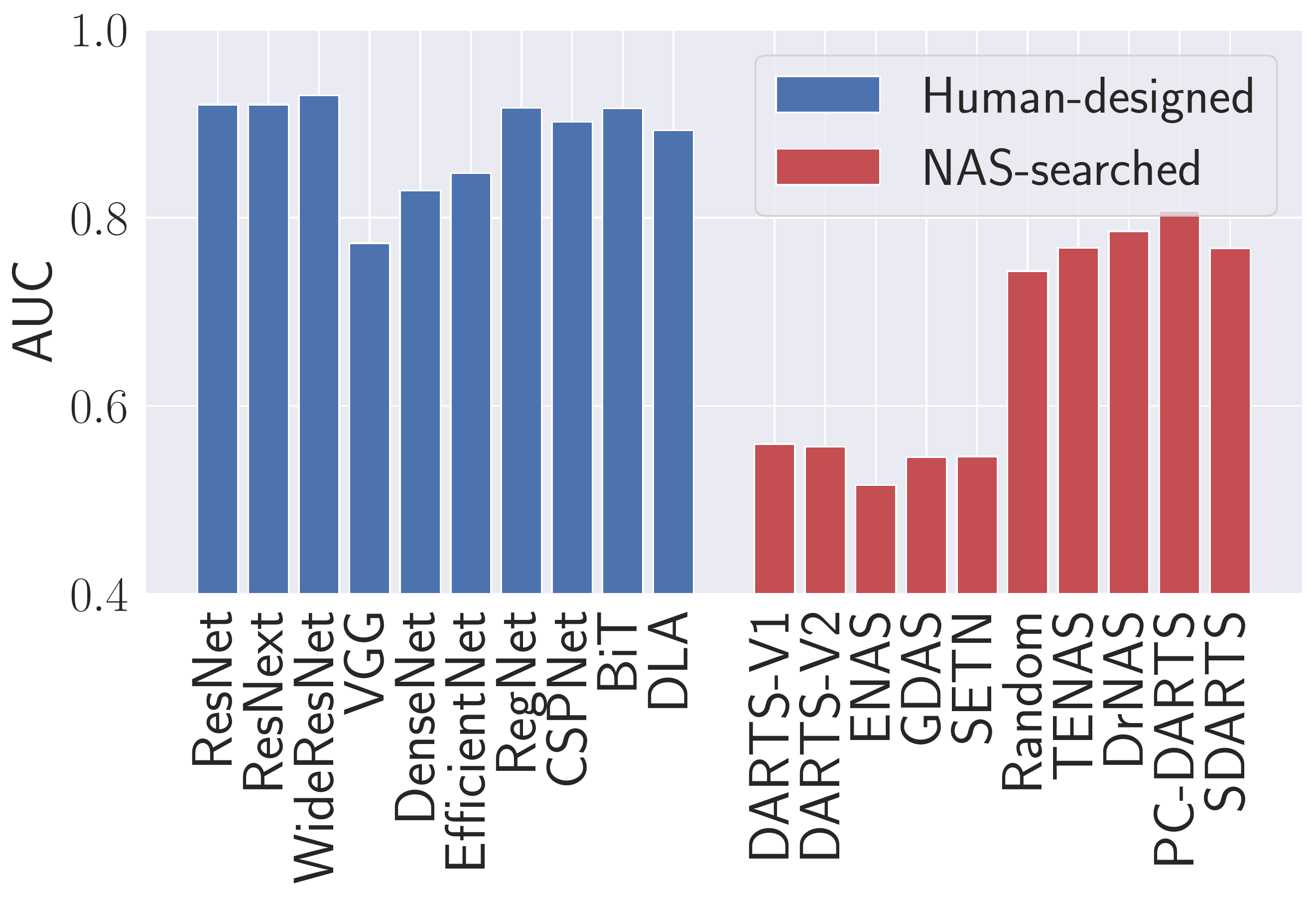}
\caption{CIFAR10}
\label{figure:whitebox_shadow_cifar10}
\end{subfigure}
\begin{subfigure}{0.6\columnwidth}
\includegraphics[width=\columnwidth]{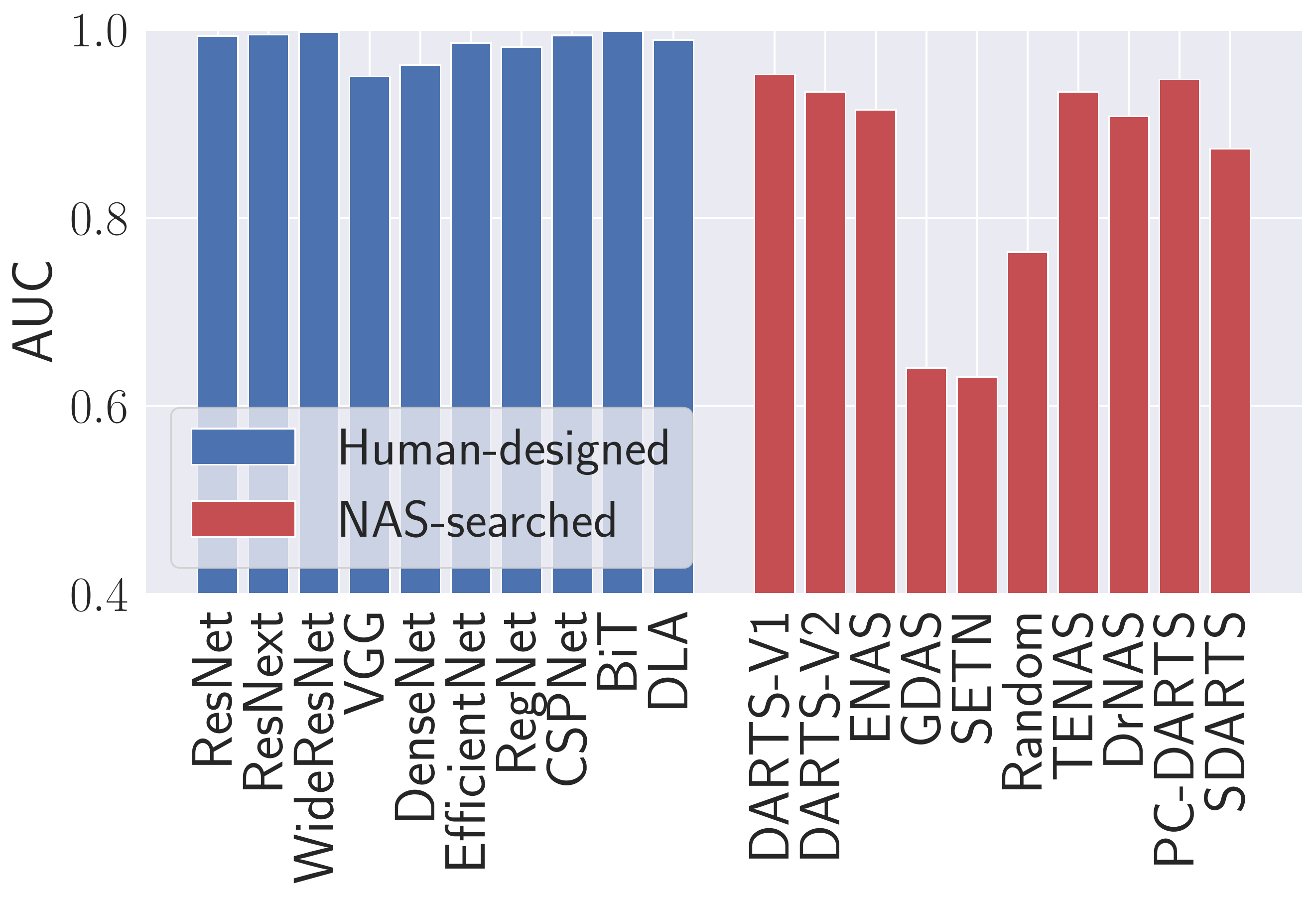}
\caption{CIFAR100}
\label{figure:whitebox_shadow_cifar100}
\end{subfigure}
\begin{subfigure}{0.6\columnwidth}
\includegraphics[width=\columnwidth]{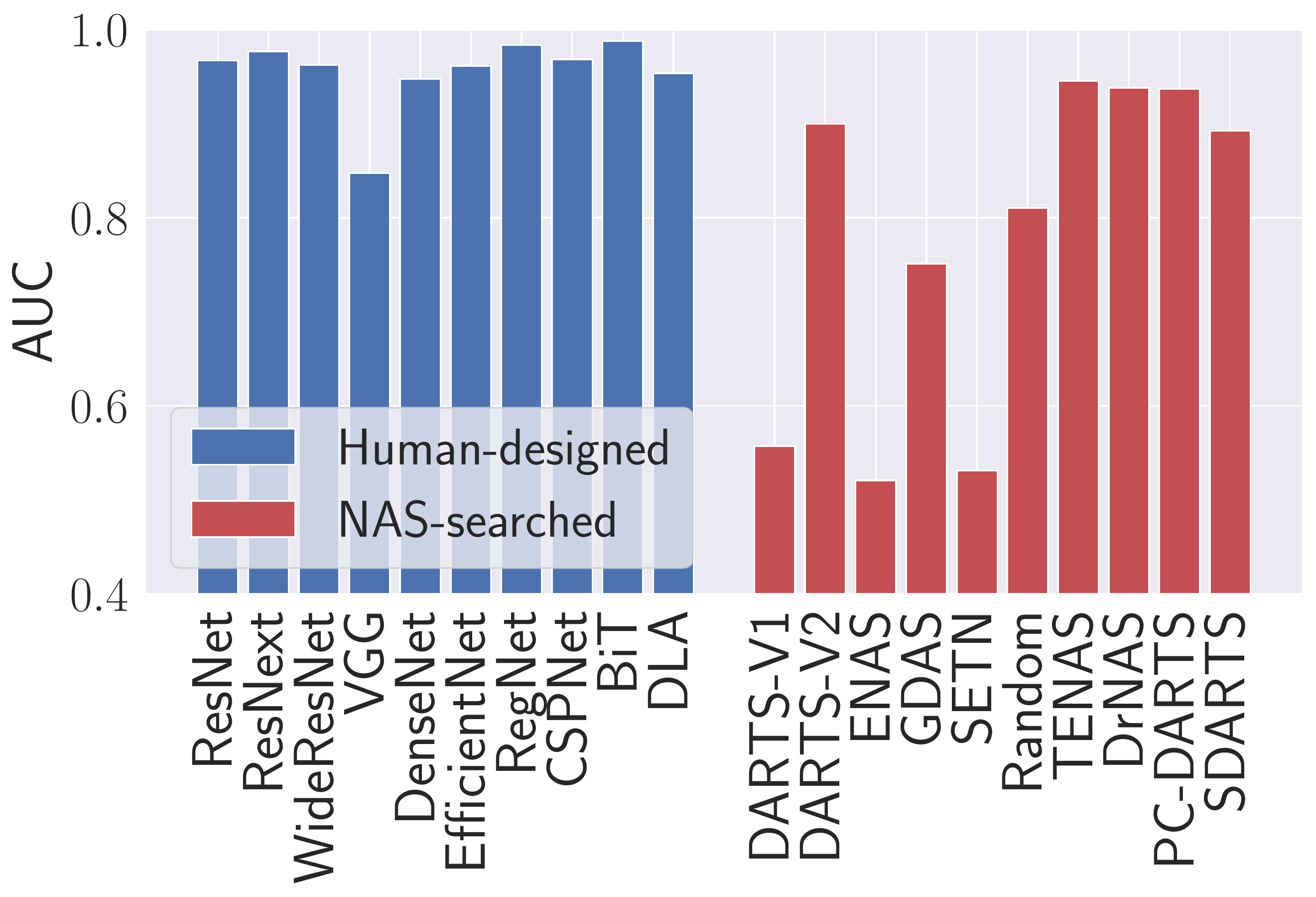}
\caption{STL10}
\label{figure:whitebox_shadow_stl10}
\end{subfigure}
\caption{The performance of MIAs with the \tuple{White\mbox{-}Box, Shadow} setting on different datasets.}
\label{figure:whitebox_shadow_mia_eval}
\end{figure*}

\mypara{Robustness Against Privacy Attacks}
We compare their experimental results under aforementioned 5 different attack settings. 
The experimental results of \tuple{Label\mbox{-}Only}, \tuple{Black\mbox{-}Box, Shadow}, and \tuple{White\mbox{-}Box, Shadow} on the CIFAR10, CIFAR100 and STL10 datasets are shown in 
\autoref{figure:label_only_mia_eval}, \autoref{figure:blackbox_shadow_mia_eval},
\autoref{figure:whitebox_shadow_mia_eval}, respectively.
Due to space limitation, we defer the results of the other two attack settings on these three datasets and all attack settings on the CelebA dataset to \autoref{app:measurement_results}.

First, we can observe that the MIAs usually have better performance under the white-box settings than the black-box settings. 
For instance, on the CIFAR10 dataset, the highest AUC score for MIAs on NAS architectures is around 0.81 under the \tuple{White\mbox{-}Box, Shadow} setting, while that under the \tuple{Black\mbox{-}Box, Shadow} is about 0.72, and that under the most constrained \tuple{Label\mbox{-}Only} settings is around 0.68. 
We believe that it is due to the abundant additional information (e.g., model weights, gradients) that the attackers extract from the white-box target model. 
Consequently, they can further manipulate such information to improve the performance of attacks. 
For example, the attacker can concatenate the gradients of hidden layers of the target white-box model with the output posteriors to serve as the input features for the attack model, which are more likely to enlarge the difference between members and non-members.

Second, given the same target dataset, the same target model tends to perform similarly in different attack settings. 
We can see that the architectures which obtain relatively high MIA AUC scores in one attack setting are still very likely to acquire relatively high MIA performance in other attack settings, and the histograms of different attack settings on the same dataset are quite similar. 
This interesting finding is also reasonable. Even though different attack settings might affect attack effectiveness, the majority of information needed by the attack is still offered by the output of the target model. 
Therefore, the same architecture tends to share similar robustness against MIAs regardless of the various attack settings.

More importantly, we find that the NAS-searched architectures tend to be more robust against various MIAs than the manual ones, which means the latter faces more serious privacy threats than the former. For example, the red bars are usually lower than the blue bars in \autoref{figure:label_only_mia_eval}.
This further validates our aforementioned conjecture.
However, we notice that our conclusion is contrary to the results from Pang et al.~\cite{PXJLW22}. 
The root cause of such divergence is due to different sampling strategies used to sample member and non-member data.
In our case, the distributions of member and non-member data have the same distribution as the original dataset.
Pang et al.~\cite{PXJLW22}, however, sample data points with high classification confidence scores from the original dataset per their official implementation.\footnote{Line 52 in \url{https://github.com/ain-soph/autovul/blob/main/projects/membership.py}}
Their sampling strategy cannot guarantee the distributions of sampled member and non-member data have the same distribution as the original dataset.

\begin{figure*}[!t]
\centering
\includegraphics[width=2\columnwidth]{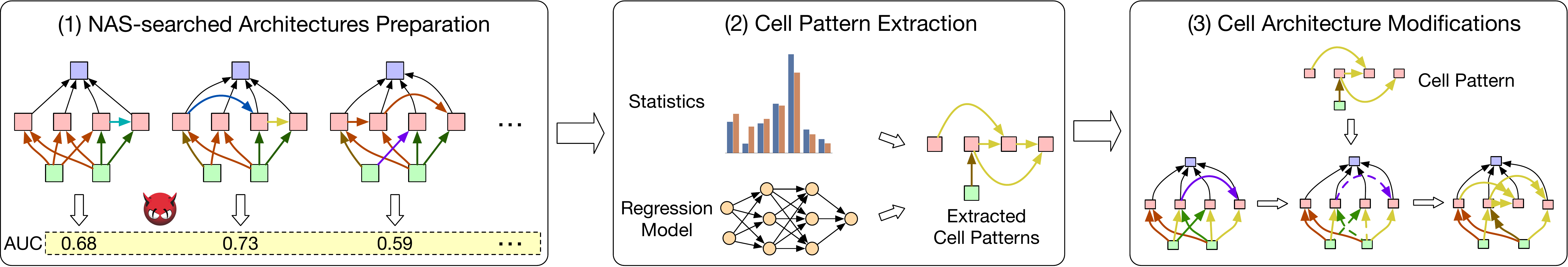}
\caption{The overview of our framework exploring the correlation between the NAS architectures and MIA performance.} 
\label{fig:framework}
\end{figure*}

% ----------------------------------------------------
\section{On Exploring Privacy-related Cell Patterns}
\label{sec:cell_pattern}
% ----------------------------------------------------

The empirical experimental results in \autoref{sec:measurement} show that the NAS architectures are usually more robust against MIAs; however, different NAS architectures still pose different levels of robustness against MIAs. 
Some NAS architectures (e.g., DARTS-V2) even appear to be more vulnerable to MIAs than the human-designed VGG on the STL10 dataset. 
Therefore, in this section, we aim to understand the relationship between the robustness against MIAs and the NAS architectures. 
Concretely, we seek to address the following questions:
(1) \emph{What are the common structures among existing NAS architectures which might impact MIAs?} 
(2) \emph{Can we find some existing architectural patterns to decrease the MIA risks of the NAS architectures when building NAS?} and
(3) \emph{If we do, can we maintain the model performance while decreasing the MIA risks?} 

% ----------------------------------------------------
\subsection{Overview}
\label{subsec:overview}
% ----------------------------------------------------

\autoref{fig:framework} illustrates the overall framework for analyzing and evaluating the correlation between the cell patterns and robustness against MIAs of NAS architectures, which consists of three steps.
\begin{itemize}
    \item \textbf{NAS-searched Architecture Preparation.}
    We first collect a large number of NAS-searched architectures with good model utility and evaluate the corresponding MIA performance on them for subsequent analysis.
    
    \item \textbf{Cell Pattern Extraction.}
    We then look deep into the internal cell structures of these NAS-searched architectures and propose a new method to extract common cell patterns that can promote or demote the MIA performance.
    
    \item \textbf{Cell Architecture Modification.}
    Finally, we use the extracted cell patterns to modify the internal cell structure of the target architecture to promote or demote the MIA performance on it.
\end{itemize}

% ----------------------------------------------------
\subsection{NAS-searched Architectures Preparation}
\label{subsec:dataset_preparation}
% ----------------------------------------------------

It is infeasible for us to train thousands of NAS-searched architectures from scratch.
Instead, we use NAS-Bench-301~\cite{SZZLKH20}, a large scale open-sourced benchmark dataset that contains 59,328 full-trained NAS-searched architectures identified by 17 representative NAS algorithms in a huge DARTS search space (i.e., $10^{18}$ possible architectures) on the CIFAR10 dataset.
We drop the redundant architectures which appear multiple times in NAS-Bench-301 and obtain a collection of NAS-searched architectures consisting of 53,558 unique architectures.

As we can see in \autoref{tab:measure_test_acc}, NAS-searched architectures with high test accuracy tend to have low overfitting levels.
Their cell patterns are more likely to have better robustness against MIAs while maintaining good model performance at the same time.
Besides, in the real world, the end users tend to choose architectures with high model performance.
As such, we select the top $5\%$ (i.e., 2,678) architectures with the highest test accuracy scores from NAS-Bench-301.
We then evaluate the MIA effectiveness under the most powerful attack setting, i.e., \tuple{White\mbox{-}Box, Partial} (see \autoref{subsec:measurement_setting} for details) for each sampled architecture to constitute an \emph{Architecture-to-MIA} dataset containing the well-performed architectures and their corresponding MIA AUC scores.
The MIA AUC scores in this new dataset range from 0.7311 to 0.8773.
We group the architectures with MIA AUC scores higher than 0.84 and lower than 0.78 into \textit{high} (denoted as $\mathcal{A}_\mathrm{high}$) and \textit{low} (denoted as $\mathcal{A}_\mathrm{low}$) MIA categories.
In total, we have 297 $\mathcal{A}_\mathrm{high}$ and 303 $\mathcal{A}_\mathrm{low}$ NAS-searched architectures for our cell pattern analysis. 

% ----------------------------------------------------
\subsection{Cell Pattern Extraction}
\label{subsec:cell_pattern_extraction}
% ----------------------------------------------------

To get a deeper understanding of the cell structures preferred by the robust or vulnerable NAS architectures, we look into the internal cell structures and try to extract some common cell patterns for specific objectives (e.g., demote the performance of MIAs). 
Here we take the $\mathcal{A}_\mathrm{high}$ and $\mathcal{A}_\mathrm{low}$ architectures for cell pattern extraction since they are the most vulnerable and robust architectures respectively in our constructed \emph{Architecture-to-MIA} dataset.

\begin{figure}[!t]
\centering
\begin{subfigure}{0.48\columnwidth}
\includegraphics[width=\columnwidth]{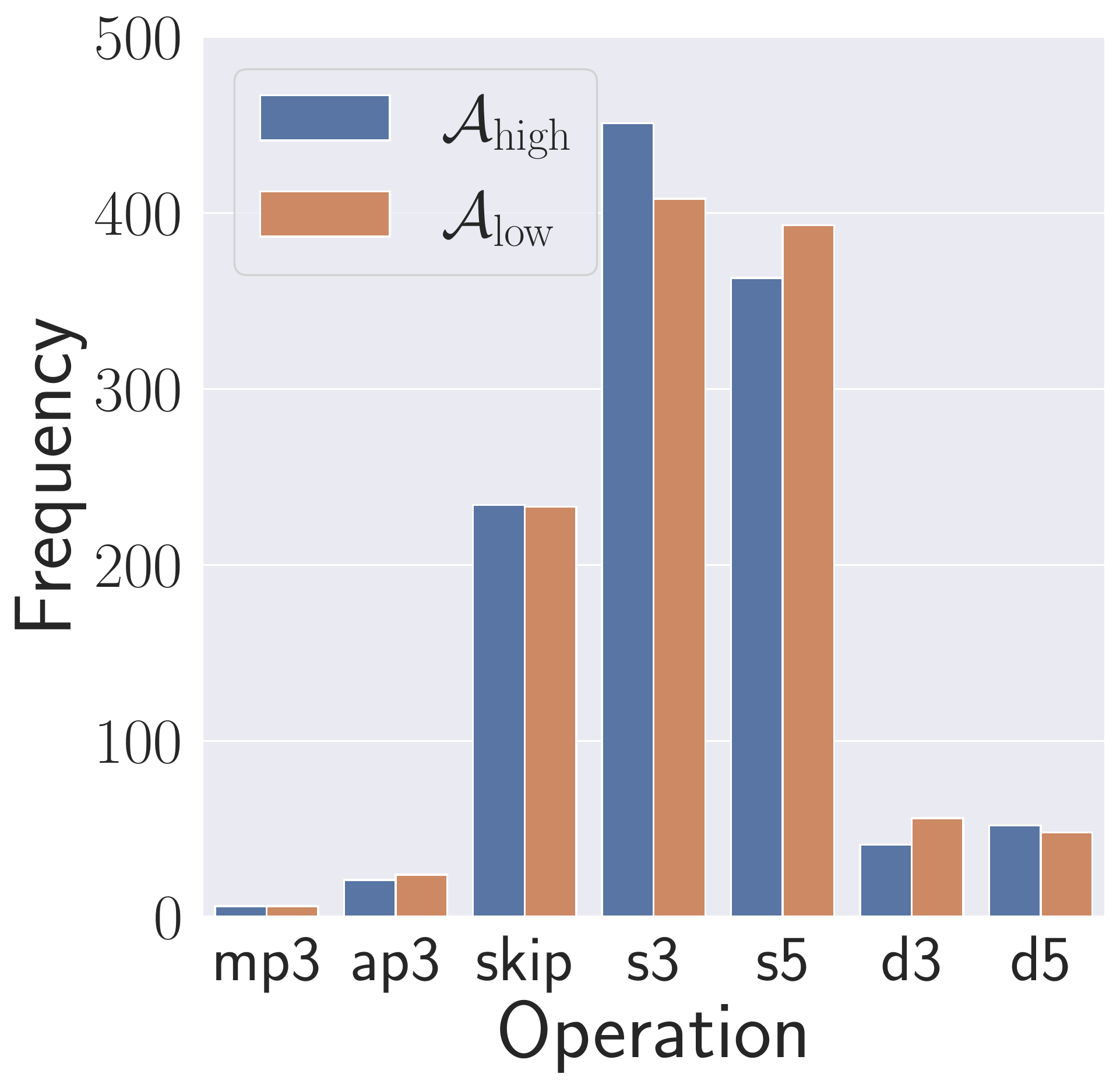}
\caption{Normal}
\label{figure:normal_cell_op_stats}
\end{subfigure}
\begin{subfigure}{0.48\columnwidth}
\includegraphics[width=\columnwidth]{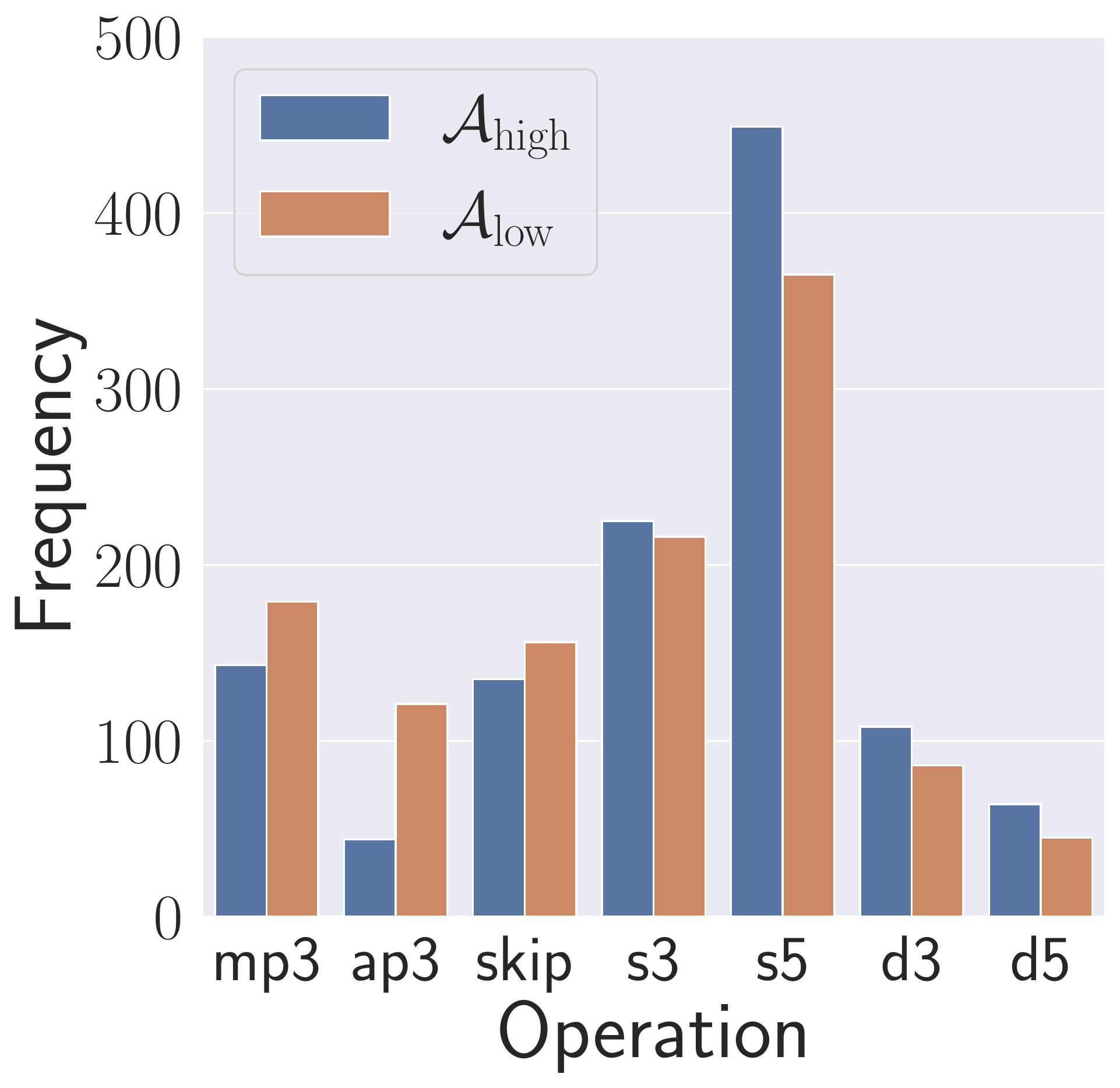}
\caption{Reduction}
\label{figure:reduction_cell_op_stats}
\end{subfigure}
\caption{Distributions of various operations in the normal and reduction cells of both $\mathcal{A}_\mathrm{high}$ and $\mathcal{A}_\mathrm{low}$ architectures.}
\label{figure:cell_ops_stats}
\end{figure}

\mypara{Operation Distributions}
We first analyze the distributions of operations in the normal and reduction cells of these two types of architectures. 
We have an interesting finding that the convolution operations seem critical for model performance while the pooling operations are preferred to mitigate MIAs.
As shown in \autoref{figure:cell_ops_stats}, the \textit{separable convolutions} and \textit{skip connections} occupy the majority of operations in both normal and reduction cells regardless of the architecture type, which means that these operations are critical to model performance on original tasks.
However, we find some changes in the distribution of operations when the architecture type moves from $\mathcal{A}_\mathrm{high}$ to $\mathcal{A}_\mathrm{low}$, especially for specific operations in the reduction cells.
Furthermore, according to previous observations in~\cite{WREL22}, the reduction cell has a relatively small impact on the overall model performance which is dominated by normal cells.
And both the $\mathcal{A}_\mathrm{high}$ and $\mathcal{A}_\mathrm{low}$ architectures sampled by us have good model performance, so it is expected that the difference between the operation distributions in the normal cells of these two types of architectures is relatively small.
Therefore, we can observe that the change of the operation distributions in the reduction cells is more obvious than that in the normal cells.
The frequency of the average pooling $3\times 3$ operation $\mathsf{ap3}$ increases drastically by $175\%$ when the architecture type changes from $\mathcal{A}_\mathrm{high}$ to $\mathcal{A}_\mathrm{low}$.
In general, as for the reduction cells, the convolution operations are preferred by the $\mathcal{A}_\mathrm{high}$ architectures, while the pooling operations (especially the average pooling operation) are favored by the $\mathcal{A}_\mathrm{low}$ architectures.
When it comes to the normal cells, even though the operation distributions of both the $\mathcal{A}_\mathrm{high}$ and $\mathcal{A}_\mathrm{low}$ architectures are quite similar to retain model performance, we can still observe that the separable convolution $3\times 3$ operation $\mathsf{s3}$ is particularly favored by the $\mathcal{A}_\mathrm{high}$ architectures.

\mypara{Operation Importance}
The information obtained from merely the distributions of various operations is far from enough to represent the common cell patterns among the well-performed architectures since the NAS architectures might comply with specific topologies to guarantee model performance.
Therefore, we go even further to extract specific cell patterns containing both the topology and operation information. 
To construct a cell pattern with specific edges, we need to determine the importance of each edge. 
Following similar strategy in~\cite{WREL22}, we use \textit{Operation Importance} (OI) to measure the impact of specific operations with specific edges. 
For a NAS cell architecture $\alpha$, the directed edge $e_{(i,j)}$ starts from node $i$ to node $j$, we assume that the edge $e_{(i,j)}$ is assigned with operation ${o_t}$, then the OI metric for $o_t$ on edge $e_{(i,j)}$ is computed as follows: 
\begin{equation}
\label{equation:operation_importance}
    \mathrm{OI}(\alpha,e_{(i,j)}:=o_t)=\frac{\sum_{k=1}^{|\mathcal{N}(\alpha,e_{(i,j)}:=o_t)|}f(\alpha_{k})}{|\mathcal{N}(\alpha,e_{(i,j)}:=o_t)|}-f(\alpha),
\end{equation}
where $f(\alpha)$ stands for the MIA AUC score of NAS cell architecture $\alpha$ and $\mathcal{N}(\alpha,e_{(i,j)}:=o_t)$ represents the neighbor cell set of the original cell architecture $\alpha$. 

We regard a cell as a neighbor cell of the original cell architecture $\alpha$ with operation $o_t$ on edge $e_{(i,j)}$ when we either replace operation $o_t$ with another operation $o_p (o_p\neq o_t)$ or change the input node $i$ of edge $e_{(i,j)}$ to another preceding node $q (q\neq i)$ in the original cell architecture $\alpha$ to obtain the current cell. 
OI compares the average MIA performance on the neighbor cells with that on the original cell architecture.
A positive value indicates that the current operation in the current edge can mitigate the MIA threats, while a negative value means that the current operation can contribute to MIA performance. 
Therefore, the attacker prefers the operation-assigned edge with a small negative value, while the defender favors an edge with a large positive value.
Note that the MIA evaluation on one cell can be time-consuming (usually takes about 2 hours in our experiments).
Even if we consider only modifying a single edge in a cell, it may affect multiple neighbor cells.
In turn, iteratively evaluating each neighbor cell after re-training from scratch would be computationally prohibitive when there are many original cells for analysis.

\mypara{GIN-based Regression Model}
To speed up the computation process and make this method feasible in reality, we use the aforementioned \emph{Architecture-to-MIA} dataset to train a regression model to directly predict the MIA performance when the cell architecture is given.
The core idea is to build a regression model to achieve the speed-up.
The input of the regression model is the architecture, and the output data is its corresponding predicted MIA AUC score.
Here we randomly divide the \emph{Architecture-to-MIA} dataset into three parts with 80\%, 10\%, and 10\% as training, validation, and testing datasets, respectively.
We test multiple regression methods (i.e., SVR~\cite{DBKSV96}, Random Forest~\cite{H95}, LGBoost~\cite{KMFWCMYL17}, XGBoost~\cite{CG16}, BANANAS~\cite{WNS21}, and GIN~\cite{XHLJ19}) to select the model which can best fit the relationship between the NAS architectures and MIA AUC scores.
The experiments use the same parameter settings for these regression models as the publicly available implementation of NAS-Bench-301.\footnote{\url{https://github.com/automl/nasbench301.git}}

\begin{table}[!t]
\centering
\caption{Performance of different regression models trained on the constructed dataset.}
\label{table:regression_model}
\scalebox{0.62}{
\begin{tabular}{l | c | c}
\toprule
\multirow{2}{*}{Model} & \multicolumn{2}{c}{Metrics}\\ \cline{2-3}
&$R^2$ & SpearmanR\\
\midrule
SVR             &       0.1097 &        0.3055 \\
Random Forest   &       0.1021 &        0.3155 \\
LGBoost         &       0.0718 &        0.2783 \\
XGBoost         & {\bf 0.1332} &        0.3394 \\
BANANAS         &      -0.0586 &        0.2003 \\
GIN             &       0.0788 &   {\bf 0.3751}\\
\bottomrule
\end{tabular}
}
\end{table}

The experimental results are show in \autoref{table:regression_model}.
We use two metrics (i.e., $R^2$ and SpearmanR) to estimate the performance of different regression models.
$R^2$(coefficient of determination) measures how well the observed results are replicated by the model according to the proportion of the outcome variance successfully explained by the model. 
SpearmanR (Spearman rank-order correlation coefficient) measures the monotonicity of the relationship between the two datasets (i.e., the ground-truth AUC scores and the predicted AUC scores in our experiments).
Higher scores of $R^2$ are preferred and the best possible value of it is 1.0, while scores with higher absolute values are favored for SpearmanR whose best score is +1/-1.
We can observe that the XGBoost model obtains the best $R^2$ score while the GIN model achieves the best SpearmanR score.
Additionally, since the architecture of a NAS cell can be represented as a DAG, and GIN model is a powerful variation of Graph Neural Networks (GNNs) which can learn plentiful information from the architectures of graph data and have surpassed many traditional techniques on graph tasks, it is very reasonable to use GIN in this scenario.
As a result, we choose the GIN model as the regression model for further experiments and analysis.
Note that we use this GIN-based regression model $\widetilde{f}$ to replace $f$ in \autoref{equation:operation_importance}. 
In turn, we can immediately estimate the MIA AUC score after we get one specific cell architecture.

\mypara{Cell Pattern Categories}
According to the goals of different cell patterns, we divide the cell patterns into two categories, i.e., MIA \textit{promotion} and MIA \textit{demotion}, where the MIA promotion cell patterns aim to improve the performance of MIAs while the MIA demotion cell patterns attempt to mitigate MIA threats.
Take the MIA demotion cell patterns for example, we prefer to choose the edges with a large positive value.
Additionally, to accumulate the effectiveness of every single edge in the cell pattern, when constructing the cell pattern one by one edge, we ensure that the selected new edge is adjacent to the current cell pattern (i.e., having nodes in common). 
In this way, all selected edges are connected as a whole graph.
Besides, we make sure that the cell patterns comply with the DARTS cell construction rule, e.g., each intermediate node has at most two input edges.

\mypara{Extraction Strategy}
\label{para:extraction_strategy}
The extraction strategy is shown in \autoref{algorithm:cell_pattern_extraction}.
To focus on the operation-signed edges which are more common among sampled architectures, we only consider and compare the OI scores of those edges appearing more than 14 times in the candidate edge set $\mathcal{E}$ when we analyze the cell patterns for the normal or reduction cells of $\mathcal{A}_\mathrm{high}$ or $\mathcal{A}_\mathrm{low}$ architectures. 
Our extraction strategy consists of 4 steps: 
\begin{itemize}
    \item \textbf{Initialization.} 
    We separately compute the operation importance for the normal and reduction cells in $\mathcal{A}_\mathrm{high}$ or $\mathcal{A}_\mathrm{low}$ architectures, and follow \autoref{algorithm:intial_start}-\ref{algorithm:intial_end} to prepare for the extraction.
    
    \item \textbf{Edge Constraint Checking.} 
    We check the number of edges in the current cell pattern in \autoref{algorithm:edge_num_check} and will terminate the extraction process in advance using \autoref{algorithm:early_stop_break} if no more new edge is successfully added to the current cell pattern graph. 
    
    \item \textbf{Construction Rule Checking.} 
    We check whether the current edge is adjacent to the current cell pattern graph and also comply with the rule in the DARTS search space using \autoref{algorithm:construct_rule_check}. 
    
    \item \textbf{Operation Importance Checking.} 
    We check whether the operation importance score of the current edge meets our requirements using \autoref{algorithm:oi_check} and will update the current cell pattern and other data recorders using \autoref{algorithm:add_edge_start}-\ref{algorithm:add_edge_end} if this edge is successfully added.
\end{itemize}
Note that the NAS cell architectures in the NAS-Bench-301 dataset have 4 intermediate nodes and up to 8 edges in a cell can be modified; thus, we thereby set $L=8$.
We set the demotion flag $\delta$ to ``True'' to search for the MIA demotion cell patterns, and ``False'' otherwise to search for the MIA promotion cell patterns.

\begin{algorithm}[!t]
\SetKw{Break}{break}
\SetCommentSty{small}
\LinesNumbered

\caption{Cell Pattern Extraction}\label{algorithm:cell_pattern_extraction}
\KwIn{Candidate edge set $\mathcal{E}$, maximum number of edges in the cell pattern $L$, demotion flag $\delta$}
\KwOut{Cell pattern graph $G_p$}
Initialize the in-degree dictionary $D_{\mathrm{in}}$ with 0 for each node\;\label{algorithm:intial_start}
Initialize the existing cell pattern graph $G_p$, edge topology set $E_p$ and node set $V_p$ as empty sets\;
\If{$\delta$}{
	Sort $\mathcal{E}$ in descending order according to OI scores\;
}
\Else{
	Sort $\mathcal{E}$ in ascending order according to OI scores\;
}
Add the input node of the first edge $\mathcal{E}_{1}$ in $\mathcal{E}$ to $V_p$\;\label{algorithm:intial_end}
\While{ $| G_p | $ < $L$}{
\label{algorithm:edge_num_check}
	 $\epsilon\gets\mathrm{False}$\;
	 $N\gets$ $|\mathcal{E}|$\;
	 \For{$i\leftarrow 1$ \KwTo $N$}{
	 \tcp{Start node $u$, end node $v$, operation $o$}{$(u,v,o)\leftarrow \mathcal{E}_{i}$\;}
	 $\tau\leftarrow \mathrm{OI}((u,v,o))$\;
	 \If{$(u \in V_p$ {\bf or} $v \in V_p)$ {\bf and} $(u,v)\notin E_p$ {\bf and} $D_{\mathrm{in}}[v]<2$}{
	 \label{algorithm:construct_rule_check}
	 \If{$(\delta$ {\bf and} $\tau>0)$ {\bf or} $(${\bf not} $\delta$ {\bf and} $\tau<0)$}{\label{algorithm:oi_check}
	 $G_p\leftarrow G_p\cup (u,v,o)$\;\label{algorithm:add_edge_start}
	 $E_p\leftarrow E_p\cup (u,v)$\;
	 $V_p\leftarrow V_p\cup \{u, v\}$\;
	 $D_{\mathrm{in}}[v]\leftarrow D_{\mathrm{in}}[v]+1$\;
	 $\mathcal{E}\leftarrow\mathcal{E}\setminus (u,v,o)$\;
	 $\epsilon\leftarrow \mathrm{True}$\;
	 \Break\label{algorithm:add_edge_end}
	 }
	 }
	 }
	 \If{{\bf not} $\epsilon$}{
	 \Break\label{algorithm:early_stop_break}
	 }
}
\end{algorithm}

% ----------------------------------------------------
\subsection{Cell Architecture Modifications}
\label{subsec:cell_arch_modify}
% ----------------------------------------------------

Through the above efforts, we can extract the cell patterns from the \emph{Architecture-to-MIA} dataset. 
Based on these cell patterns, we could go further to modify the internal cell structure of the target architecture to promote or demote MIA performance on it.

\begin{figure*}[!t]
\centering
\includegraphics[width=1.83\columnwidth]{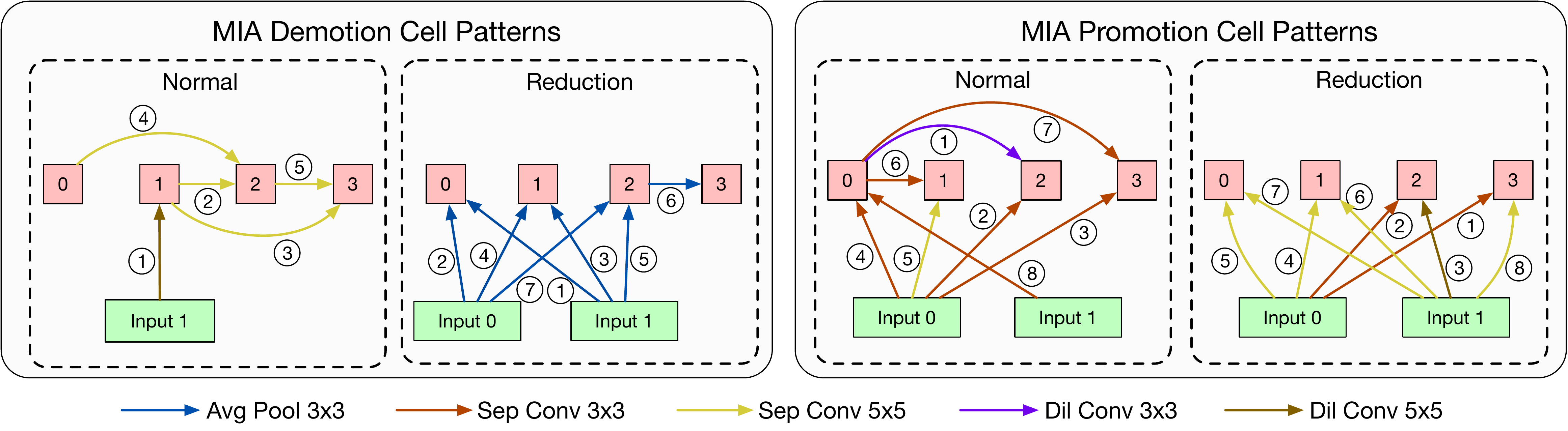}
\caption{The extracted cell patterns for both MIA demotion and promotion on both normal and reduction cells. The numbers in circles near edges indicate the order in which the corresponding edges were selected for the cell pattern in the cell. 
} 
\label{figure:full_cell_patterns}
\end{figure*}

\mypara{Extracted Patterns}
\autoref{figure:full_cell_patterns} illustrates the extracted patterns, including both MIA demotion and promotion cell patterns extracted from normal cells and reductions, respectively.
We have several interesting findings according to the extracted cell patterns. 
\begin{itemize}
    \item First, separable convolution and dilated separable convolution operations are preferred by the normal cells in both the MIA demotion and promotion cell patterns. As we have discussed before, this is because the model performance is mainly determined by the normal cells and some common operations are necessary for maintaining good model performance.
    
    \item Second, reduction cells prefer average pooling operations when demoting MIAs but favor convolution operations for MIA promotion. This is also consistent with our findings in the operation distributions of different architectures (see \autoref{subsec:cell_pattern_extraction}).
    
    \item Third, the edges in the MIA demotion cell pattern on normal cells are mainly connected between intermediate nodes, while those of other cell patterns are mainly connected to the input nodes.
    We speculate the reason is that MIA demotion cell patterns use this way to impede direct information transmission for normal cells.
    According to the statistical results in \autoref{figure:cell_ops_stats} and the extracted MIA promotion cell patterns in \autoref{figure:full_cell_patterns}, convolution operations seem to be beneficial to MIAs but the normal cells have to use them to retain good model performance even in the MIA demotion cell pattern.
    To weaken the negative impact of convolutions on privacy, the MIA demotion cell patterns prefer to place these operations in some positions not directly connected to the input nodes to avoid the information with high fidelity being leaked through these operations.
    On the contrary, max-pooling operations seem to be helpful to mitigate MIAs, so they are mainly directly connected between the input nodes and the intermediate nodes to protect the original information.
    As for the MIA promotion cell patterns, convolution operations are largely connected to the input nodes to facilitate extracting information with high fidelity.
\end{itemize}

\begin{figure}[!t]
\centering
\includegraphics[width=0.48\textwidth]{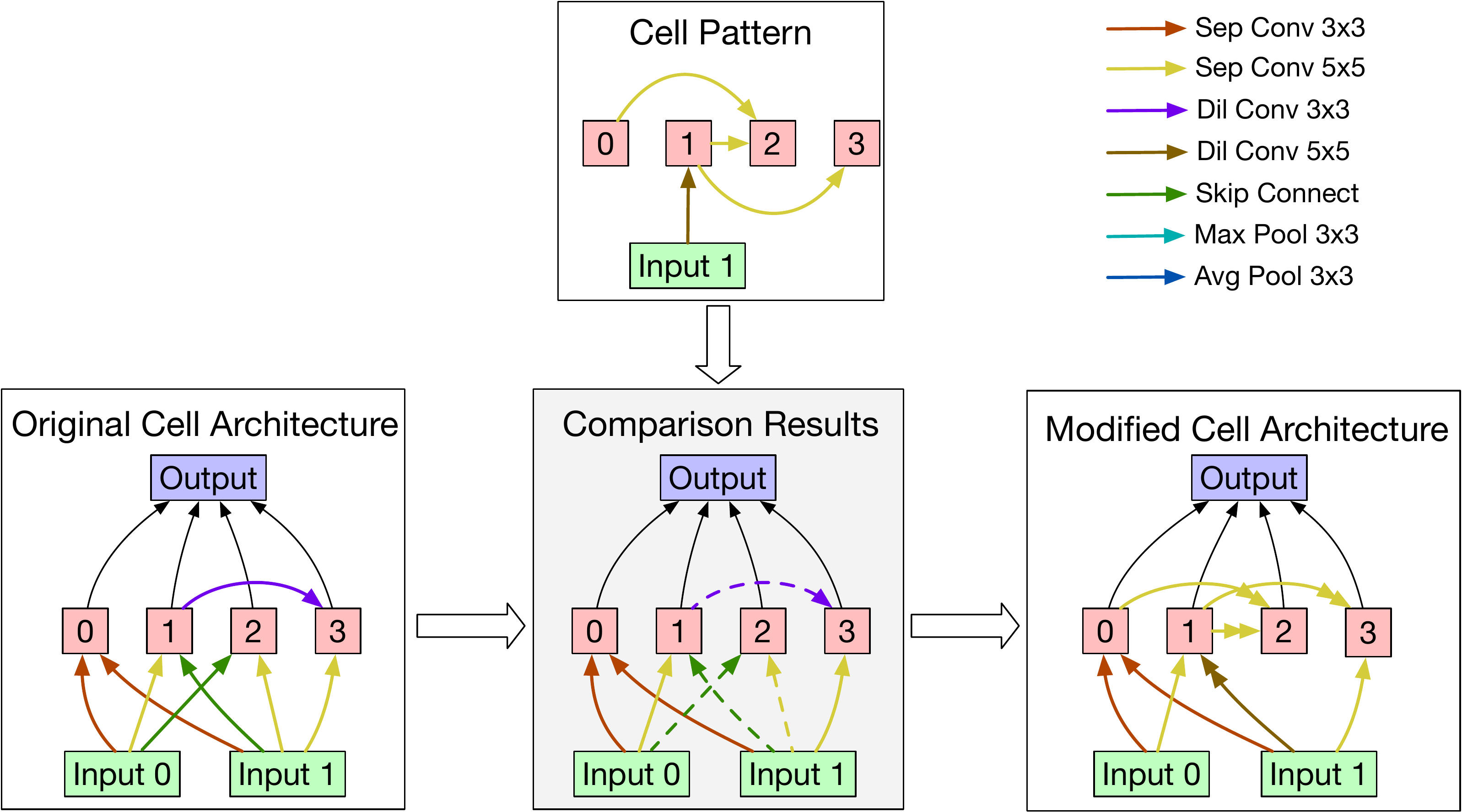}
\caption{An example of cell architecture modifications based on the cell pattern. The arrows with dashed lines represent candidate edges to be replaced, while the filled double arrows with solid lines stand for the newly added edges as modifications.} 
\label{fig:modification_illustration}
\end{figure}

\mypara{Cell Architecture Modifications}
Our architectural modifications to the target architectures work as follows.
First, according to our goal (e.g., promoting MIA performance on the target architecture) and the constraints on our modifications, we constitute corresponding cell patterns from \autoref{figure:full_cell_patterns} to guide modifications on specific types of cells. 
For example, when we want to achieve MIA demotion goal, and we are limited to perturb only the normal cell architectures and determine the structure in a cell for up to 4 edges, we select 4 edges from the MIA demotion cell pattern on normal cells following the sequence of the corresponding circled numbers (shown in the left side of \autoref{figure:full_cell_patterns}). 
We then get the required cell pattern as shown in the upper part of \autoref{fig:modification_illustration}.

Second, upon obtaining the cell pattern, we compare the internal structure of both the cell pattern and the normal cell of the target architecture.
If an edge $e_{(i,j)}$ with the assigned operation $o_{t}$ in the cell pattern does not exist in the target normal cell, we  replace another existing edge $e_{(k,j)}$ in the target normal cell with the same end node of $e_{(i,j)}$. 
If there already exists the edge $e_{(i,j)}$ but with a different operation $o_{t'}(o_{t'}\neq o_{t})$ in the target normal cell, we simply alter its operation $o_{t'}$ to $o_{t}$ later.
The arrows with dashed lines in \autoref{fig:modification_illustration} represent the original edges ready to be changed.   

Finally, we make the modifications to the replacement candidate edges, and replace these edges with the corresponding edges in the cell pattern.
The filled double arrows with solid lines in \autoref{fig:modification_illustration} stand for the newly added  edges to the target normal cell. 
After taking these three steps, the target normal cell is successfully modified to comply with the cell pattern.

% ----------------------------------------------------
\section{Cell Pattern Evaluation}
\label{sec:cell_pattern_evaluation}
% ----------------------------------------------------

% ----------------------------------------------------
\subsection{Experimental Setup}
\label{subsec:experimental_setup}
% ----------------------------------------------------

\mypara{Cell Pattern Modification Configuration}
We evaluate three categories of cell modifications, namely \textit{Only-Reduction}, \textit{Only-Normal} and \textit{Dual} modifications. 
\emph{Only-Reduction} modifications are exclusively made to the reduction cell of the target NAS architecture.
Similarly, \emph{Only-Normal} modifications are  made only to the normal cell of the target NAS architecture. 
\emph{Dual} modifications indicate that the modifications are made to both reduce and normal cells of the target architecture.
For all these modifications, we use the cell patterns identified in \autoref{figure:full_cell_patterns}.

\mypara{Modification Budget}
Note that the cell patterns extracted in \autoref{figure:full_cell_patterns} tend to include as many edges as possible to either demote or promote MIA performance.
However, in the real world, we may not be allowed to modify that many edges in a cell since it would significantly limit the space searchable by various NAS algorithms.
That is, the more edges determined by the cell patterns, the fewer edges useable by the NAS algorithm, since the overall amount of edges is fixed. 
For example, we have 8 edges in total for the DARTS cell (see \autoref{fig:cell_nas}).
If the cell pattern has 6 edges, a NAS algorithm has only 2 edges to search.
We thereby set a modification budget $m$ to limit the number of edges we can select from one cell pattern.
In turn, our final cell patterns used in our experiments contain at most the first $m$ edges from the full cell patterns as shown in \autoref{figure:full_cell_patterns}.
We use $m=\{3, 4, 5\}$ to evaluate the effectiveness of cell pattern modifications.

\mypara{Runtime Configuration}
Unless otherwise mentioned, we randomly sample 10 NAS-searched architecture instances from our \emph{Architecture-to-MIA} dataset as the target NAS architectures and average their results.
We set $m=4$ and conduct the experiments on the CIFAR10 dataset by default.

% ----------------------------------------------------
\subsection{Effectiveness of the Cell Patterns}
\label{subsec:effect_cell_pattern}
% ----------------------------------------------------

\begin{figure}[!t]
\centering
\begin{subfigure}{0.48\columnwidth}
\includegraphics[width=\columnwidth]{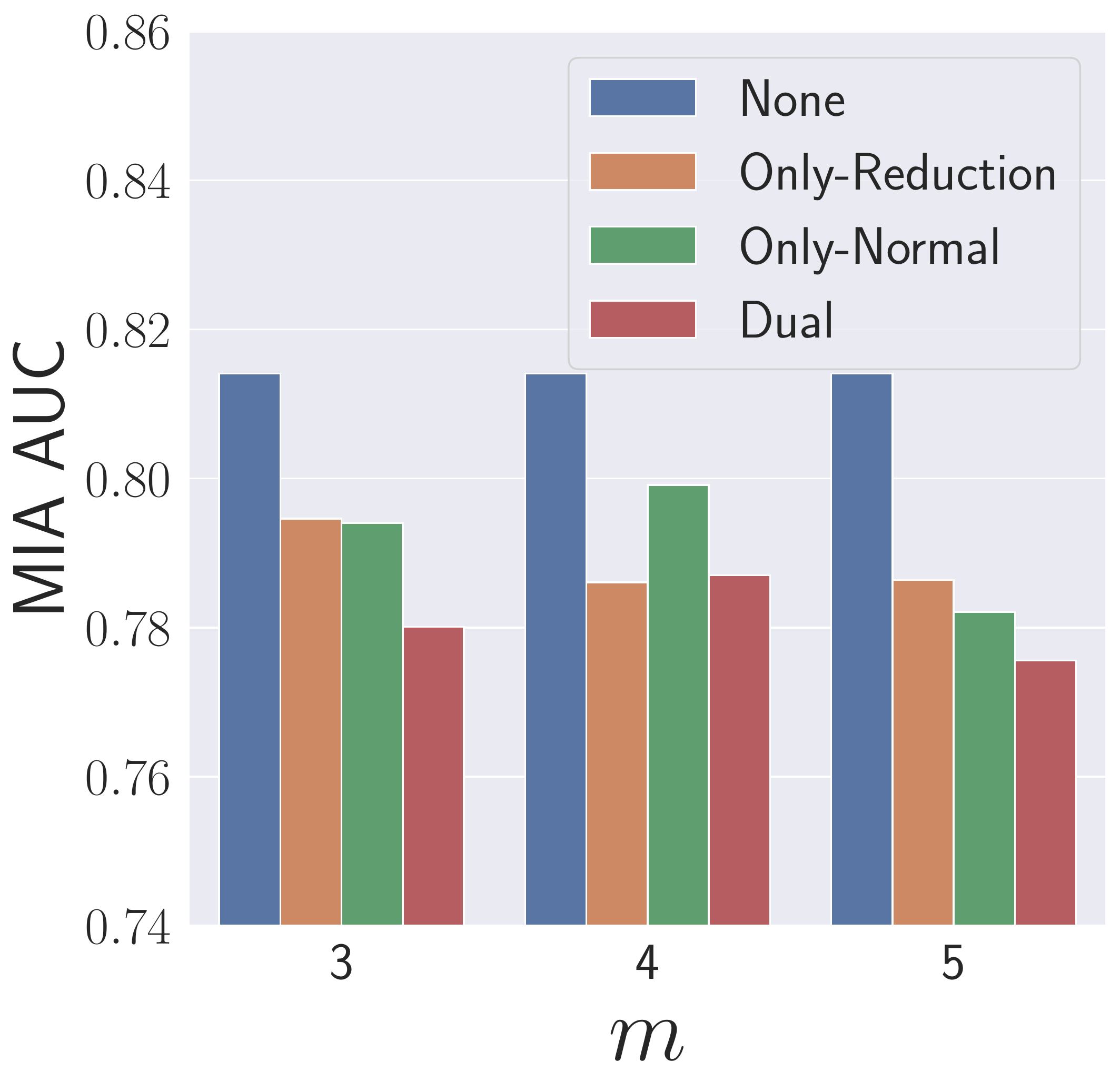}
\caption{Demotion}
\label{figure:mia_auc_demotion}
\end{subfigure}
\begin{subfigure}{0.48\columnwidth}
\includegraphics[width=\columnwidth]{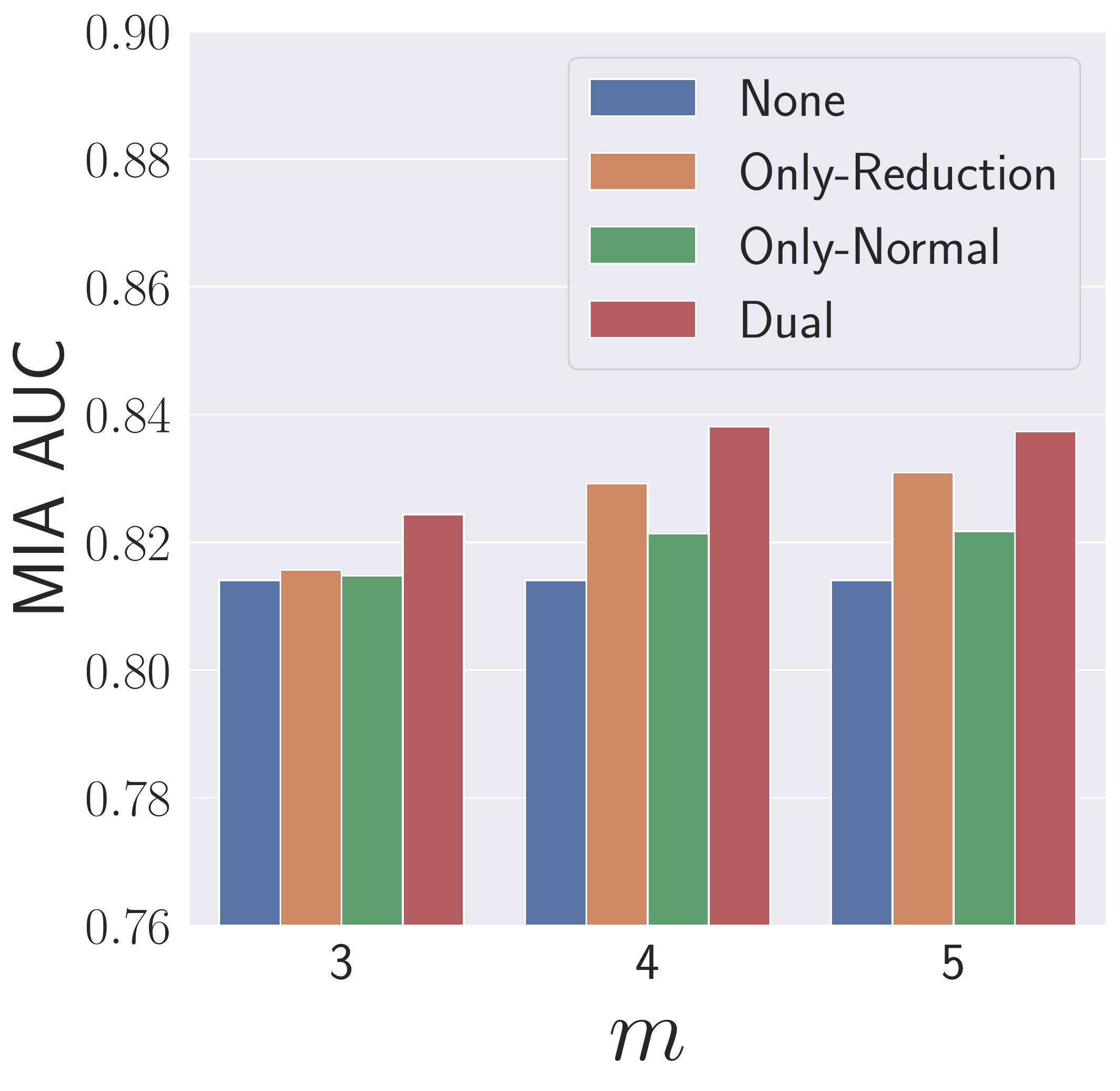}
\caption{Promotion}
\label{figure:mia_auc_promotion}
\end{subfigure}
\caption{MIA performance with various cell patterns under different cell pattern constraints. ``None'' means no architectural modification is applied to the target architectures.}
\label{figure:mia_auc_cell_pattern_eval}
\end{figure}

\begin{figure}[!t]
\centering
\begin{subfigure}{0.48\columnwidth}
\includegraphics[width=\columnwidth]{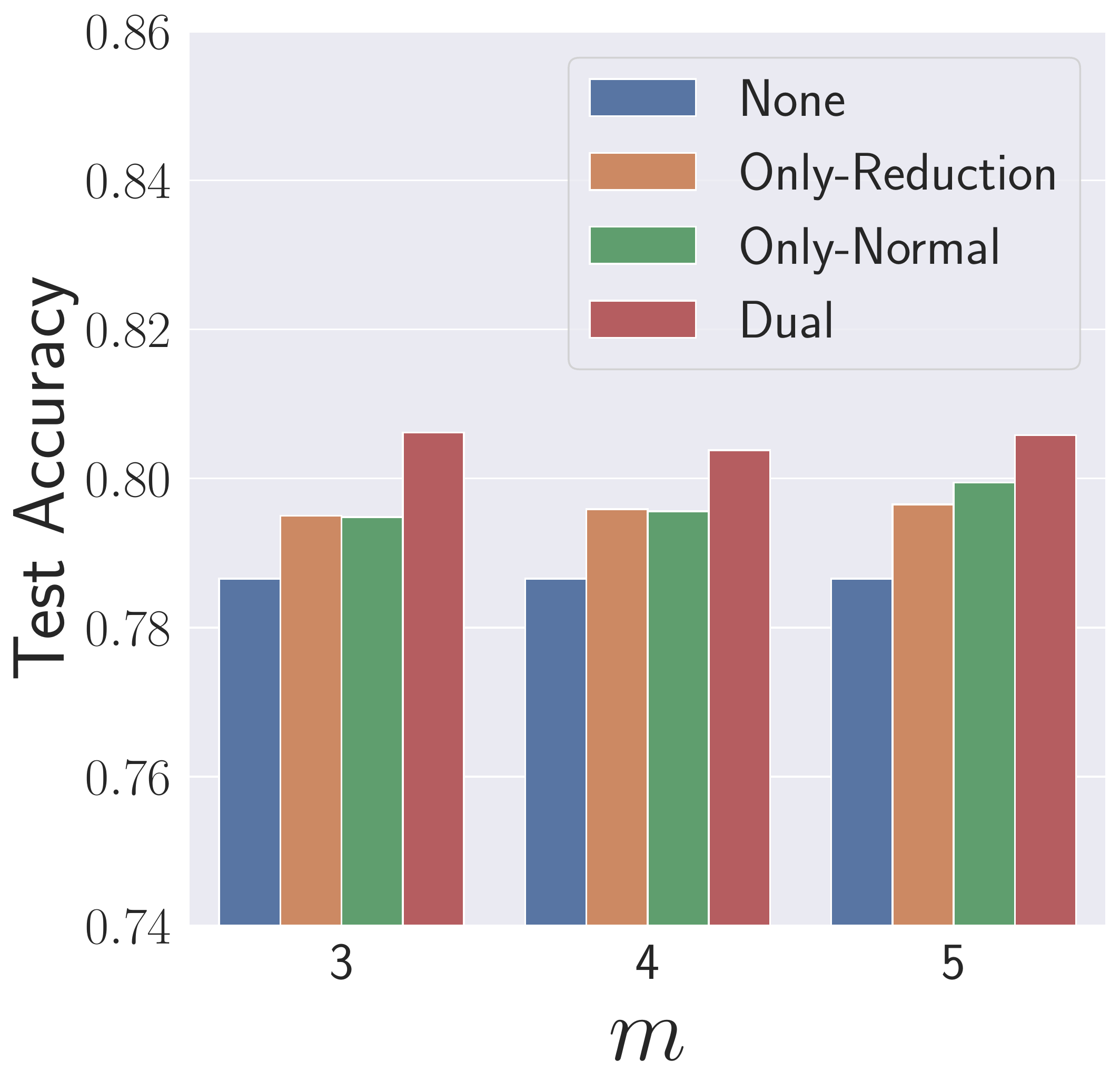}
\caption{Demotion}
\label{figure:test_acc_demotion}
\end{subfigure}
\begin{subfigure}{0.48\columnwidth}
\includegraphics[width=\columnwidth]{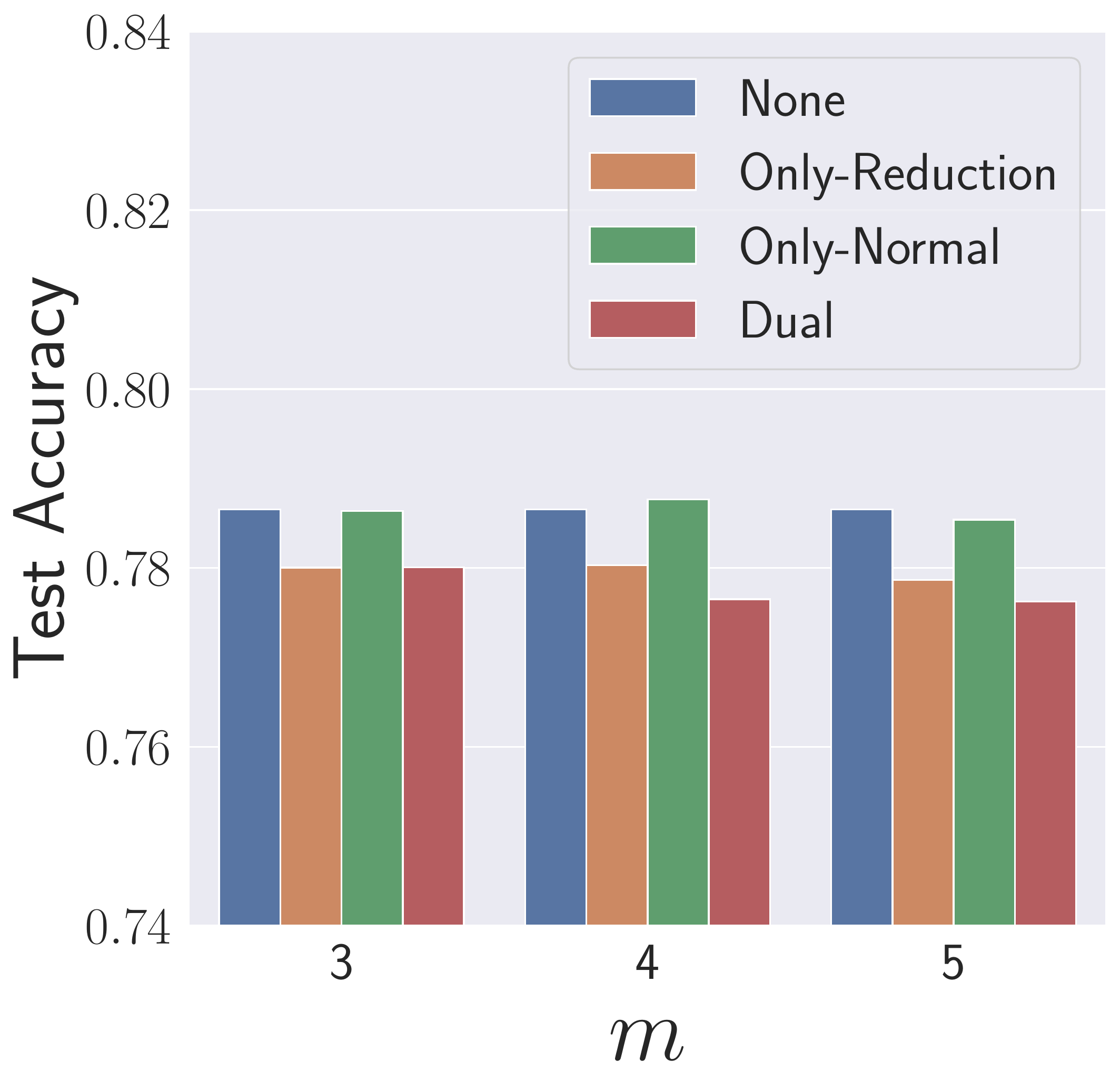}
\caption{Promotion}
\label{figure:test_acc_promotion}
\end{subfigure}
\caption{Model utility with various cell patterns under different cell pattern constraint values. 
``None'' means no architectural modifications are applied to the target architectures.}
\label{figure:test_acc_cell_pattern_eval}
\end{figure}

We evaluate the MIA performance under the \tuple{White\mbox{-}Box, Partial} setting on the CIFAR10 dataset.
The experimental results for both MIA performance and model utility (i.e., the test performance on the testing dataset for the target model) are shown in \autoref{figure:mia_auc_cell_pattern_eval} and \autoref{figure:test_acc_cell_pattern_eval} respectively. 

The first observation is that our cell patterns can successfully demote or promote MIAs as intended in all cases.
Take Only-Reduction modification and modification budget $m=4$ for instance, the average MIA AUC score of the target architectures drops from 0.8141 to 0.7861.
Though a small MIA AUC score decrease from the absolute value perspective, however, such drop can reduce the privacy risks since the real-world models usually requires a huge amount of user data to train.

\begin{figure}[!t]
\centering
\begin{subfigure}{0.48\columnwidth}
\includegraphics[width=\columnwidth]{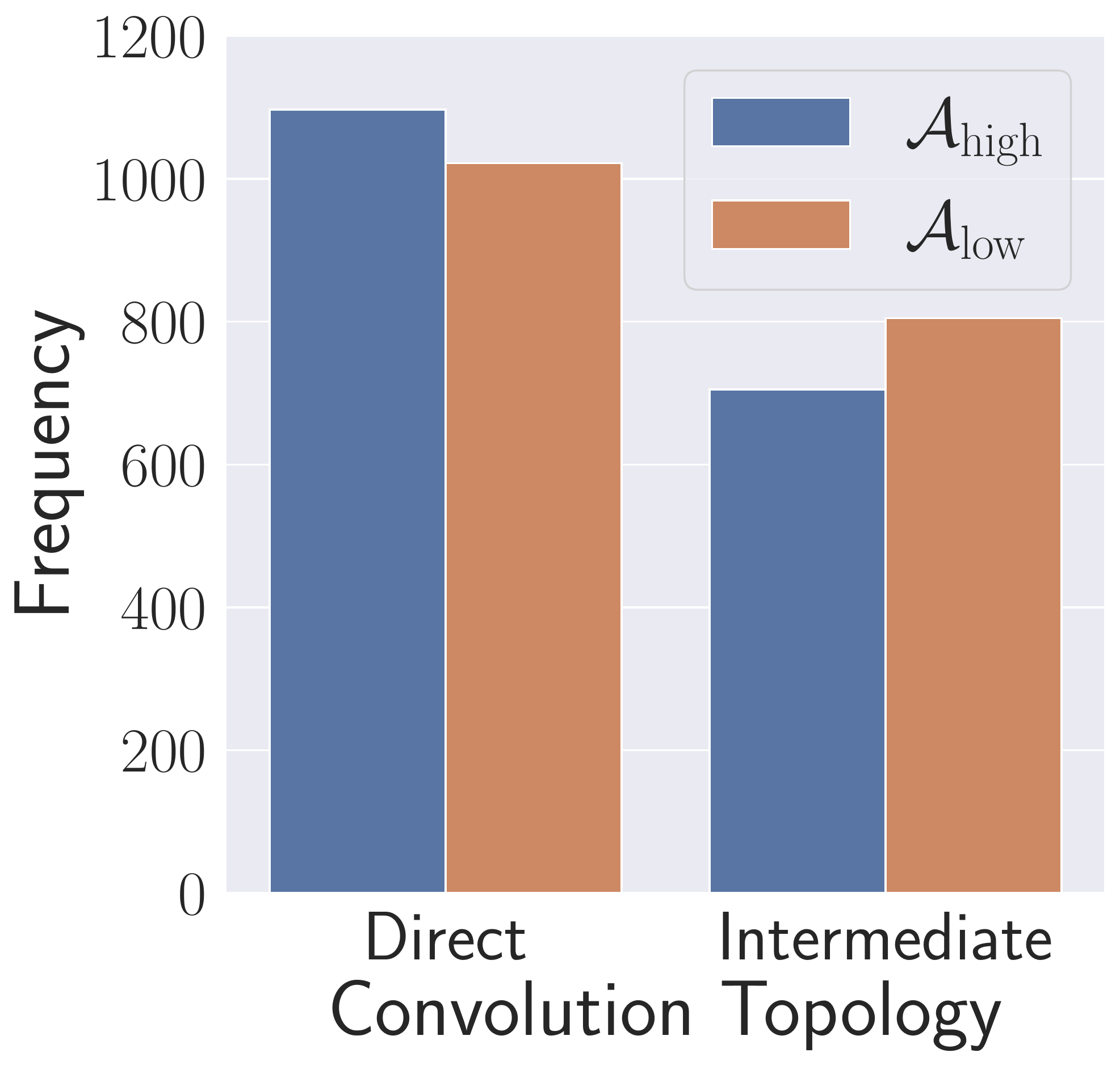}
\caption{Normal}
\label{figure:convs_stats_normal}
\end{subfigure}
\begin{subfigure}{0.48\columnwidth}
\includegraphics[width=\columnwidth]{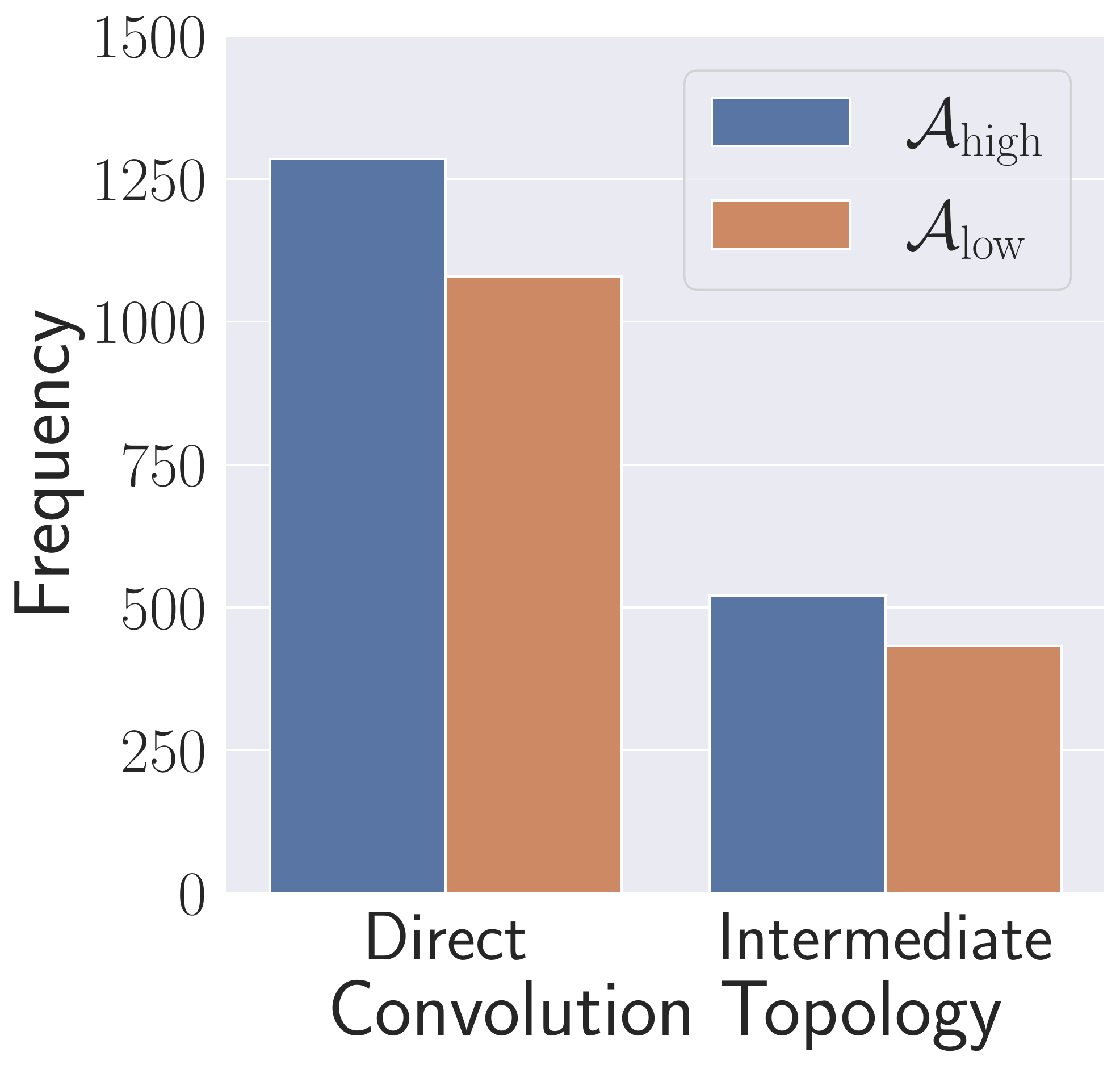}
\caption{Reduction}
\label{figure:convs_stats_reduce}
\end{subfigure}
\caption{Distributions of convolutions with two types of topology in the normal and reduction cells of both $\mathcal{A}_\mathrm{high}$ and $\mathcal{A}_\mathrm{low}$ architectures.}
\label{figure:convs_stats}
\end{figure}

Our second observation is that it is relatively easier to demote MIA AUC scores of NAS-searched architectures than to promote them.
Our hypothesis is that the convolution topology difference between $\mathcal{A}_\mathrm{low}$ and $\mathcal{A}_\mathrm{high}$ may contribute to this phenomenon.
To this end, we further count the number of the convolution operations with two different topologies --- the \emph{direct} topology where the convolution operations connect to the input nodes and the \emph{intermediate} topology where the convolution operations connect two intermediate nodes.
For instance, in the MIA demotion cell pattern on the reduction cells in \autoref{figure:full_cell_patterns}, the edge marked with the circled number 6 stands for the intermediate topology, while all the other edges belong to the direct topology.
The statistical results for convolution topologies are shown in \autoref{figure:convs_stats}. 
As we can see in \autoref{figure:convs_stats}, the frequency of \emph{direct} convolution topology is always higher than that of the \emph{intermediate} convolutions. 
And also, the normal cells in the $\mathcal{A}_\mathrm{low}$ architectures contain more \emph{intermediate} convolution topology than those of the $\mathcal{A}_\mathrm{high}$ architectures.
Recall that \autoref{subsec:cell_arch_modify} shows that the MIA promotion cell pattern on both normal and reduction cells contains many direct convolutions, while the MIA demotion cell patterns on both the normal and reduction cells contain few \emph{direct} convolutions.
Adding more direct convolutions (i.e., promotion) to $\mathcal{A}_\mathrm{high}$ and $\mathcal{A}_\mathrm{low}$ would have less impact on MIA performance due to the fact that the frequency of \emph{direct} convolution topology is always high in both cases.
On the other hand, reducing direct convolutions (i.e., demotion) is more effective in decreasing the MIA AUC scores, hence more evident in $\mathcal{A}_\mathrm{high}$ and $\mathcal{A}_\mathrm{low}$.

Our third observation is that cell patterns with more edges tend to have a larger impact on MIA robustness.
For example, when we increase the modification budget $m$ from 3 to 5 in \autoref{figure:mia_auc_promotion}, the MIA performance of the target architectures under the Only-Normal MIA promotion modifications ascend from 0.8148 to 0.8217.
Note that, even when the cell patterns have merely $m=3$ edges, the Only-Normal MIA demotion modifications can still effectively decrease the MIA performance from 0.8141 to 0.7940 in \autoref{figure:mia_auc_demotion}. 

Besides, in the MIA demotion scenario, the model utility is even improved after the robustness of the target architecture has been strengthened, which is a desired and satisfying result. 
The reason behind this phenomenon is that the MIA performance is correlated with the overfitting level of the target model~\cite{LWHSZBCFZ22}. 
After we apply the MIA demotion cell patterns, the overfitting level of the target model decreases. 
The reduced overfitting level helps the target model generalize better and promote the model utility at the test time. 
For instance, after the Only-Reduction MIA demotion modifications with $m=4$, the MIA performance in \autoref{figure:mia_auc_demotion} drops from 0.8141 to 0.7861, while the corresponding test accuracy in \autoref{figure:test_acc_demotion} increases from 0.7866 to 0.7959. 
The overfitting level of the target model drops from 0.2134 to 0.2041 in this case.
Note that, Only-Normal modifications usually have a relatively small impact on the model performance of the target architectures. 
As we have discussed in \autoref{subsec:cell_arch_modify}, the cell patterns on normal cells have tried to mitigate the side effects on model performance and mainly contain convolution operations. 

Finally, Dual modifications usually have the largest impact on the model utility and MIA robustness. 
For instance, when the modification budget limited to $m=3$ in 
\autoref{figure:mia_auc_demotion}, Only-Normal and Only-Reduction MIA demotion modifications can demote the MIA AUC score of the target architectures from 0.8141 to 0.7940 and 0.7946 respectively, while Dual modifications can further reduce the MIA AUC score to 0.7801. 
At the same time, the test accuracy of the architecture in \autoref{figure:test_acc_demotion} is improved from 0.7866 to 0.7948, 0.7950, and 0.8061 respectively for Only-Normal, Only-Reduction, and Dual modifications.
Note that, we also conduct ablation studies on the impact of the number of cells and the number of intermediate nodes in \autoref{app:ablation_studies} to get a deeper understanding of the cell patterns. 
Particularly, we find that deeper networks tend to be more robust while wider networks tend to be more vulnerable to MIAs.

% ----------------------------------------------------
\subsection{Loss Contour Analysis}
\label{sec:loss_contour_analysis}
% ----------------------------------------------------

\begin{figure}[!t]
\centering
\begin{subfigure}{0.52\columnwidth}
\includegraphics[width=\columnwidth]{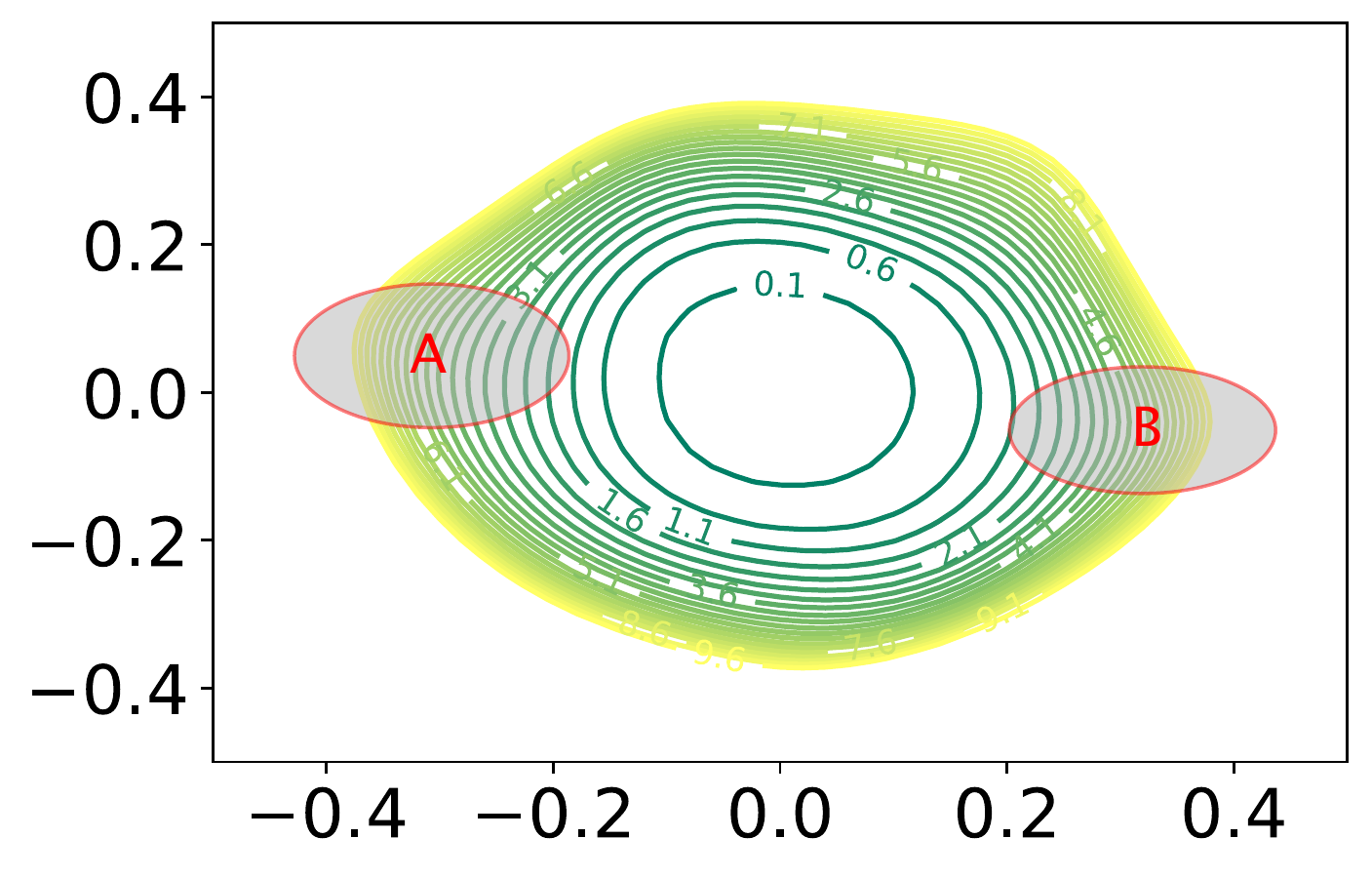}
\caption{Original}
\label{figure:loss_origin_nb301}
\end{subfigure}
\begin{subfigure}{0.49\columnwidth}
\includegraphics[width=\columnwidth]{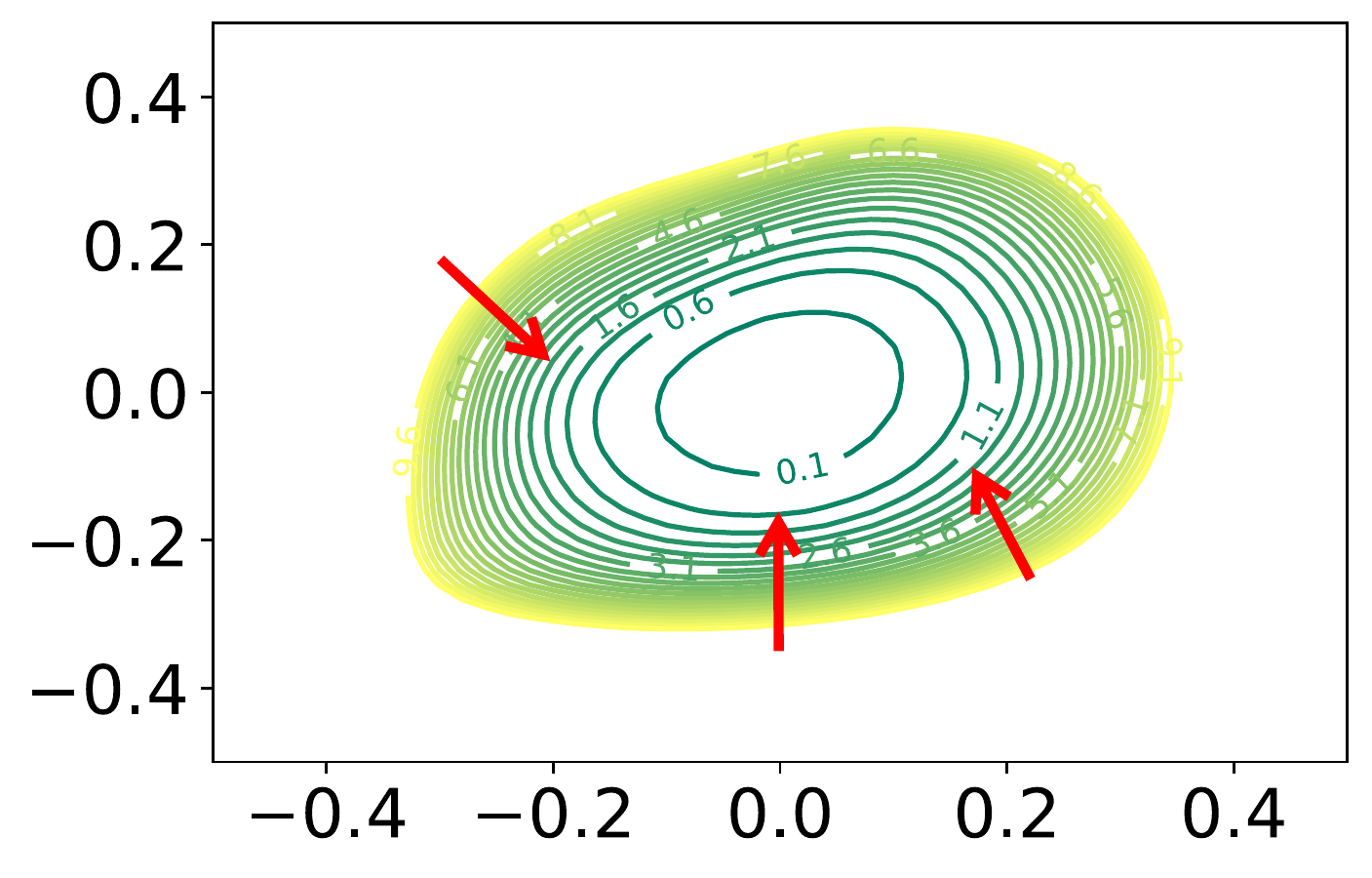}
\caption{After Demotion}
\label{figure:loss_after_demotion_nb301}
\end{subfigure}
\begin{subfigure}{0.49\columnwidth}
\includegraphics[width=\columnwidth]{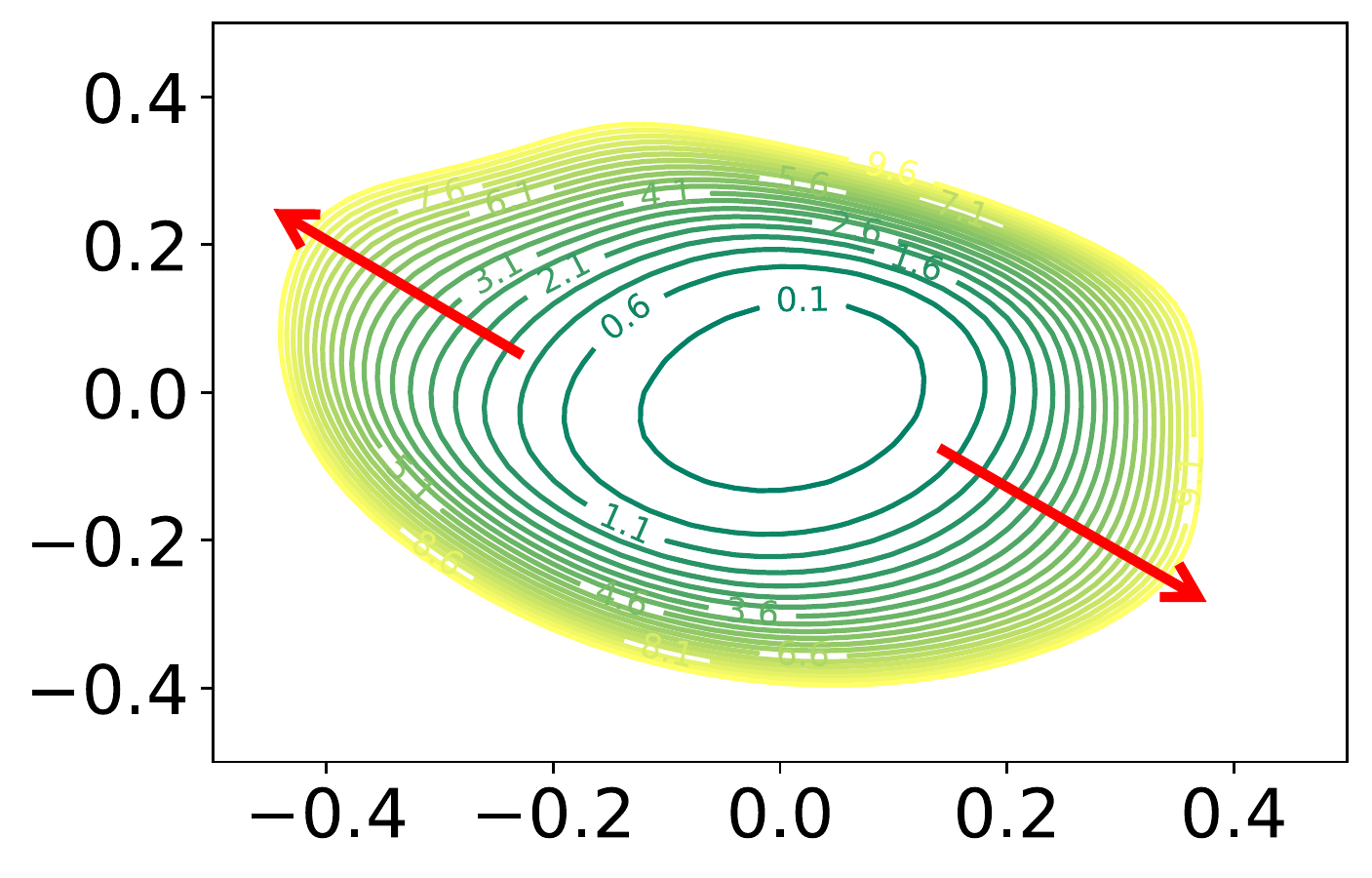}
\caption{After Promotion}
\label{figure:loss_after_promotion_nb301}
\end{subfigure}
\caption{The contours before and after MIA demotion or promotion modifications of the target architecture.}
\label{figure:loss_contour_nb301}
\end{figure}

To further investigate the impact of MIA demotion or promotion modifications, we leverage loss contour of the target architecture to understand how those architectures behave after the modifications have been made.
We sample one architecture already evaluated in \autoref{subsec:effect_cell_pattern}, and plot its training loss contour with regarding to model weight parameters changes using the code implementation\footnote{\url{https://github.com/tomgoldstein/loss-landscape}} of Li et al.'s work~\cite{LXTSG18}. 
Both MIA demotion and promotion modifications are Only-Reduction ones, and the experiments are conducted on the CIFAR10 dataset. 
The loss contours are shown in \autoref{figure:loss_contour_nb301}. 
The circled gray areas $A$ and $B$ in \autoref{figure:loss_origin_nb301} represent the ``trustworthy'' areas for MIAs in the loss contour of the original target architecture. 
We notice that the loss contour tends to be sparse in $A$ and $B$. 
It means that small perturbations (e.g., training randomness) on the model weights do not affect the loss values of the training dataset much, offering ``trustworthy'' posteriors of training members for the attacker to discriminate those of non-members.
In \autoref{figure:loss_after_demotion_nb301} and \autoref{figure:loss_after_promotion_nb301}, we use arrows to demonstrate the changing trends of the loss contours near the trustworthy areas compared to the original loss contour after the corresponding modifications.
We observe that the loss contour near the trustworthy areas of the original loss contour tends to be denser after the MIA demotion modifications. 
In comparison, the trustworthy areas tend to be sparser after the MIA promotion modifications.
The potential reason is that, when the loss contour in these areas becomes denser, little perturbations on the model weights may  make the loss values change significantly. 
The consequence is that many data samples are pushed close to the decision boundary of the current model.
In this way, the attacker can only get less confident posteriors from the target model, which hinders the performance of MIAs.
And both the MIA demotion and promotion modifications affect the robustness against MIAs by changing the shapes of the loss contour of the target model.
Further, our cell patterns have little impact on the model performance, so the overall flatness of the loss contour of the target model has not been changed much. 

% ----------------------------------------------------
\subsection{Transferability}
\label{subsec:transferability}
% ----------------------------------------------------

Our cell patterns are extracted based on the MIA evaluation results with the most knowledgeable \tuple{White\mbox{-}Box, Partial} attack setting on the sampled architectures from NAS-Bench-301, a NAS architecture dataset searched in the DARTS search space on the CIFAR10 dataset.
Here we want to check whether our cell patterns can transfer to other scenarios.

\begin{figure}[!t]
\centering
\begin{subfigure}{0.48\columnwidth}
\includegraphics[width=\columnwidth]{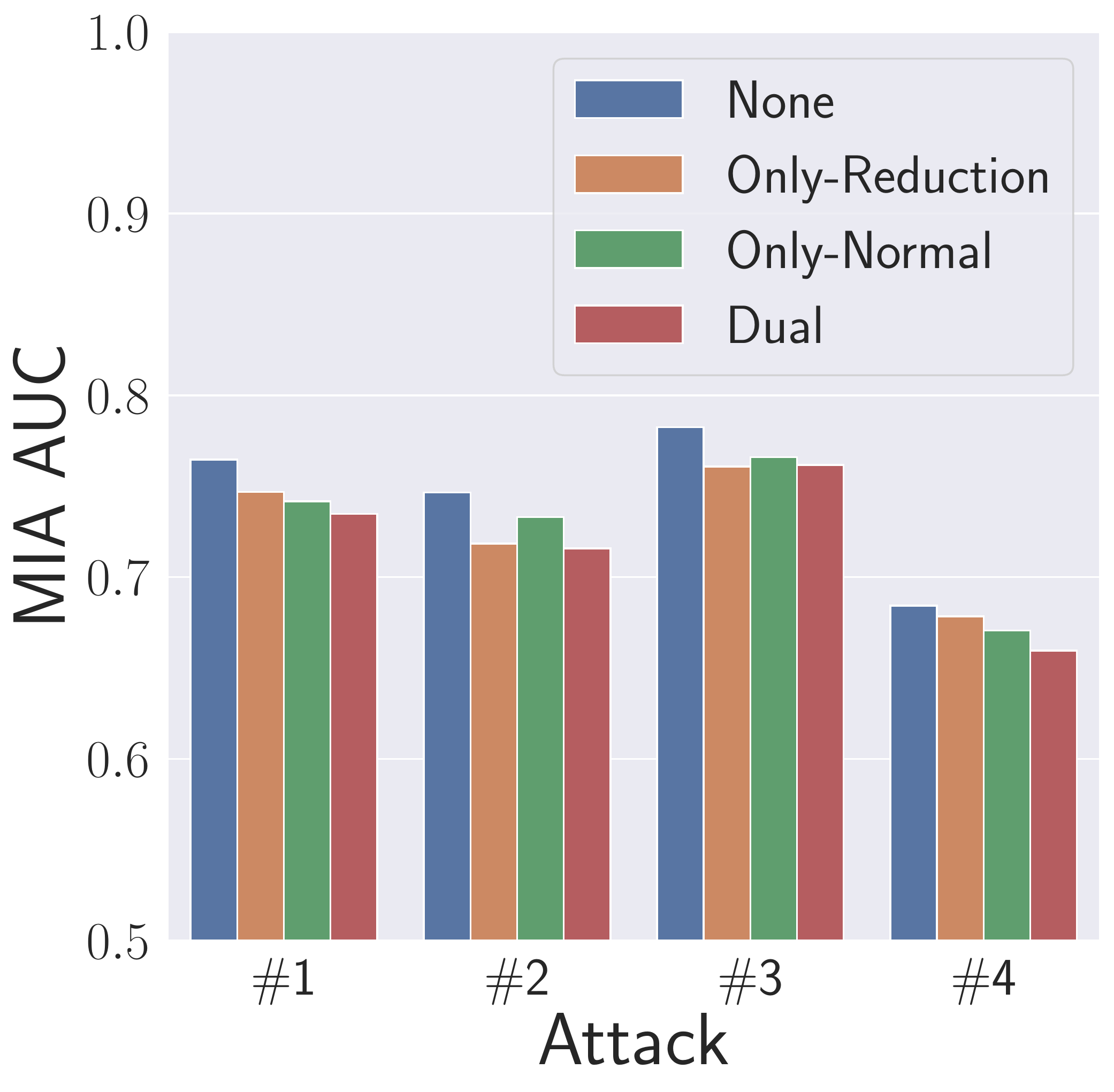}
\caption{Demotion}
\label{figure:mia_auc_transfer_attacks_demotion}
\end{subfigure}
\begin{subfigure}{0.48\columnwidth}
\includegraphics[width=\columnwidth]{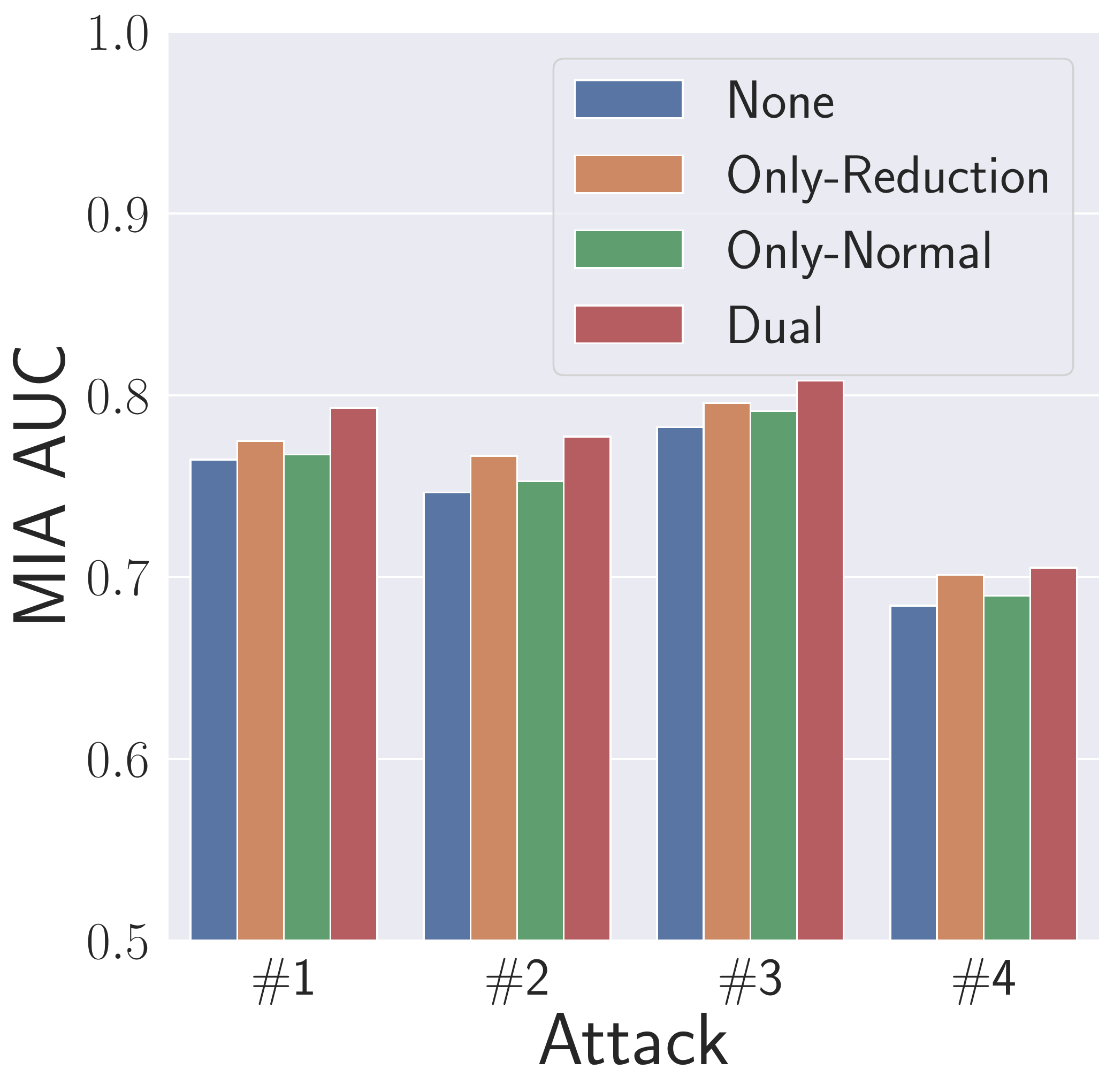}
\caption{Promotion}
\label{figure:mia_auc_transfer_attacks_promotion}
\end{subfigure}
\caption{Transferring MIA performance of the cell patterns on various MIAs. \#1, \#2, \#3 and \#4 represent the \tuple{Black\mbox{-}Box, Shadow}, \tuple{Black\mbox{-}Box, Partial},  \tuple{White\mbox{-}Box, Shadow} and \tuple{Label\mbox{-}Only} attack settings respectively.}
\label{figure:mia_auc_transfer_attacks}
\end{figure}

\mypara{Transferability among Different Attack Settings}
Our cell patterns are extracted based on the MIA evaluation results with the \tuple{White\mbox{-}Box, Partial} setting, here we would like to see whether our cell patterns also work for other attacks with different attack settings.
We change the attack setting for MIAs and re-evaluate the target architectures before and after the MIA demotion and promotion modifications, and the results are shown in \autoref{figure:mia_auc_transfer_attacks}.
We could see that our cell patterns are still effective for all these attacks in all cases, which demonstrates the high transferability of our cell patterns.

\mypara{Transferability among Different Datasets}
Our cell patterns are extracted from the architectures searched on the CIFAR10 dataset, here we want to explore whether our cell patterns are also effective on other datasets.
Here we apply our cell patterns to 4 architectures using the last four NAS algorithms in \autoref{sec:nas_algos} and searched on CIFAR100, STL10 and CelebA datasets respectively, and evaluate the MIA performance and model utility before and after the architectural modifications.

\begin{figure}[!t]
\centering
\begin{subfigure}{0.49\columnwidth}
\includegraphics[width=\columnwidth]{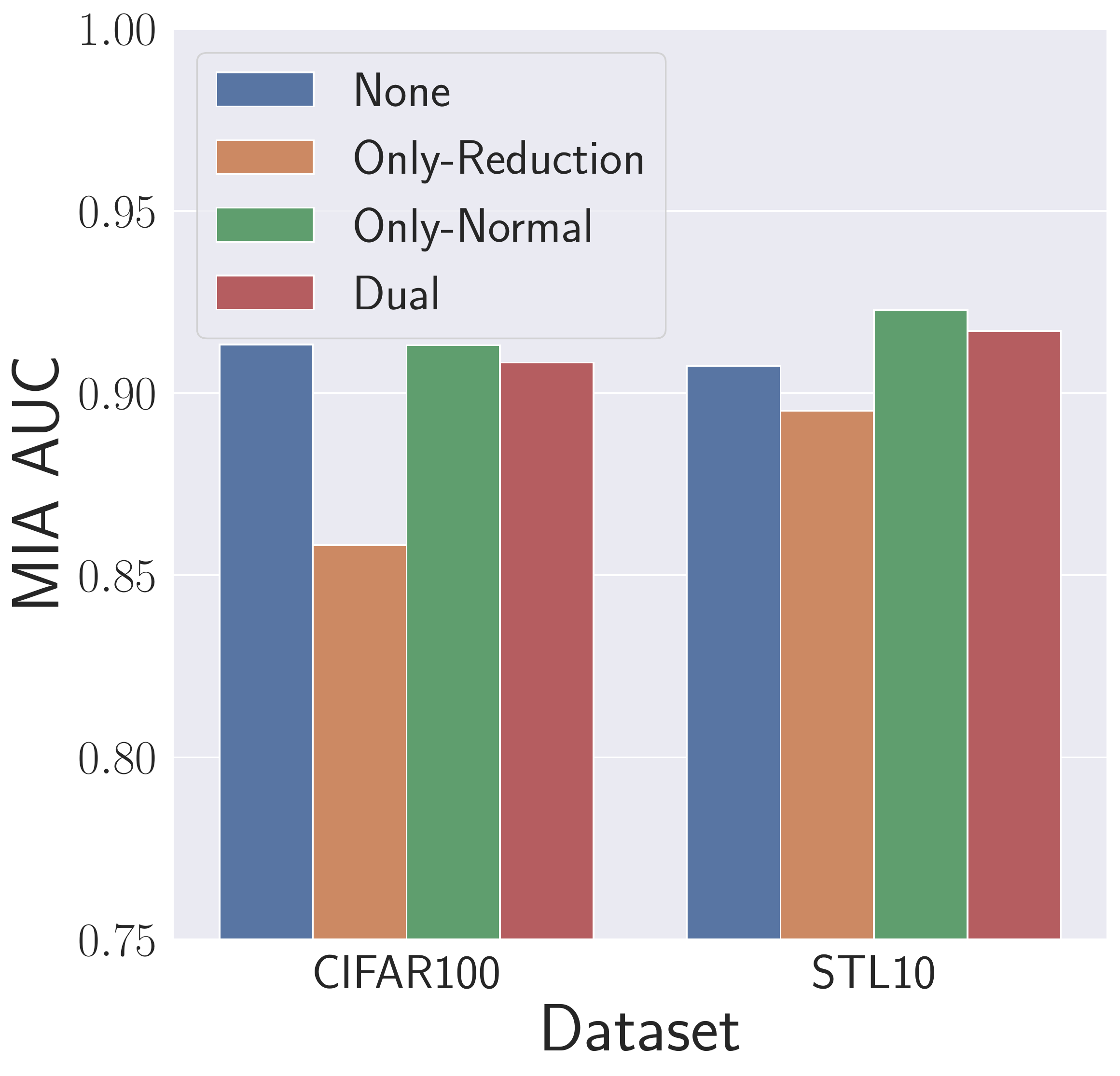}
\caption{Demotion}
\label{figure:mia_auc_transfer_datasets_demotion}
\end{subfigure}
\begin{subfigure}{0.49\columnwidth}
\includegraphics[width=\columnwidth]{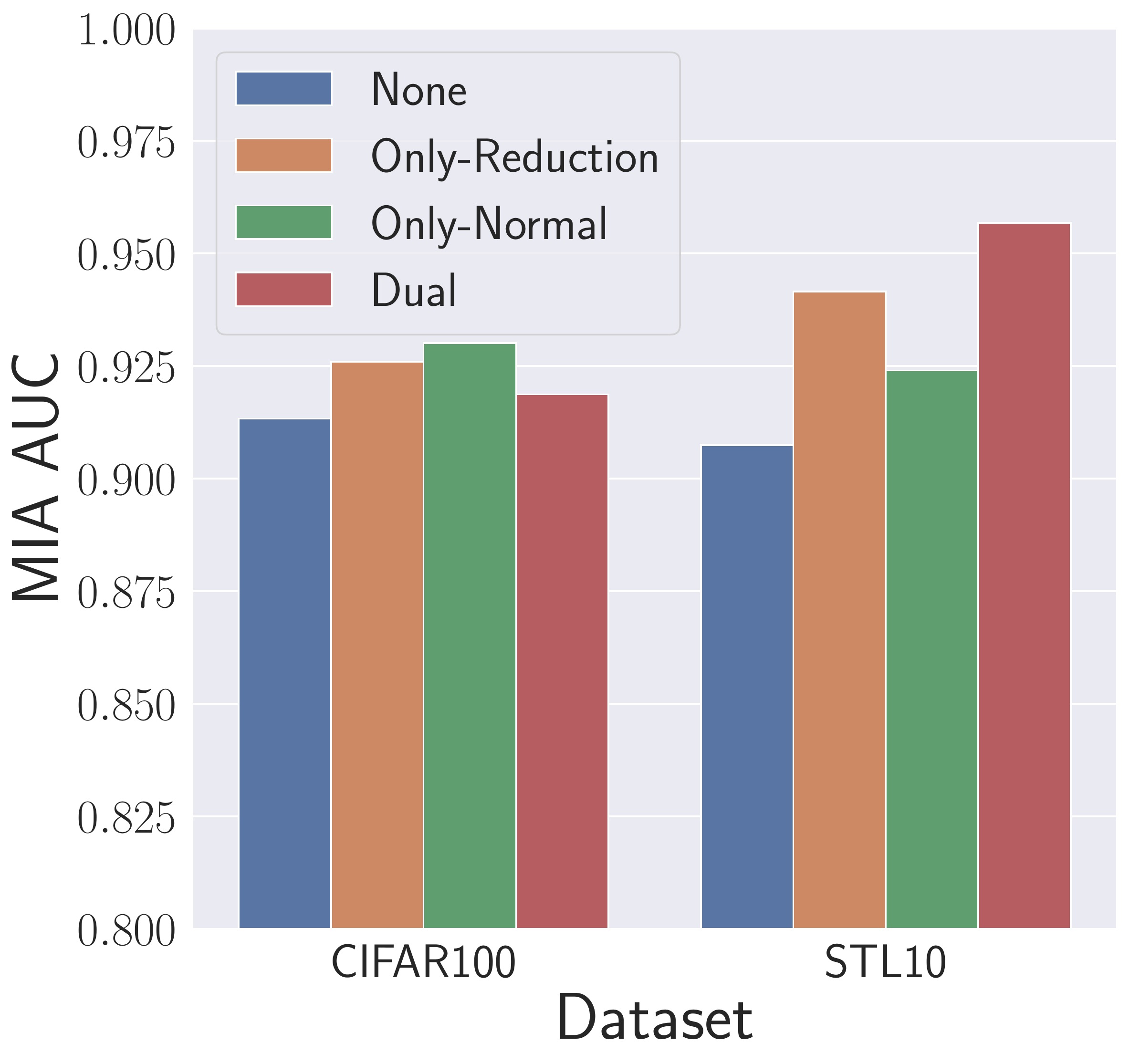}
\caption{Promotion}
\label{figure:mia_auc_transfer_datasets_promotion}
\end{subfigure}
\caption{Transferring MIA performance for the cell patterns on different datasets.}
\label{figure:mia_auc_transfer_datasets}
\end{figure}

\begin{figure}[!t]
\centering
\begin{subfigure}{0.49\columnwidth}
\includegraphics[width=\columnwidth]{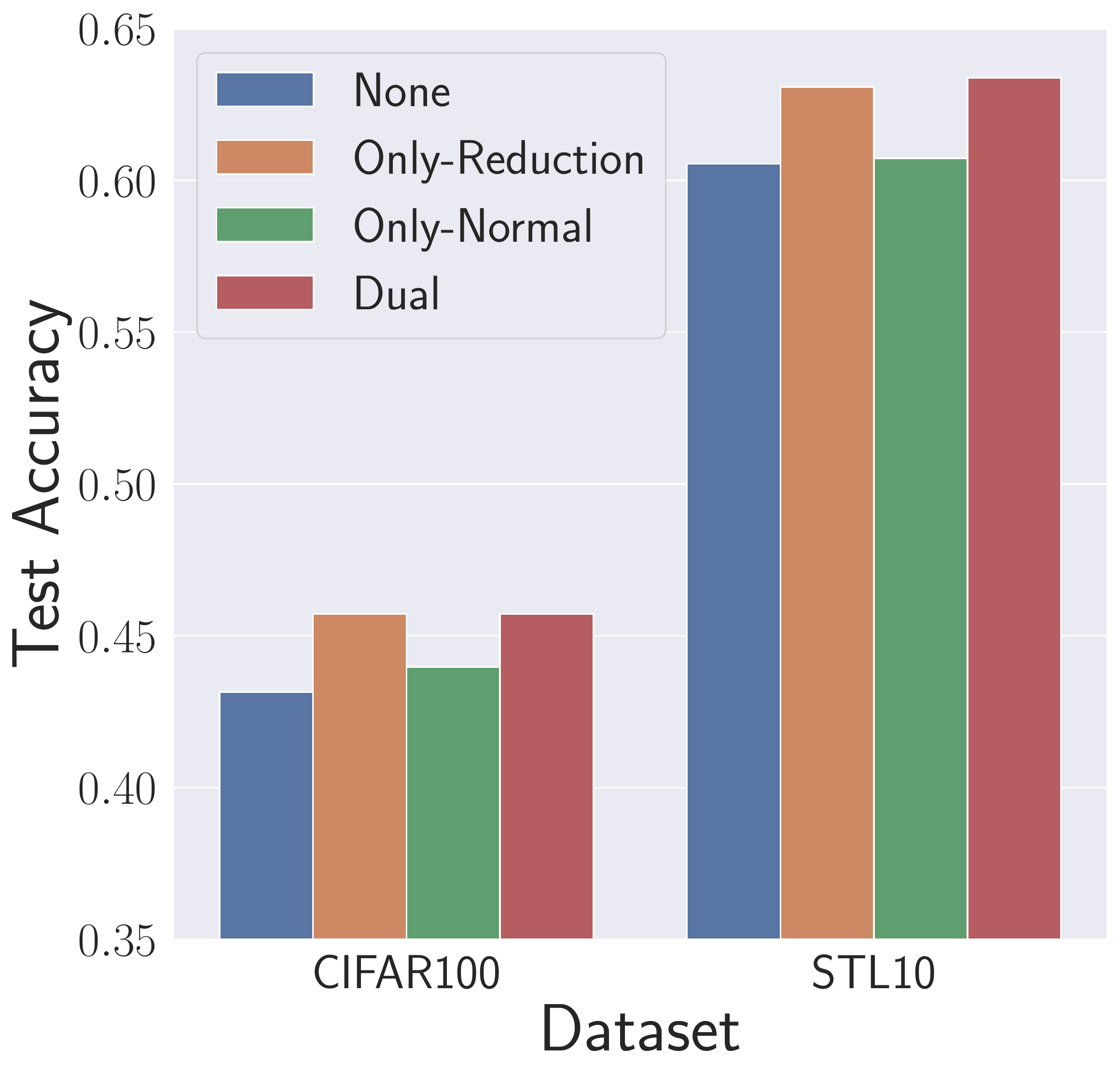}
\caption{Demotion}
\label{figure:test_acc_transfer_datasets_demotion}
\end{subfigure}
\begin{subfigure}{0.49\columnwidth}
\includegraphics[width=\columnwidth]{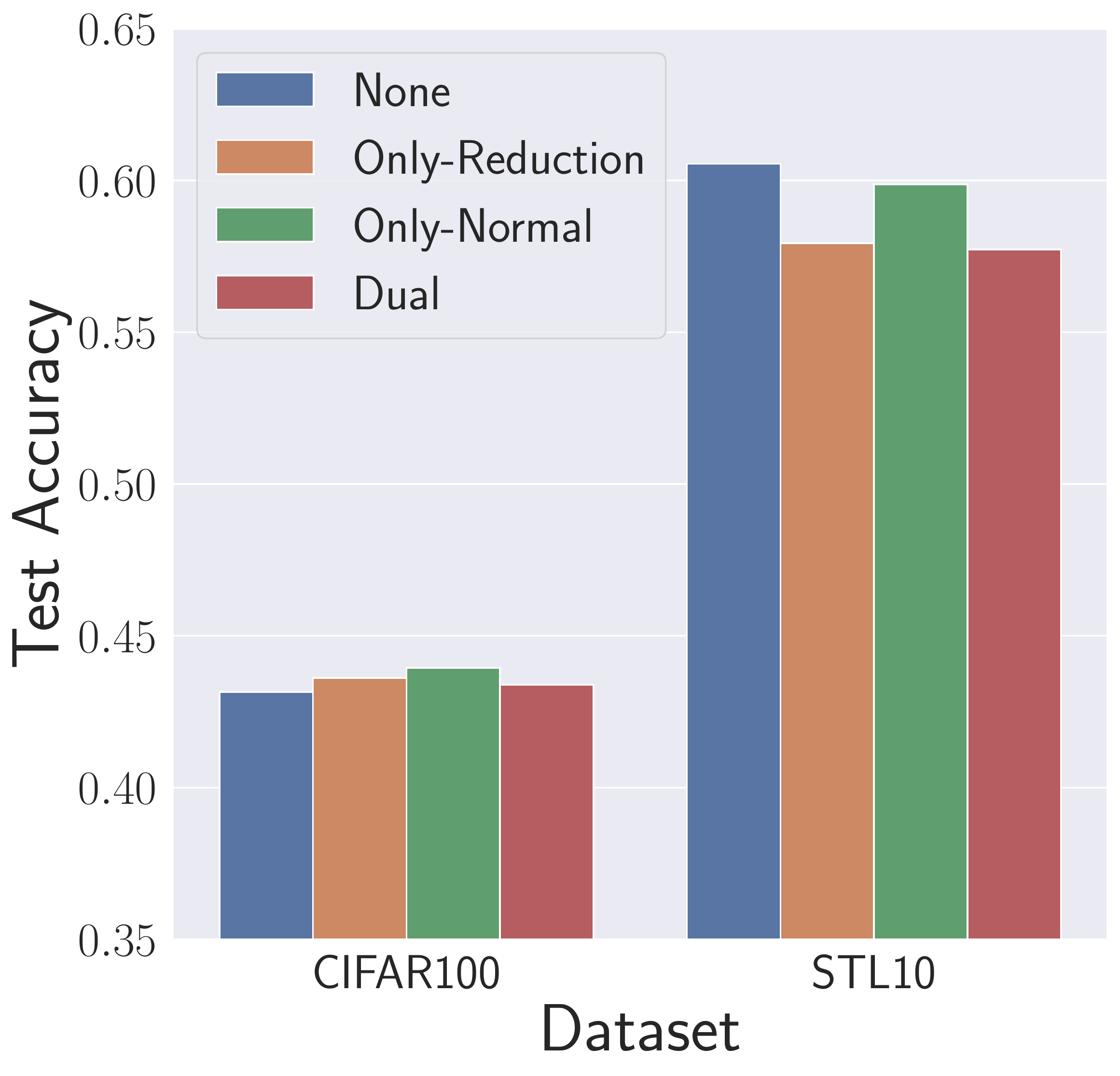}
\caption{Promotion}
\label{figure:test_acc_transfer_datasets_promotion}
\end{subfigure}
\caption{Model utility under the transferred cell patterns on different datasets.}
\label{figure:test_acc_transfer_datasets}
\end{figure}

The experimental results on the CIFAR100 and STL10 datasets are shown in \autoref{figure:mia_auc_transfer_datasets} and \autoref{figure:test_acc_transfer_datasets}. We defer the evaluation results and corresponding analysis on the CelebA dataset to \autoref{appendix:transfer_celeba}. 
It is observed that our cell patterns can still achieve the desired MIA demotion or promotion goals in most cases, which means our cell patterns are also transferable to different datasets.
Besides, we can observe that the Only-Reduction modifications tend to have the best performance, while the Only-Normal modifications are inferior to the former.
The possible reason is that the model performance of cell-based NAS architectures is mainly determined by the normal cells~\cite{WREL22}, and the architectures searched on different datasets tend to have different normal cell architectures, which hinders the transferring of the cell patterns on normal cells.

\smallskip
Note that, since our cell patterns are extracted from the DARTS search space, we also test the transferring effectiveness of our cell patterns in the NAS-Bench-201 search space in \autoref{sec:transfer_search_space}. And it is observed that our cell patterns can still partially transfer to the NAS-Bench-201 search space.

% ----------------------------------------------------
\subsection{Enhancing Existing Defenses}
\label{subsec:enhancing_existing_defenses}
% ----------------------------------------------------

Existing defenses against MIAs mainly focus on improving the robustness of the target model by masking confidence scores~\cite{SSSS17,LZ21,JSBZG19}, regularization~\cite{SSSS17,NSH18,KD21}, or differential privacy~\cite{JE19,SH21,CYZF20}, etc. 
In contrast, our MIA demotion cell patterns are performed in the model architecture design phase and are different from previous defense strategies. 
Naturally, we wonder if our approach can not only improve the model robustness against MIAs by itself but also complement existing defense techniques when using together?

To this end, we further conduct experiments to estimate whether our MIA demotion cell patterns can enhance existing defense strategies.
We choose three representative defense methods which could be applied on both black-box and white-box MIA settings, i.e., data augmentation (DA)~\cite{KD21}, label smoothing (LS)~\cite{SVISW16}, and differential privacy (DP)~\cite{DMNS06}.
DA increases the number of training samples by exerting slight modifications to the original data samples and can decrease the overfitting level. 
LS is a regularization method to mitigate the overconfidence of the target model. 
DP adds well-calibrated perturbations to the training process of the target model.

\begin{figure}[!t]
\centering
\begin{subfigure}{0.48\columnwidth}
\includegraphics[width=\columnwidth]{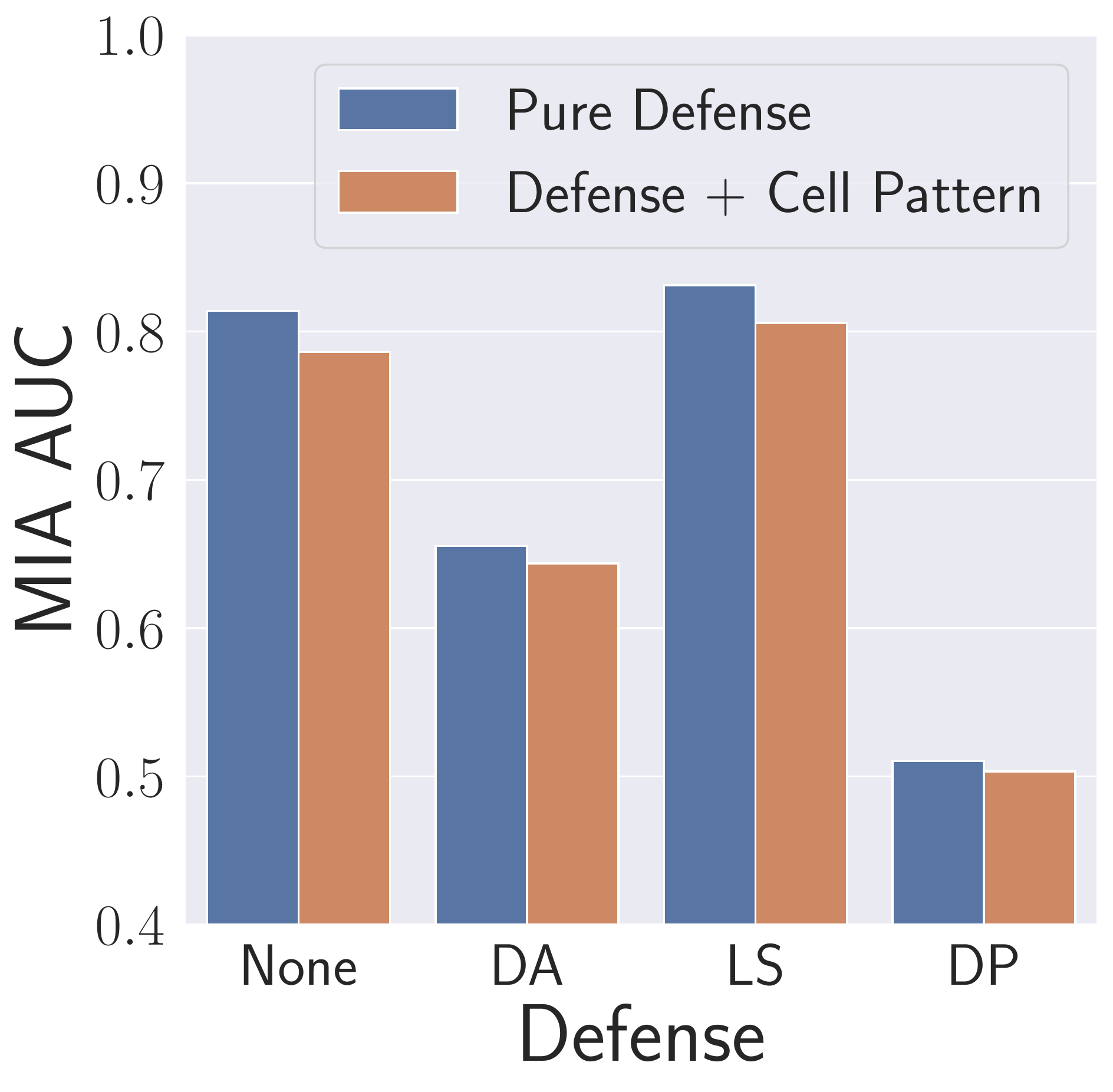}
\caption{MIA Performance}
\label{figure:mia_auc_defenses}
\end{subfigure}
\begin{subfigure}{0.48\columnwidth}
\includegraphics[width=\columnwidth]{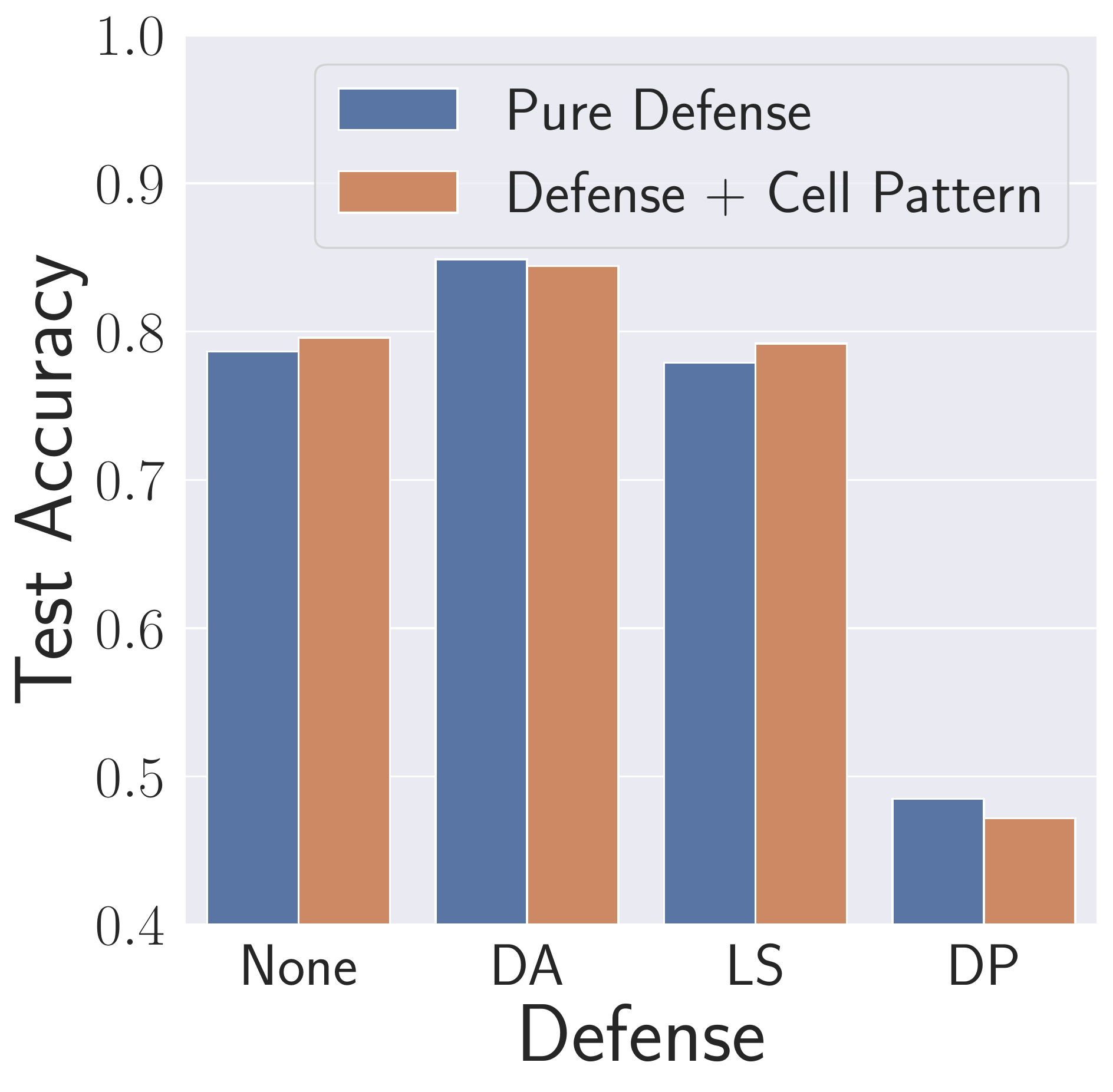}
\caption{Model Utility}
\label{figure:test_acc_defenses}
\end{subfigure}
\caption{MIA effectiveness and model utility under defenses.}
\label{figure:defenses_eval}
\end{figure}

The experimental results are shown in \autoref{figure:defenses_eval}.
We can see that, our MIA demotion cell patterns can successfully enhance the performance of these existing defense strategies in all cases in \autoref{figure:mia_auc_defenses} with a slight impact on the model utility in \autoref{figure:test_acc_defenses}.
To further validate whether our cell patterns can enhance existing defenses even when transferring to other attack settings, we conduct the transferability experiments of other attack settings under defense in \autoref{sec:transfer_defense}.
Overall, our MIA demotion cell patterns are shown to be able to promote the effectiveness of the existing defense methods in almost all cases for various attack settings, setting a safer lower bound for the existing defense strategies.

\subsection{Guidelines for Improving Robustness}
\label{subsec:guidelines}

According to previous experimental results and analysis, we can conclude some general and meaningful guidelines for designing and using more robust cell-based architectures against MIAs:
\begin{enumerate}[leftmargin=*]
    \item \textit{Convolution operations} are important for model performance; however, the architectures with relatively high MIA performance tend to have more convolution operations than others.
    Therefore, we should use moderate number of convolution operations to strike a tradeoff between model utility and robustness.
    \item Use less convolutions directly connected to the input nodes.
    \item If more convolution operations are indeed necessary, refer to the second guideline. 
    In this case, constructing a deeper network architecture is recommended.
    \item Limit the width of the cell, which usually can be easily done by limiting the number of intermediate nodes in a cell.
\end{enumerate}

% ----------------------------------------------------
\section{Related Work}
\label{sec:related_work}
% ----------------------------------------------------

\mypara{Membership Inference Attack}
Membership inference attack (MIA) is a privacy attack method to infer whether a given data sample is present in the training dataset of the target machine learning models~\cite{HSSDYZ21}. 
It is firstly proposed by Shokri et al. ~\cite{SSSS17} and has developed many variants for various scenarios.
Shokri et al. ~\cite{SSSS17} use multiple shadow models trained on the shadow dataset to mimic the behavior of the target model, then train an attack model with the output posteriors of the shadow models to predict if a sample is used to train the target model. 
Salem et al.~\cite{SZHBFB19} relax the key assumptions of Shokri et al.~\cite{SSSS17} and propose model and data independent attack methods to effectively predict memberships under various attack settings.
Since then, membership inference attacks have been applied to many domains (e.g., computer vision~\cite{YGFJ18,LZ21,NSH19}, graph data~\cite{HWWBSZ21}, unlearning systems~\cite{CZWBHZ21,CZWBHZ22}, and even recommender system~\cite{ZRWRCHZ21}) and different target models (e.g., classification model~\cite{CZWBHZ21}, generative model~\cite{CYZF20}, embedding model~\cite{LJQG21} and multi-exit model~\cite{LLHYBZ222}) with different attack techniques.
To mitigate the privacy threats of membership inference attacks, many defense strategies have been proposed, including confidence masking~\cite{LZ21}, regularization~\cite{SSSS17, NSH18}, differential privacy~\cite{CYZF20,ZWLHC18,ZWHLBHCZ21,WCZSCLLJ21,DZBLJCC21} and knowledge distillation~\cite{SH21}.
Our MIA demotion cell patterns utilize the inherent robustness of specific structures to mitigate the vulnerability of the target model.
Those patterns are model-agnostic and complementary to existing defense work to further enhance the robustness of the target model.

\mypara{Security and Privacy of NAS}
The security and privacy threats of NAS architectures have not been well-studied and previous research also leads to contradictory findings. 
Guo et al.~\cite{GYXLL20} reveal several insightful observations (e.g., convolution operations to direct connection edge, densely connected patterns, etc.) that can improve the adversarial robustness of NAS-searched architectures.
Their research leads to some recent research work improving the robustness of NAS-searched architectures against adversarial perturbations to improve the accuracy of those models~\cite{DAMGB21,LYWX21}.
On the other hand, Oymak et al.~\cite{OLS21} demonstrate that the train-validation accuracy gap decreases rapidly when the validation data is mildly large (i.e., achieving a low overfitting level), which in turn makes NAS-searched architectures robust against MIAs. 
Yet, in the most recent work, Pang et al. ~\cite{PXJLW22} show that NAS architectures are more vulnerable to existing security and privacy attacks than the human-designed architectures including MIAs. 
However, their research only evaluates limited scenarios for these attacks. 
In our work, we conduct a comprehensive measurement study of the privacy risks of NAS architectures.
We eliminate the experimental bias introduced in the previous research and compare the MIA performance on both NAS-searched architectures and human-designed architectures using all known MIA attack scenarios.
We find that NAS-searched architectures are generally more robust against MIAs but such robustness varies from architecture to architecture.

% ----------------------------------------------------
\section{Conclusion}
% ----------------------------------------------------

In this paper, we conduct comprehensive measurement experiments for MIAs on both NAS-searched and human-designed architectures, and show that NAS-searched architectures tend to be more robust against MIAs. 
Furthermore, to analyze the hidden cell patterns affecting the robustness against MIAs, we design a general framework to extract the cell patterns based on the evaluation results on sampled NAS architectures. 
We use this framework to extract both MIA demotion and promotion cell patterns from existing well-performed NAS architectures.
The experimental results show that our cell patterns can successfully demote or promote the MIA performance on the target architectures and can transfer to various scenarios. 
Additionally, our MIA demotion cell patterns are complementary to existing defense techniques and can further enhance the performance of the latter.
Finally, we offer some guidelines to design more robust NAS architectures against MIAs in the future.

% ----------------------------------------------------
\section*{Acknowledgments}
% ----------------------------------------------------

We thank our shepherd Xi He and all anonymous reviewers for their constructive comments.
This work is partially funded by the Helmholtz Association within the project ``Trustworthy Federated Data Analytics'' (TFDA) (funding number ZT-I-OO1 4) and supported by NSFC under Grant 62132011.

\bibliographystyle{plain}
\bibliography{normal_generated_py3}

\newpage
% ----------------------------------------------------
\appendix
% ----------------------------------------------------

% ----------------------------------------------------
\section{Details of the Measurement Experiments}
\label{app:details_measurement}
% ----------------------------------------------------

\mypara{Datasets}
The CIFAR10~\cite{CIFAR} dataset contains 60,000 color images belonging to 10 different classes, and each class has 6,000 images. 
The CIFAR100~\cite{CIFAR} dataset is an image dataset consisting of 60,000 color images from 100 different classes, with 600 images in each class. 
The STL10~\cite{CNL11} dataset is also a color image dataset containing 13,000 labeled images belonging to 10 classes, and each class contains 1,300 images.
The CelebA~\cite{LLWT15} dataset is a human face dataset containing 202,599 images and each image is associated with 40 binary attributes. The 8-class classification task for the CelebA dataset in our experiments is to classify these images into 3 attributes (i.e., Smiling, MouthSlightlyOpen, and HeavyMakeup).

\mypara{NAS Algorithms}
DARTS~\cite{LSY19} is the first differential NAS algorithm converting both the architectural parameters and model weights as continuous variables to update them with gradient descent. 
DARTS-V1~\cite{LSY19} and DARTS-V2~\cite{LSY19} utilize the first-order and the second-order approximation for architecture gradients in DARTS respectively. 
ENAS~\cite{PGZLD18} uses the parameter sharing mechanism among child networks to speed up the evaluation of candidate architectures in the search process. 
GDAS~\cite{DY19} utilizes a differentiable architecture sampler to effectively sample meaningful sub-graphs in the searching procedure. 
SETN~\cite{DY192} proposes an evaluator and a template network to promote the quality of the sampled candidate architectures for evaluation. 
A random algorithm randomly samples candidate architectures and evaluates their performance.
TENAS~\cite{CGW21} identifies two training-free indicators to rank the quality of candidate architectures and achieves a fast neural architecture search without gradient descent. 
DrNAS~\cite{CWCTH21} formulates the differential architecture search method as a Dirichlet distribution problem. 
PC-DARTS~\cite{XXZCQTX20} uses edge normalization to reduce redundancy in exploring the network space and computation cost. 
SDARTS~\cite{CH20} uses a perturbation-based regularization to smooth the loss landscape of DARTS-based architectures to improve stability and generalizability.

\mypara{Human-designed Architectures}
ResNet~\cite{HZRS16} uses the residual learning framework to alleviate the training of extremely deep neural networks. 
ResNext~\cite{XGDTH17} adds the size of the set of residual transformations to be considered to ResNet. 
WideResNet~\cite{ZK16} is a variant of ResNet with shallower and wider architectures. VGG~\cite{SZ15} uses very small convolution filters to constitute significantly deep convolutional networks. 
DenseNet~\cite{HLMW17} builds dense connections between each layer and its all preceding layers. 
EfficientNet~\cite{TL19} scales model width, depth, and resolution with a set of fixed scaling coefficients. 
RegNet~\cite{RKGHD20} adds a regulator module to ResNet to aggregate extra complementary features. 
CSPNet~\cite{WLYWCH20} integrates feature maps for the gradient information from start to end stages to reduce computation cost. 
BiT~\cite{KBZPYGH20} is a transfer learning model that pre-trains on a large supervised source dataset and fine-tunes the weights on the given target task.
DLA~\cite{YWSD18} utilizes deeper aggregation structures to better merge the information across layers.

\mypara{Training Details}
We train all 20 target architectures for 100 epochs on the training dataset $\mathcal{D}_{\mathrm{target}}^{\mathrm{train}}$. 
We use stochastic gradient descent (SGD) as the optimizer and cross-entropy as the loss function. 
As for the optimizer, we set its momentum as 0.9, its weight decay coefficient as 5E-4, and its learning rate as 1E-2 for the first 50 epochs, 1E-3 for the 51st-75th epochs, and 1E-4 for the remaining epochs. 
We use the same training settings to train shadow models on $\mathcal{D}_{\mathrm{shadow}}^{\mathrm{train}}$. 
Note that we set the architecture of the shadow model as a simple convolution network with 3 convolution layers and 2 max-pooling layers sequentially combined under the \tuple{Black\mbox{-}Box, Shadow} setting. 
Since the target architecture is already known by the attacker, we set the shadow model with the same architecture as the target model under the \tuple{White\mbox{-}Box, Shadow} setting. 
Additionally, we use the open-sourced PyTorch implementation\footnote{\url{https://github.com/D-X-Y/AutoDL-Projects}} which uses the NAS-Bench-201 search space to search for the architectures for the first 6 NAS algorithms in \autoref{sec:nas_algos}. 
We utilize the official source code implementations using DARTS search space for the other 4 NAS algorithms.

% ----------------------------------------------------
\section{More Measurement Results}
\label{app:measurement_results}
% ----------------------------------------------------

\begin{figure*}[!t]
\centering
\begin{subfigure}{0.6\columnwidth}
\includegraphics[width=\columnwidth]{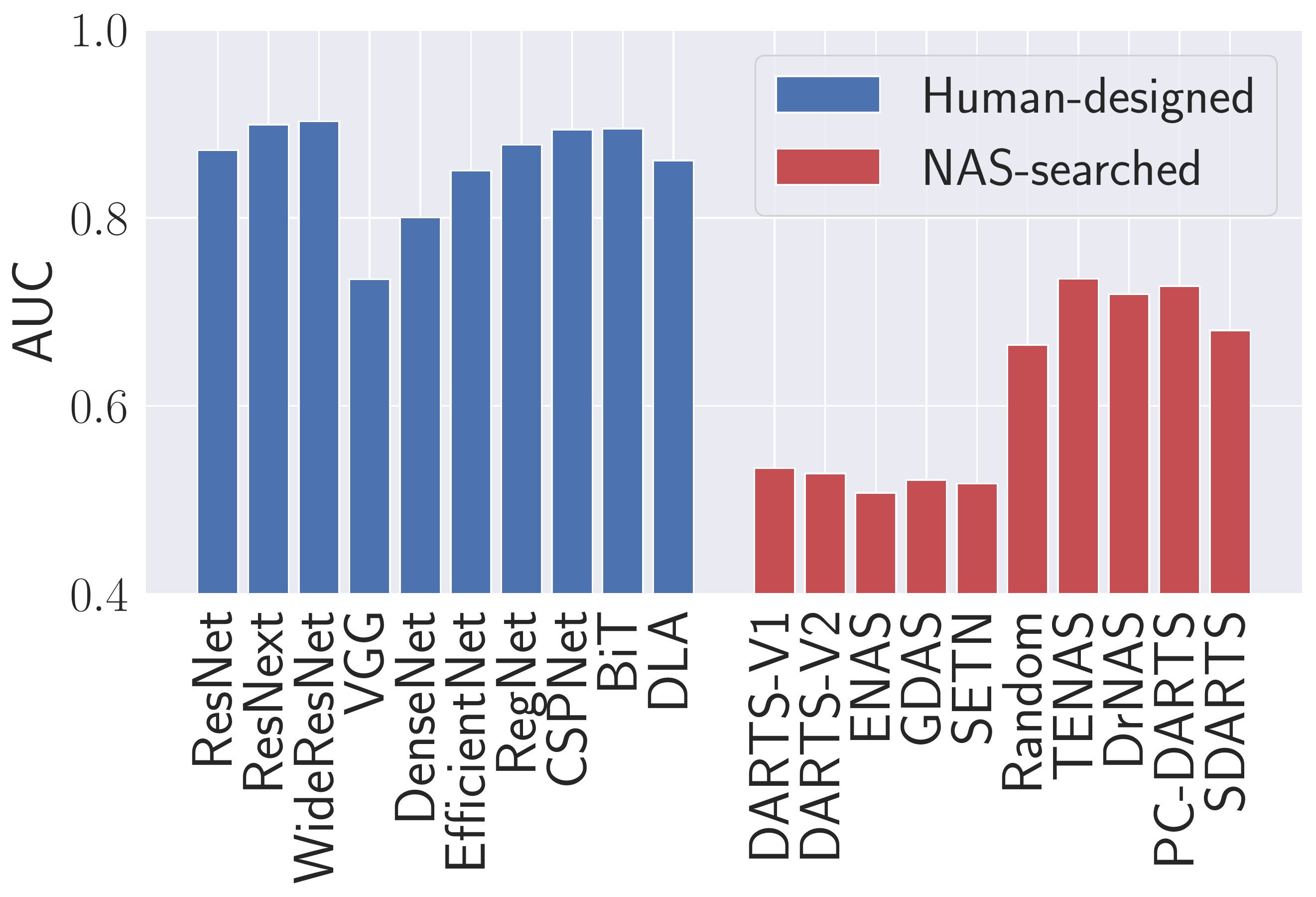}
\caption{CIFAR10}
\label{figure:blackbox_partial_cifar10}
\end{subfigure}
\begin{subfigure}{0.6\columnwidth}
\includegraphics[width=\columnwidth]{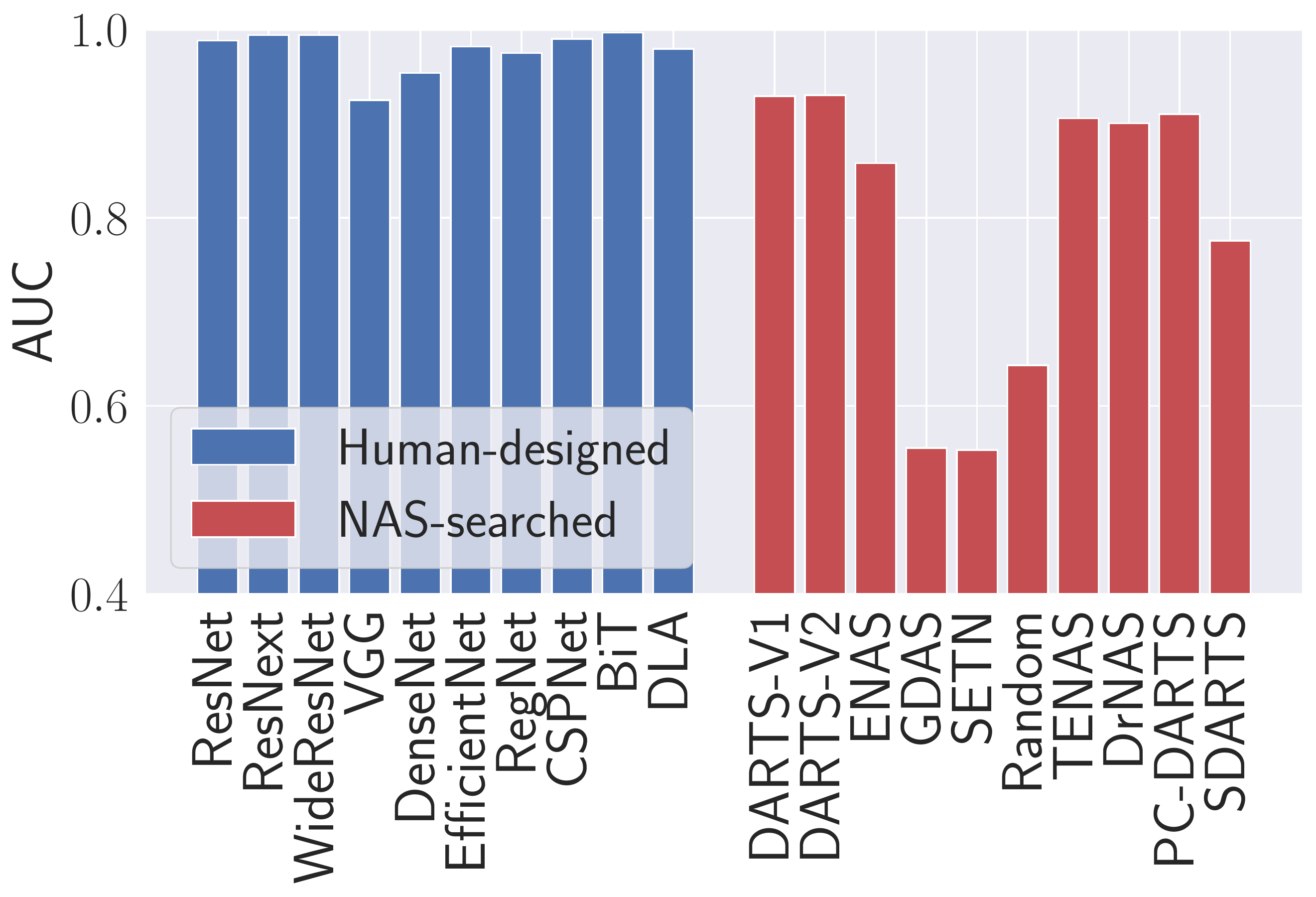}
\caption{CIFAR100}
\label{figure:blackbox_partial_cifar100}
\end{subfigure}
\begin{subfigure}{0.6\columnwidth}
\includegraphics[width=\columnwidth]{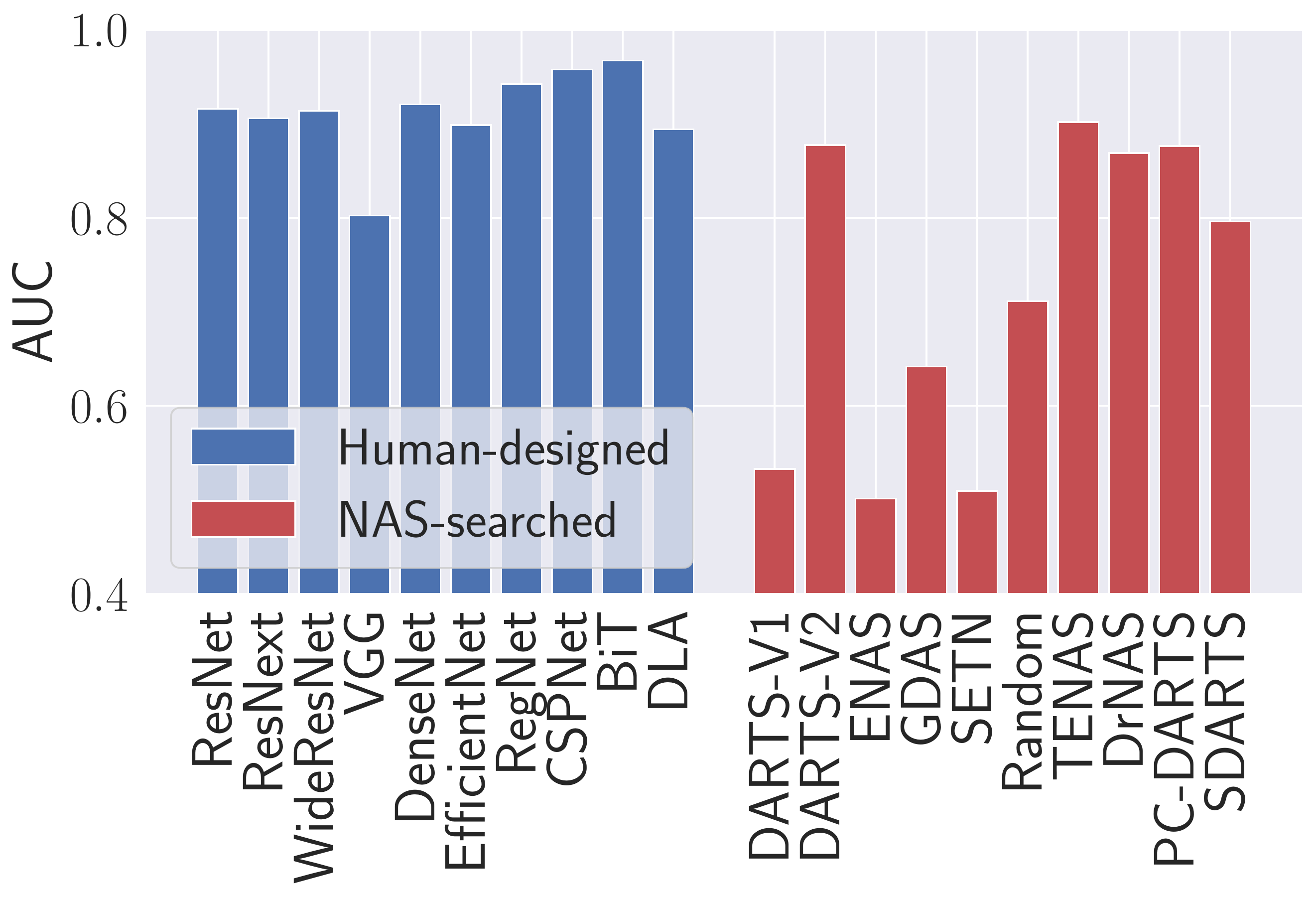}
\caption{STL10}
\label{figure:blackbox_partial_stl10}
\end{subfigure}
\caption{The performance of MIAs with the \tuple{Black\mbox{-}Box, Partial} setting on different datasets.}
\label{figure:blackbox_partial_mia_eval}
\end{figure*}

The results of \tuple{Black\mbox{-}Box, Partial} and \tuple{White\mbox{-}Box, Partial} for the CIFAR10, CIFAR100 and STL10 datasets are shown in \autoref{figure:blackbox_partial_mia_eval} and \autoref{figure:whitebox_shadow_mia_eval}, respectively.
The conclusion is consistent with that of \autoref{subsec:measurement_results}.

We also conduct measurement experiments on the human face dataset CelebA. 
This dataset is a large-scale face attributes dataset and is different from CIFAR10, CIFAR100, and STL10 which are for object recognition. 
The model utility evaluation results of the original models for different architectures on the CelebA dataset are shown in \autoref{tab:measure_test_acc}. 
We can observe that, on the CelebA dataset, the NAS-searched architectures still have higher test accuracy than the human-designed ones, and the overfitting levels of the former are usually lower than that of the latter. 
For instance, DARTS-V1 has an overfitting level of 0.027 while ResNet has 0.256.

\begin{figure*}[!t]
\centering
\subfloat[\tuple{Black\mbox{-}Box, Shadow}]{\includegraphics[width=0.6\columnwidth]{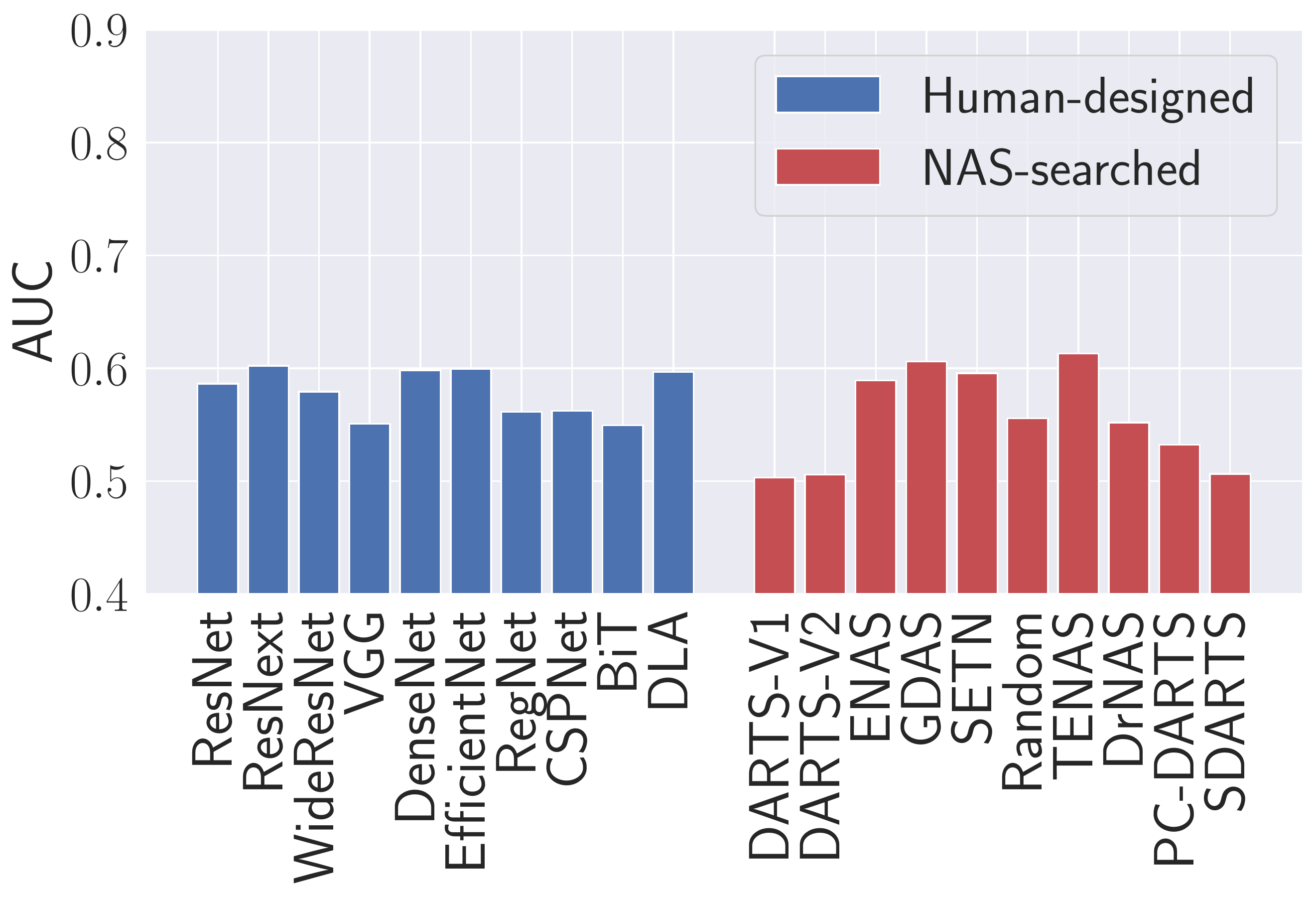}}\,
\subfloat[\tuple{Black\mbox{-}Box, Partial}]{\includegraphics[width=0.6\columnwidth]{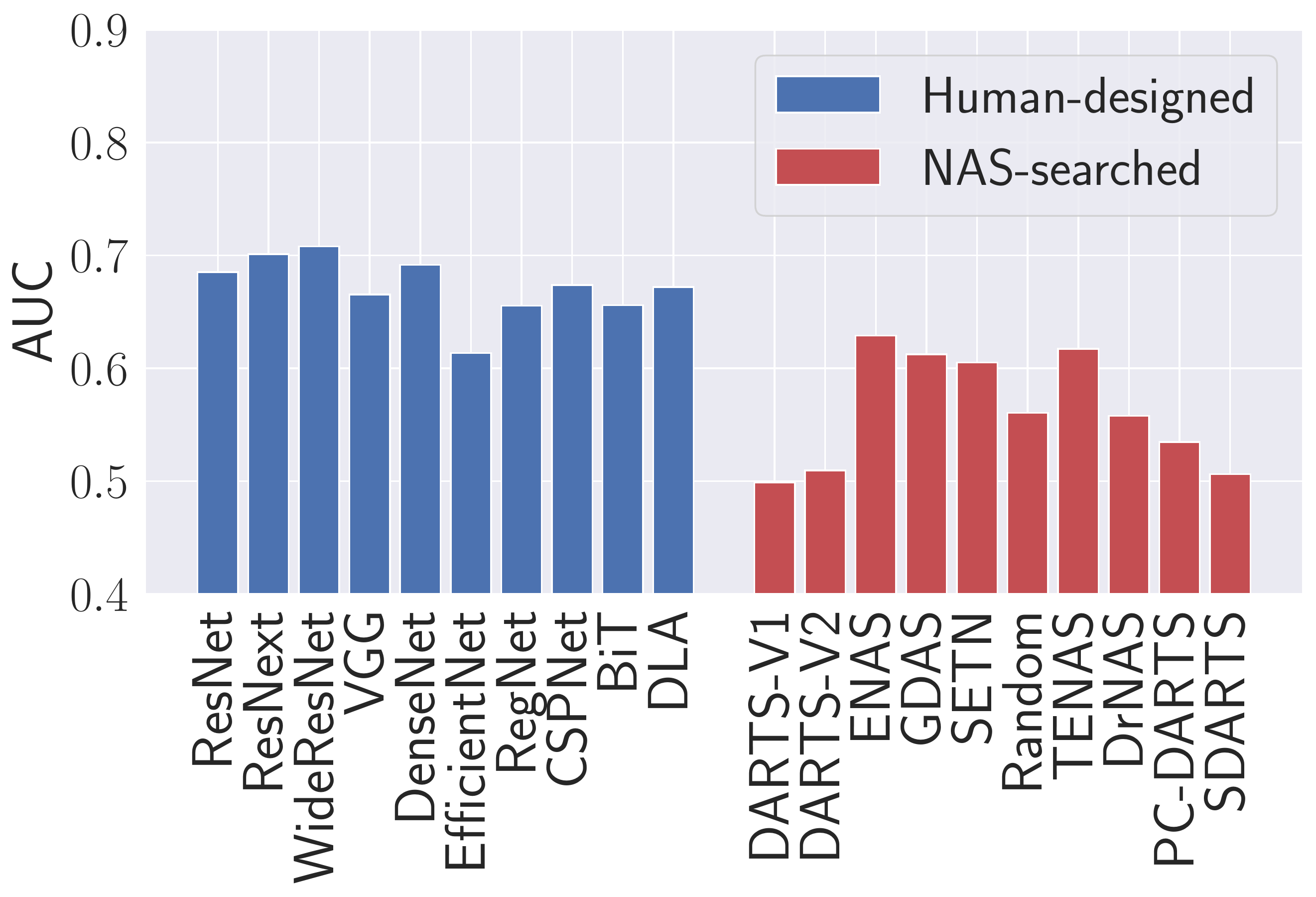}}\,
\subfloat[\tuple{Label\mbox{-}Only}]{\includegraphics[width=0.59\columnwidth]{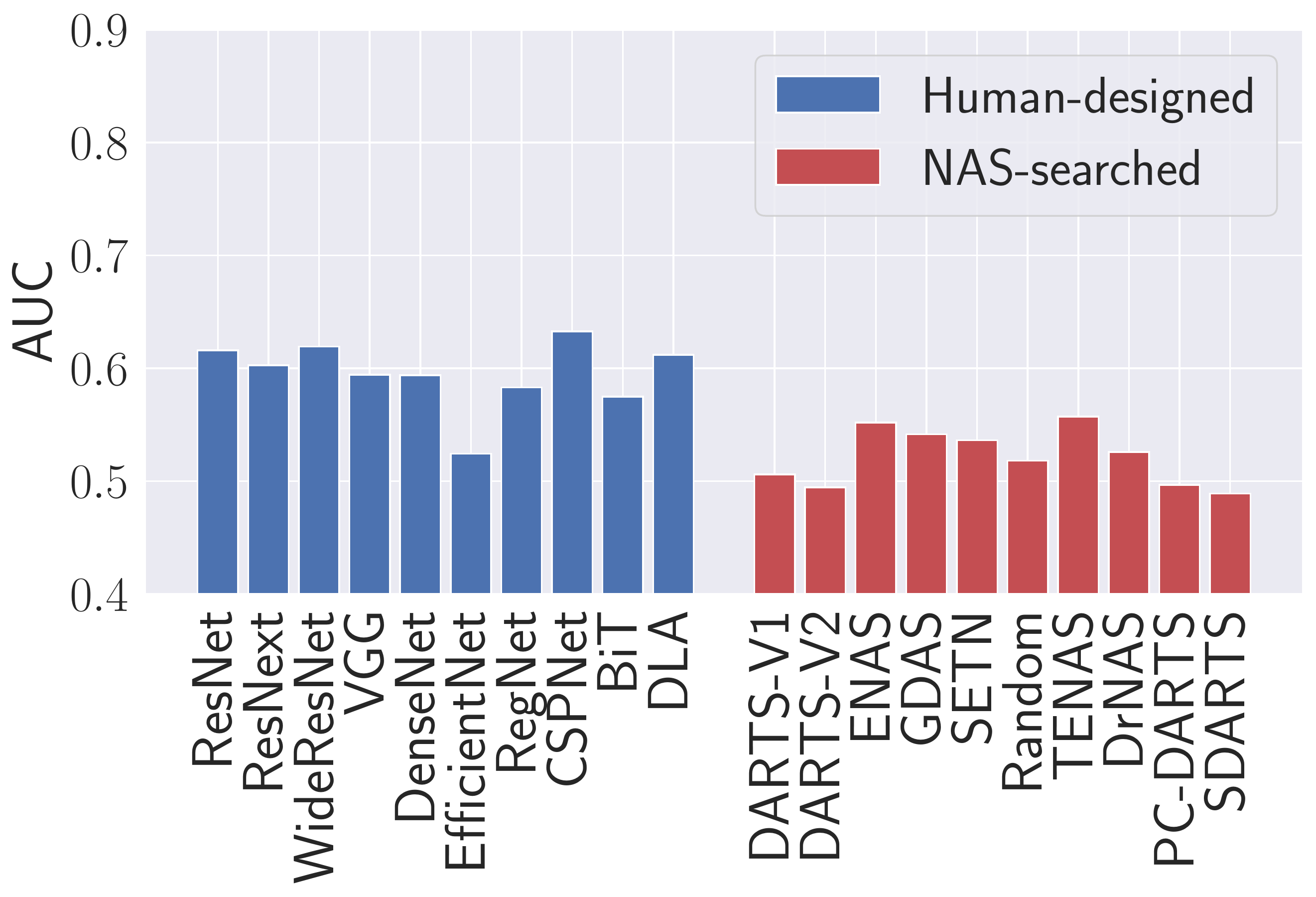}}
\\
\subfloat[\tuple{White\mbox{-}Box, Shadow}]{\includegraphics[width=0.6\columnwidth]{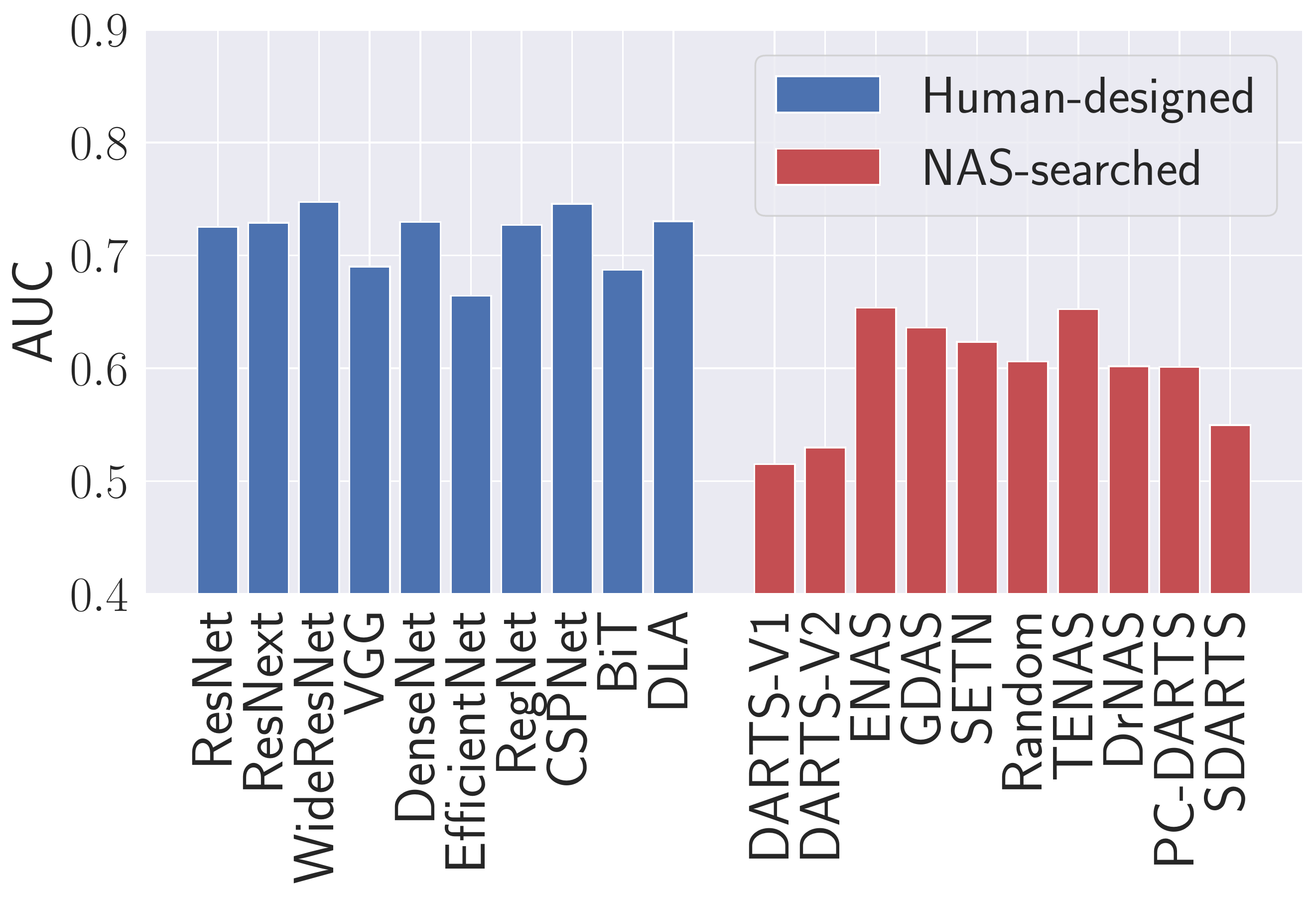}}\,
\subfloat[\tuple{White\mbox{-}Box, Partial}\label{figure:mia_auc_whitebox_partial_celeba}]{\includegraphics[width=0.6\columnwidth]{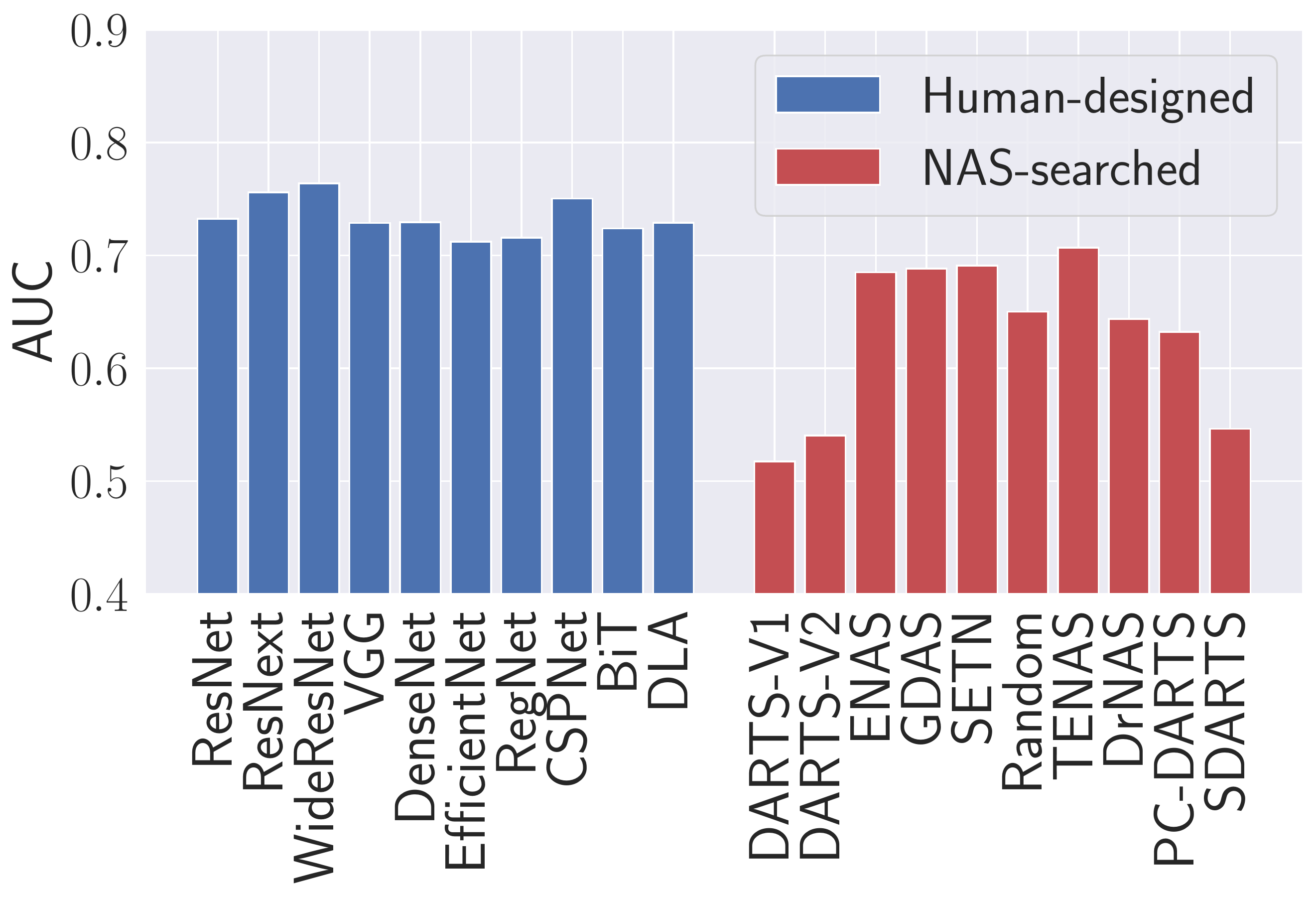}}
\caption{MIA performance with different attack settings on the CelebA dataset.}
\label{fig:celeba_mia_eval}
\end{figure*}

The evaluation results for MIAs under all five attack settings on the CelebA dataset are shown in \autoref{fig:celeba_mia_eval}. 
We can see that the AUC scores for MIAs on the NAS-searched architectures still tend to be smaller than that on the human-designed ones with various attack settings. 
It indicates that the former tends to be more robust against MIAs than the latter on the CelebA dataset. 
For example, in \autoref{figure:mia_auc_whitebox_partial_celeba}, the highest MIA AUC score for the NAS-searched architectures on the CelebA dataset with the \tuple{White\mbox{-}Box, Partial} MIA setting is 0.7068, even lower than the smallest MIA AUC score for the human-designed ones (i.e., 0.7121) in this case. 
Overall, we draw similar observations on the CelebA dataset as those from the aforementioned datasets.

\begin{figure*}[!t]
\centering
\begin{subfigure}{0.6\columnwidth}
\includegraphics[width=\columnwidth]{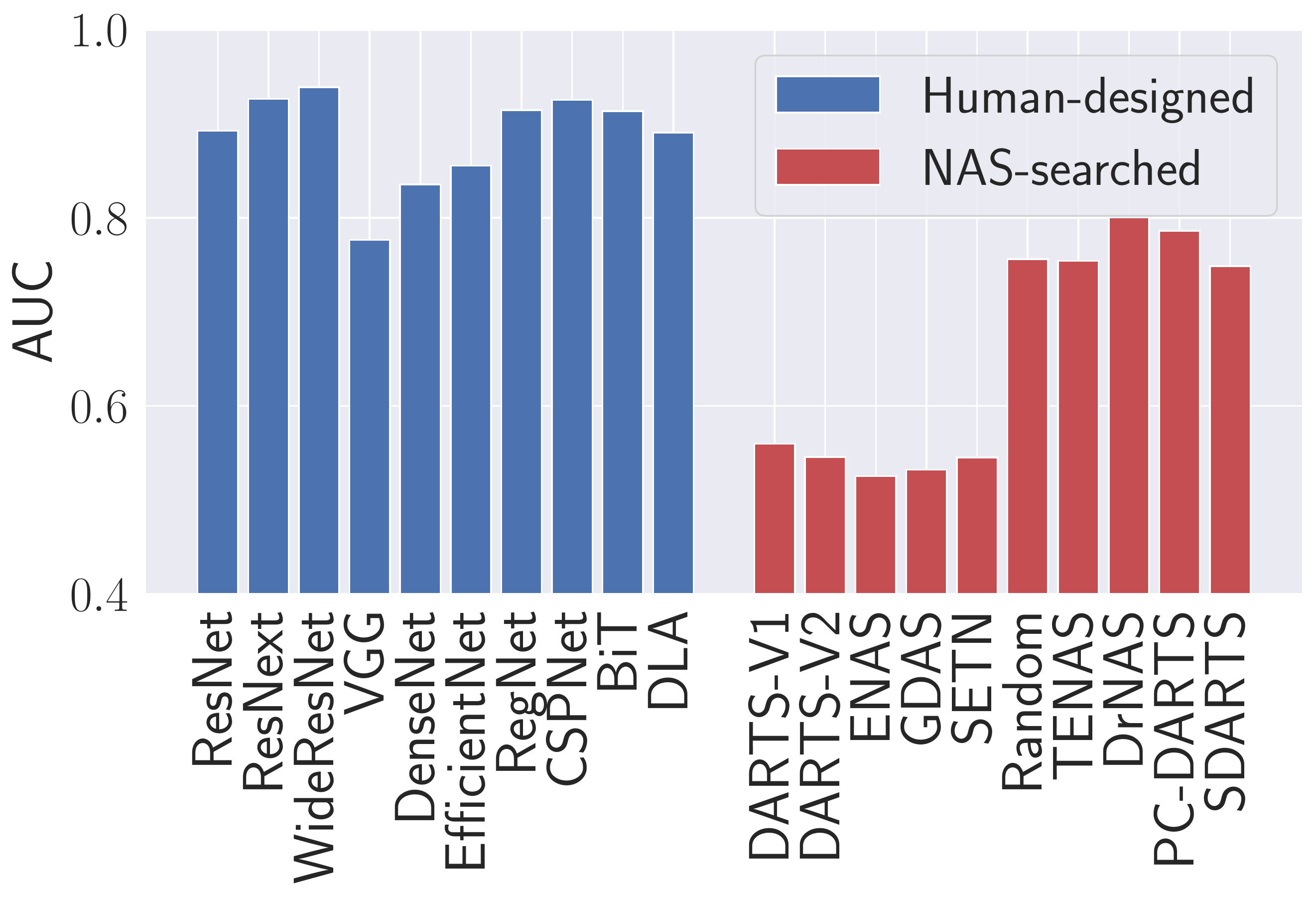}
\caption{CIFAR10}
\label{figure:whitebox_partial_cifar10}
\end{subfigure}
\begin{subfigure}{0.6\columnwidth}
\includegraphics[width=\columnwidth]{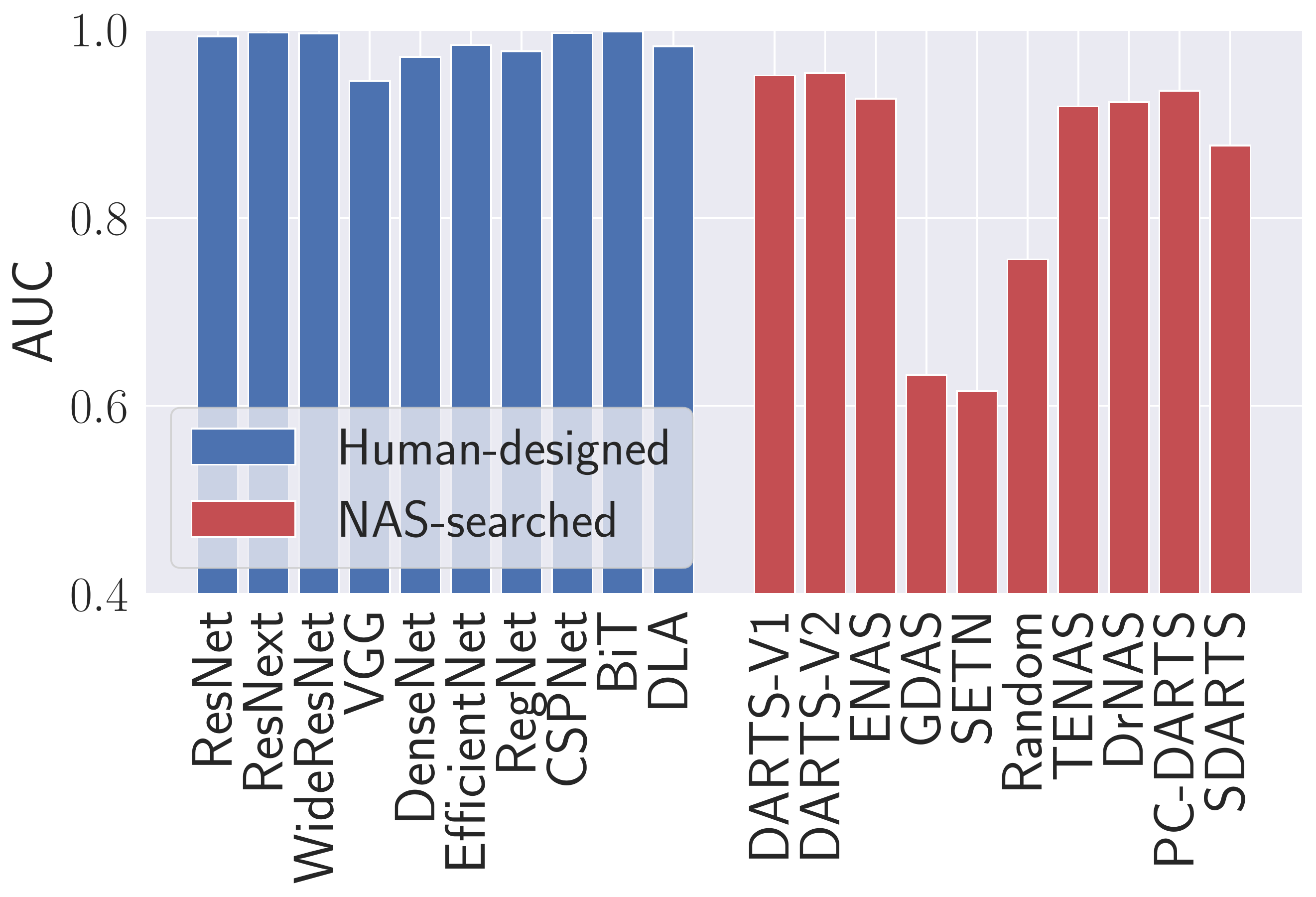}
\caption{CIFAR100}
\label{figure:whitebox_partial_cifar100}
\end{subfigure}
\begin{subfigure}{0.6\columnwidth}
\includegraphics[width=\columnwidth]{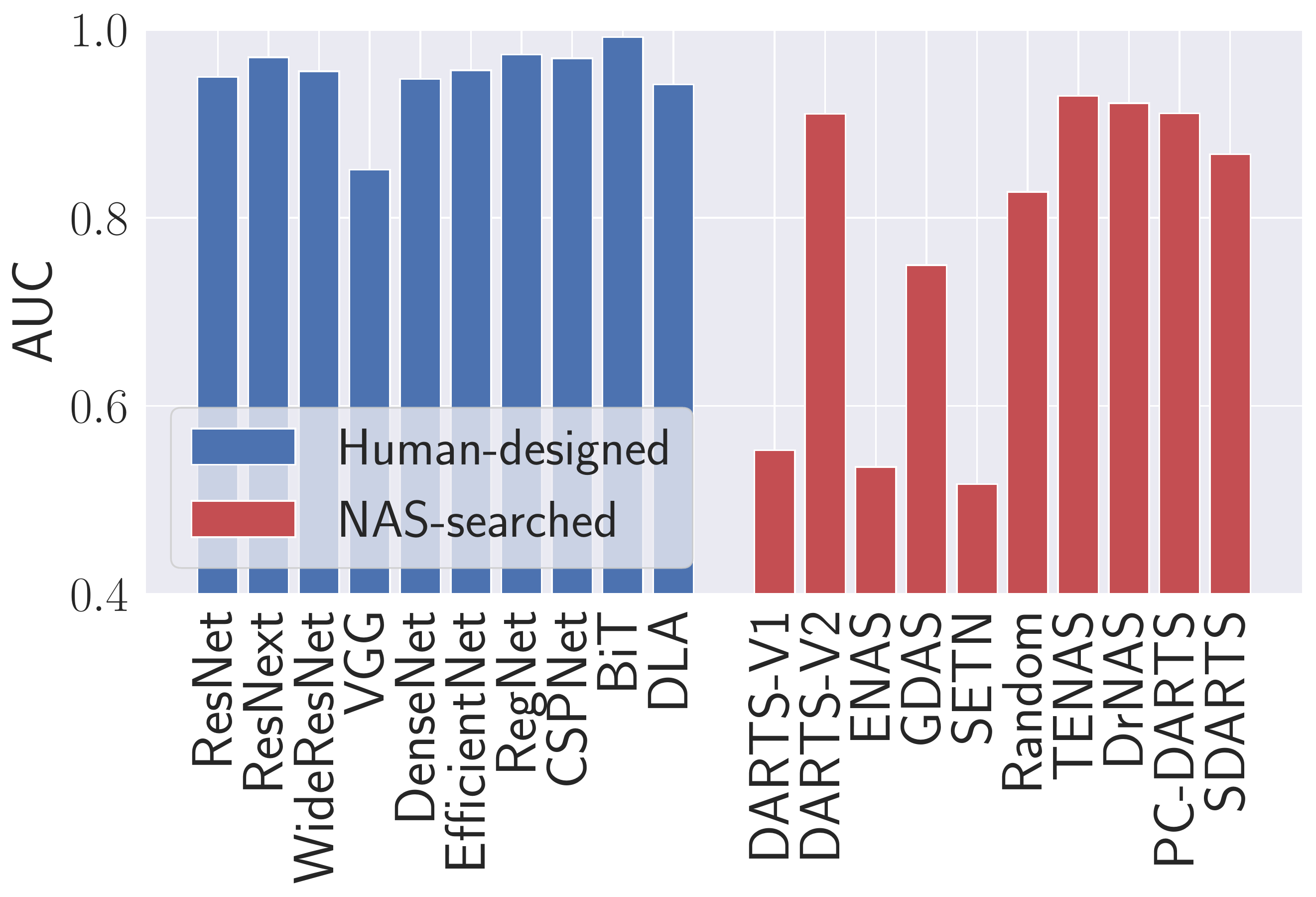}
\caption{STL10}
\label{figure:whitebox_partial_stl10}
\end{subfigure}
\caption{The performance of MIAs with the \tuple{White\mbox{-}Box, Partial} setting on different datasets.}
\label{figure:whitebox_partial_mia_eval}
\end{figure*}

% ----------------------------------------------------
\section{NAS-Bench-201 Search Space}
\label{app:nb201}
% ----------------------------------------------------

\begin{figure}[!t]
\centering
\includegraphics[width=0.43\textwidth]{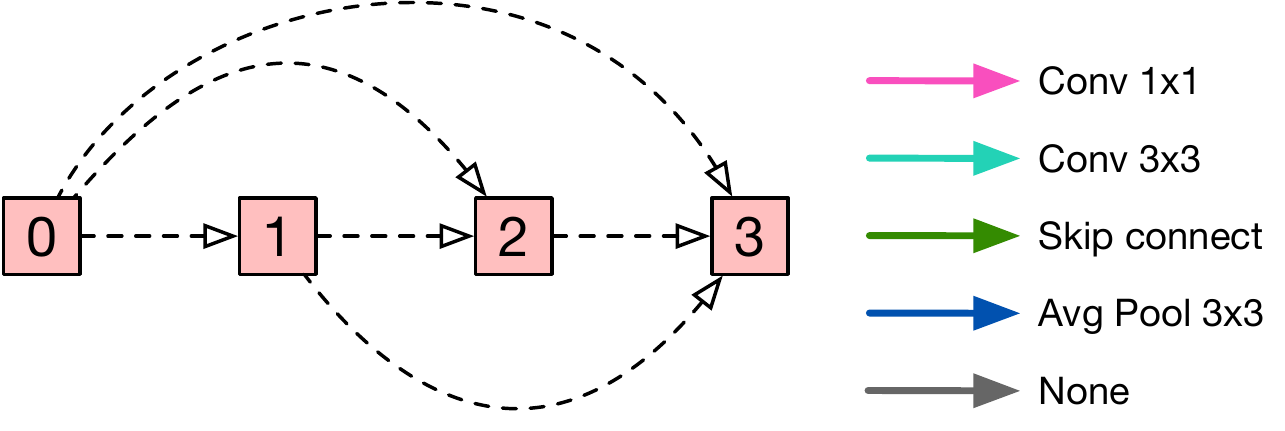}
\caption{The NAS-Bench-201 cell search space. 
The arrows with dashed lines represent the candidate edges whose topology is already predefined while the operation for each edge needs to be searched.} 
\label{figure:cell_nb201}
\end{figure}

NAS-Bench-201~\cite{DY20} search space defines a much simpler cell architecture with fewer candidate operations. ~\autoref{figure:cell_nb201} shows the general architecture of a single cell in the NAS-Bench-201 search space. 
There is only one type of cell in the NAS-Bench-201 search space, and all the nodes in a cell also belong to the same category. 
The candidate operation for each edge is sampled from a set of $K=5$ predefined operations where the ``None'' operation actually means no connection for that edge.

% ----------------------------------------------------
\section{Ablation Studies of Cell Patterns}
\label{app:ablation_studies}
% ----------------------------------------------------

Here we evaluate some factors which may affect the performance of the cell patterns.

\begin{figure}[!t]
\centering
\begin{subfigure}{0.48\columnwidth}
\includegraphics[width=\columnwidth]{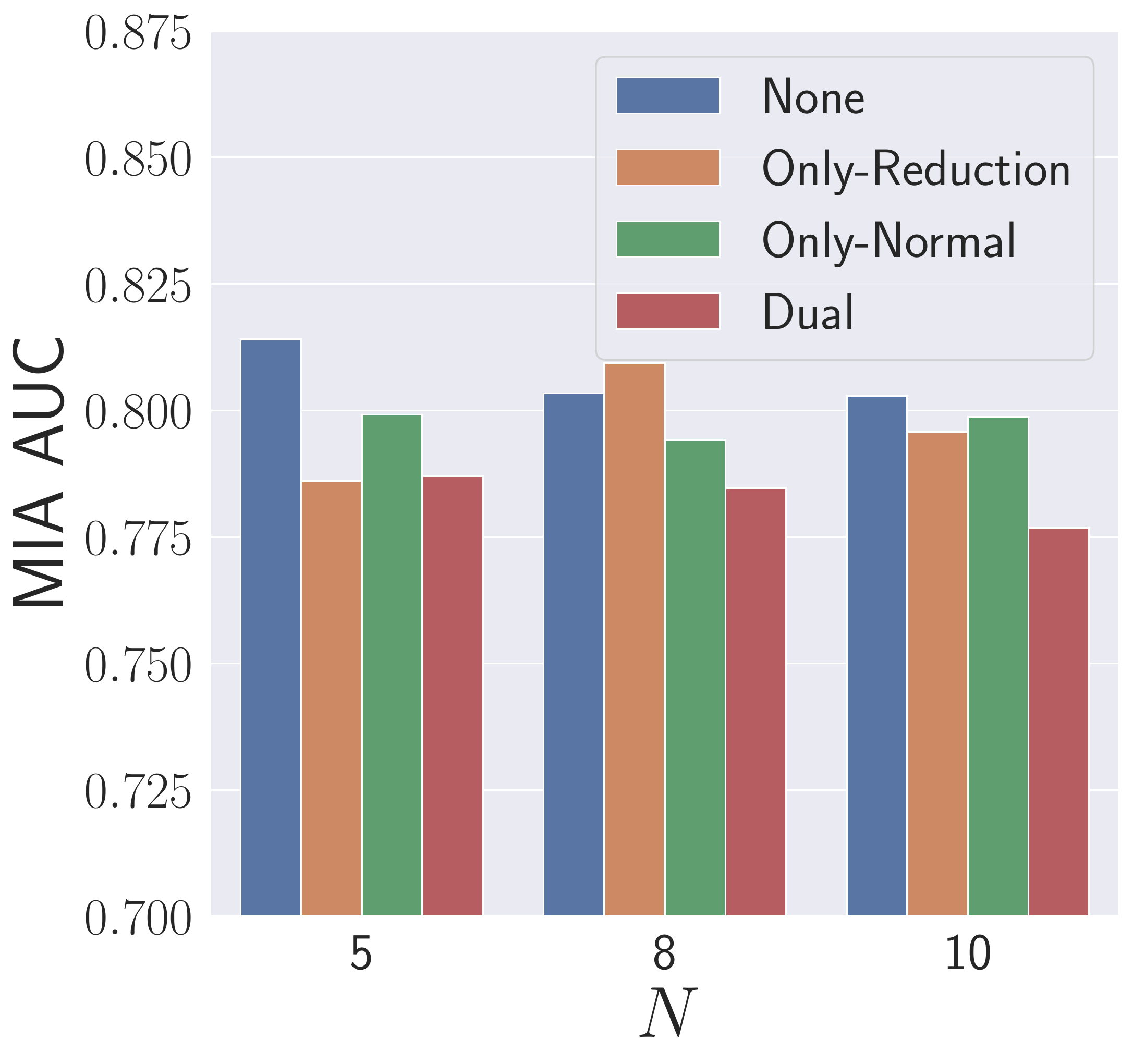}
\caption{Demotion}
\label{figure:mia_auc_ncells_demotion}
\end{subfigure}
\begin{subfigure}{0.48\columnwidth}
\includegraphics[width=\columnwidth]{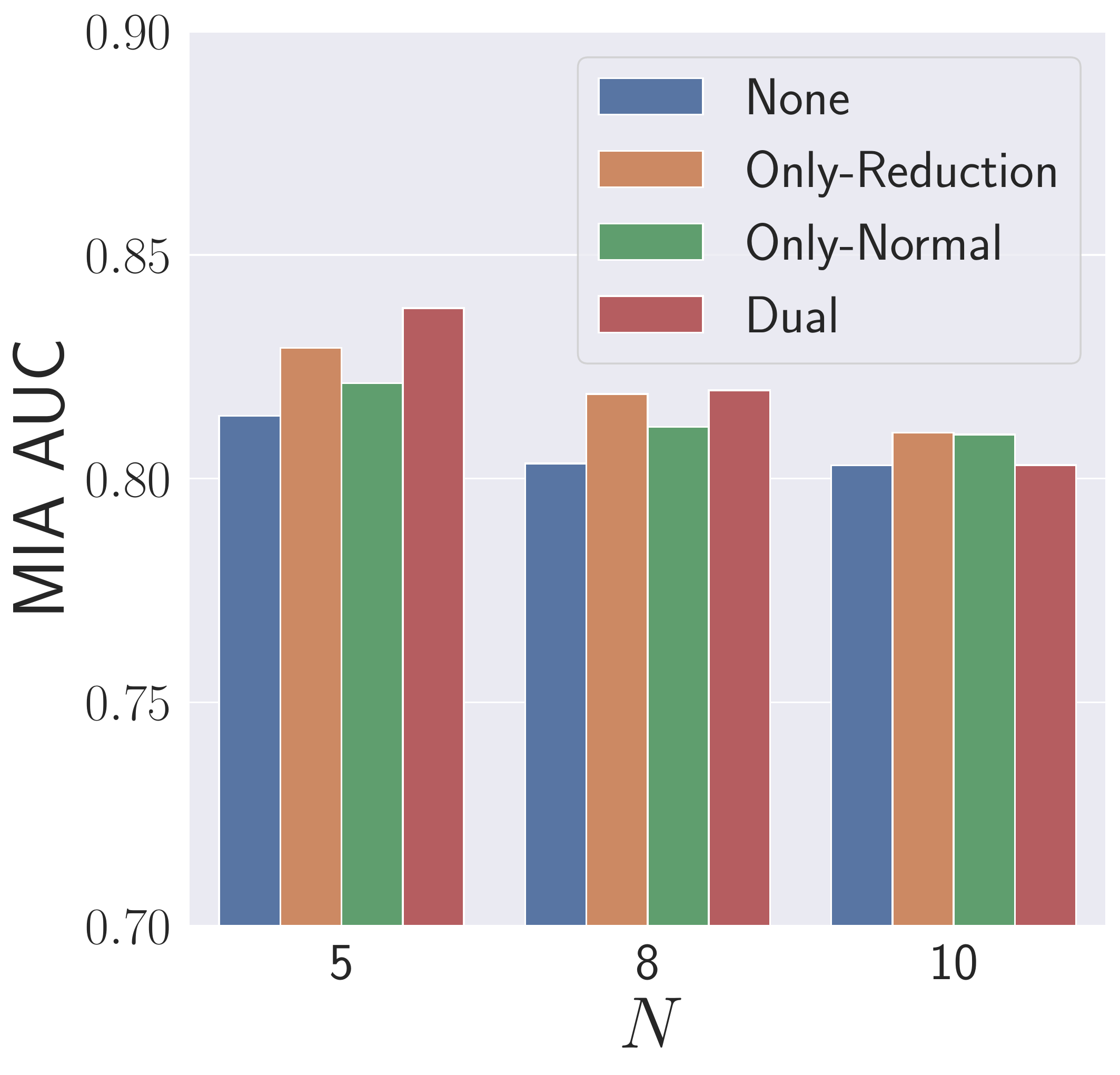}
\caption{Promotion}
\label{figure:mia_auc_ncells_promotion}
\end{subfigure}
\caption{MIA performance when the target model has different number of cells.}
\label{figure:mia_auc_ncells}
\end{figure}

\begin{figure}[!t]
\centering
\begin{subfigure}{0.48\columnwidth}
\includegraphics[width=\columnwidth]{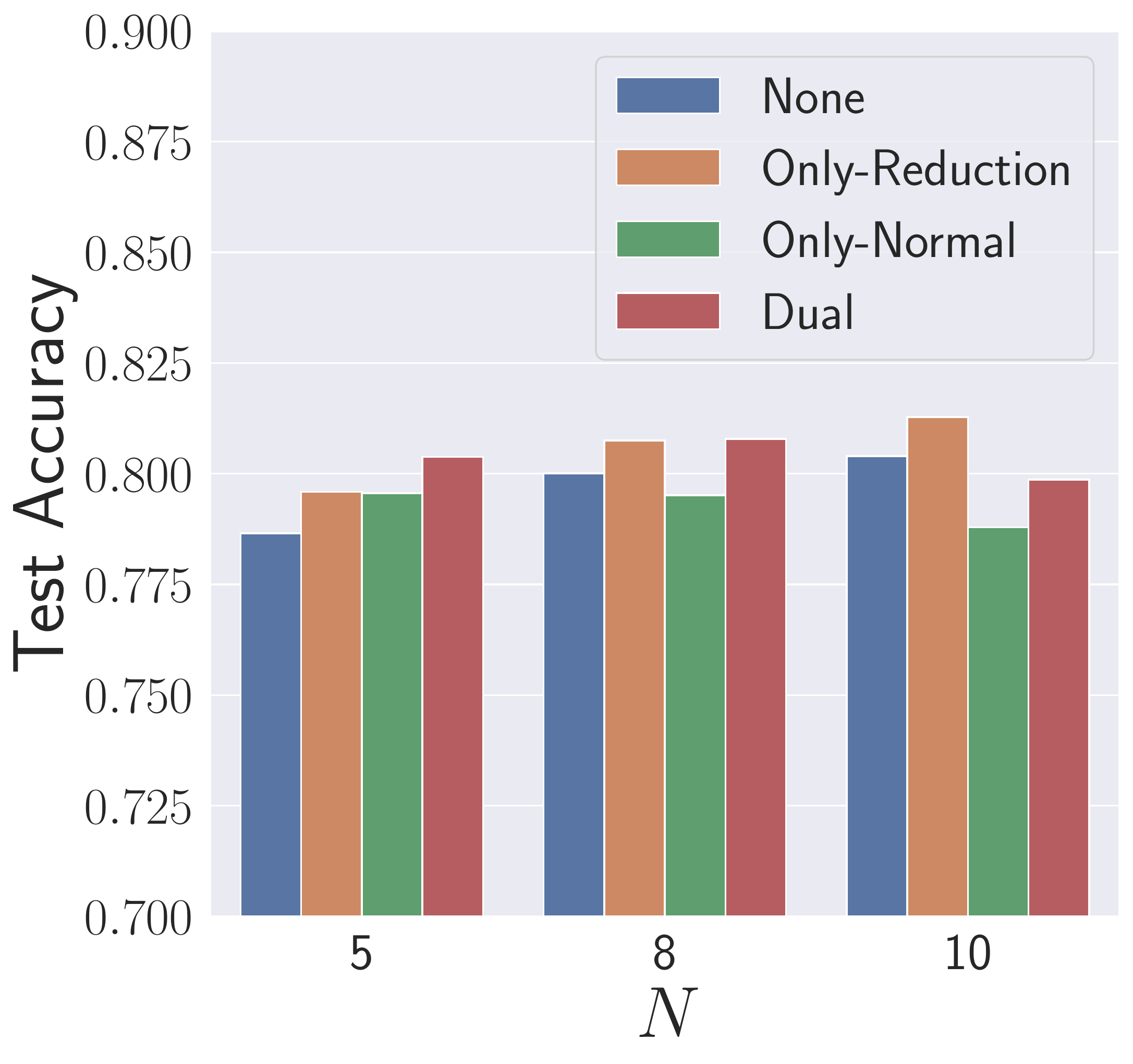}
\caption{Demotion}
\label{figure:test_acc_ncells_demotion}
\end{subfigure}
\begin{subfigure}{0.48\columnwidth}
\includegraphics[width=\columnwidth]{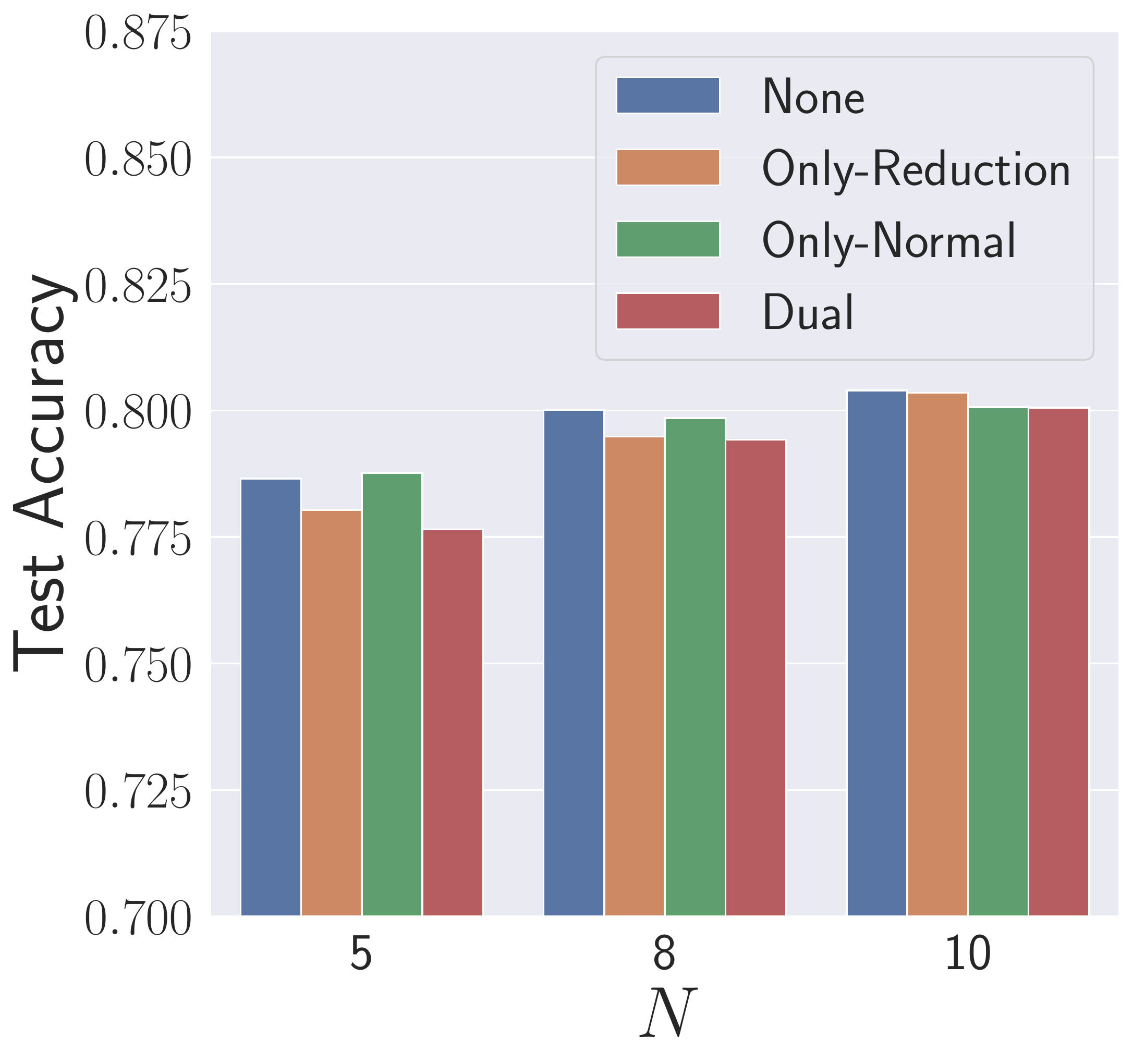}
\caption{Promotion}
\label{figure:test_acc_ncells_promotionn}
\end{subfigure}
\caption{Model utility when the target model has different number of cells.}
\label{figure:test_acc_ncells}
\end{figure}

\mypara{The Impact of the Number of Cells}
When applying cell-based NAS architectures, we can fix the number of repeated cells $N$ to 5 by default in our previous experiments.
We further investigate whether a different $N$ will affect the effectiveness of our extracted cell patterns.
Here we set $N$ to different values and compare the corresponding MIA performance and model utility.
The experimental results are shown in ~\autoref{figure:mia_auc_ncells} and ~\autoref{figure:test_acc_ncells}.
It is straightforward to see that our MIA cell patterns successfully promote or demote the target architectures in most cases.
And also, we can observe that when the number of cells $N$ increases, the test accuracy of the original target model goes up. 
At the same time, the MIA AUC score of the original target model drops. Our results show that the robustness against MIAs and model utility  increases with $N$, indicating that a deeper network can lead to better model performance and stronger robustness.
Moreover, there are usually only two reduction cells in the whole cell-based NAS architecture. 
When we increase the total number of cells $N$, we also increase the number of normal cells.
As such, the effectiveness of the reduction cells  is expected to be weakened as $N$ increases.
Yet, the Only-Reduction cell pattern modifications attain comparable or even better performance than the other two types of modifications in most cases. 
Our results demonstrate that reduction cells are still an important tool to improve model robustness.
As for the normal cells, they increase in number when $N$ becomes larger. 
The impact of the cell patterns on them is expected to be stronger. However, the increased depth of the target model architectures would also exert more transformations on the original data information. 
It becomes increasingly hard to extract information with high fidelity for MIAs in this situation.
As a result, the performance of Only-Normal modifications has not been promoted much even when $N$ increases.

\begin{figure}[!t]
\centering
\begin{subfigure}{0.48\columnwidth}
\includegraphics[width=\columnwidth]{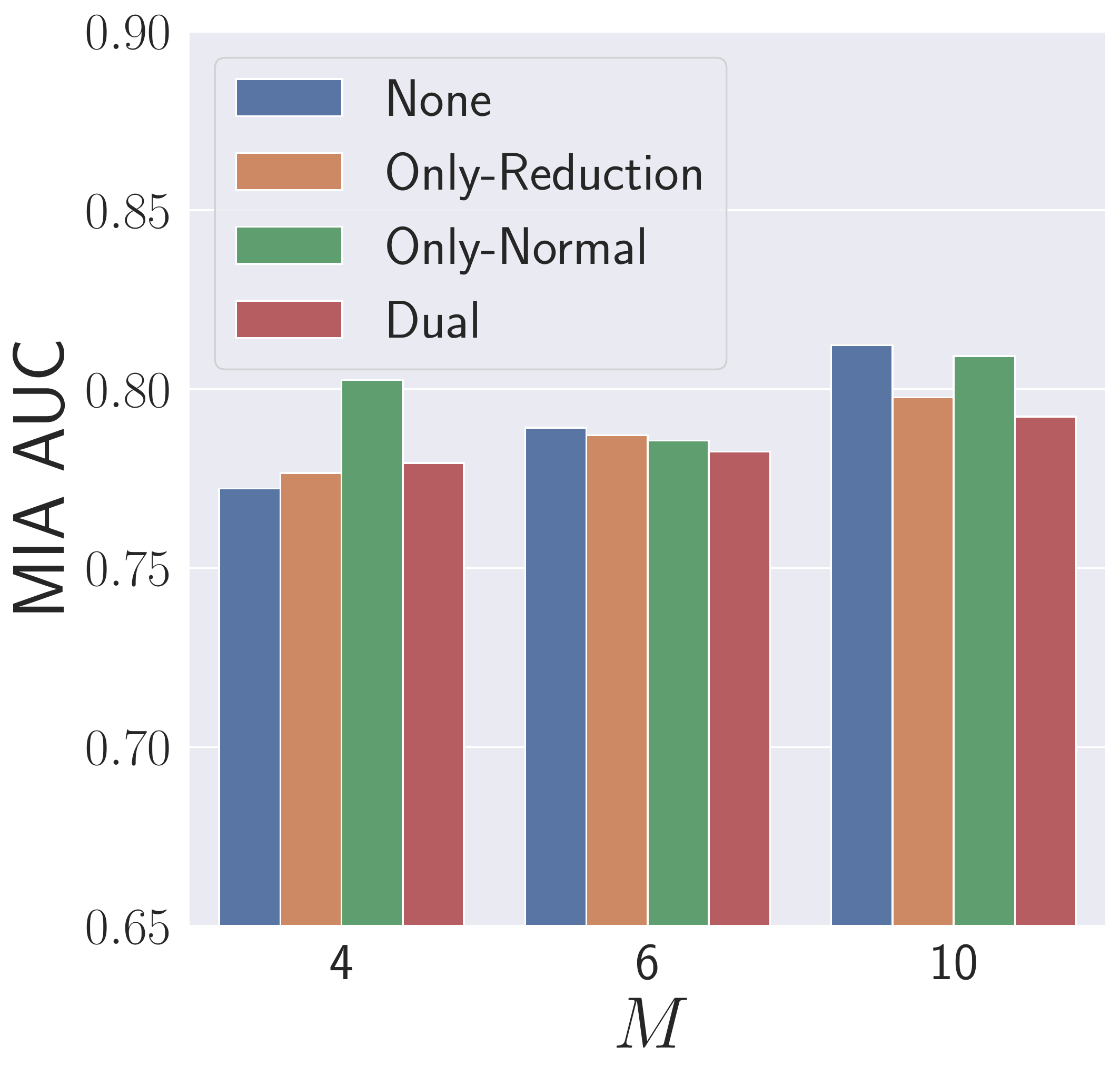}
\caption{Demotion}
\label{figure:mia_auc_inter_nodes_demotion}
\end{subfigure}
\begin{subfigure}{0.48\columnwidth}
\includegraphics[width=\columnwidth]{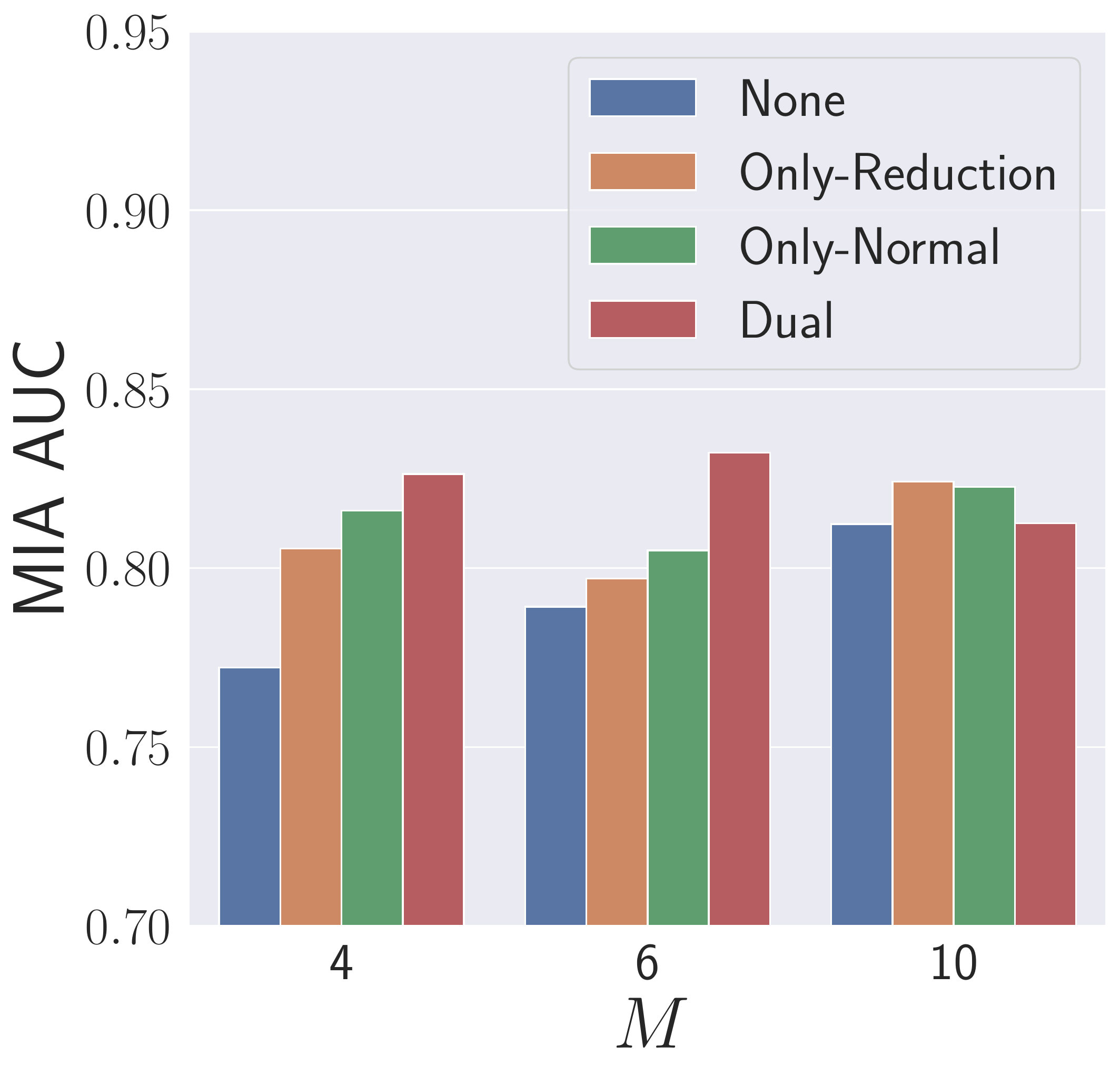}
\caption{Promotion}
\label{figure:mia_auc_inter_nodes_promotion}
\end{subfigure}
\caption{MIA performance when the cell has different number of intermediate nodes.}
\label{figure:mia_auc_intermediate_nodes}
\end{figure}

\begin{figure}[!t]
\centering
\begin{subfigure}{0.48\columnwidth}
\includegraphics[width=\columnwidth]{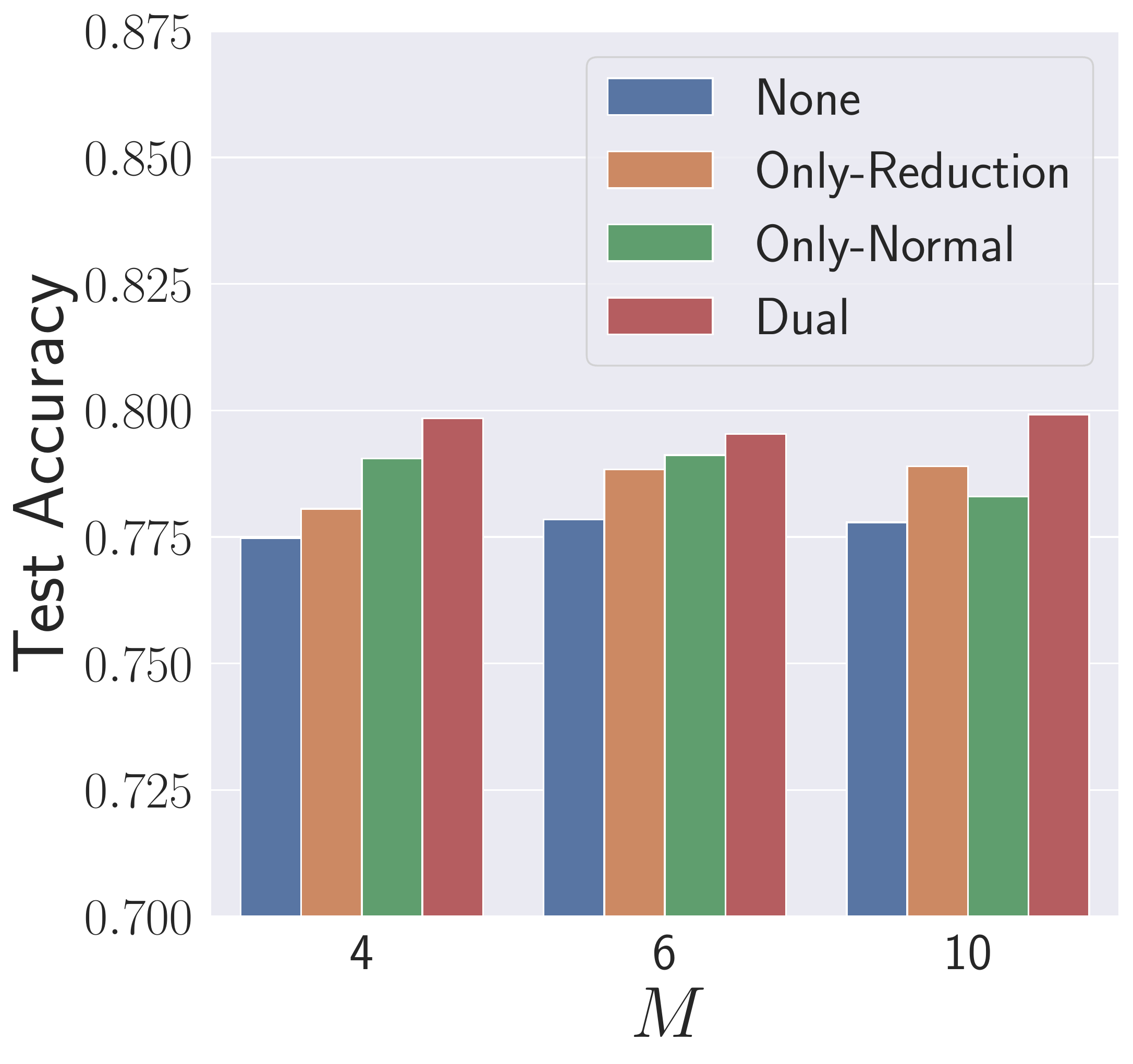}
\caption{Demotion}
\label{figure:test_acc_inter_nodes_demotion}
\end{subfigure}
\begin{subfigure}{0.48\columnwidth}
\includegraphics[width=\columnwidth]{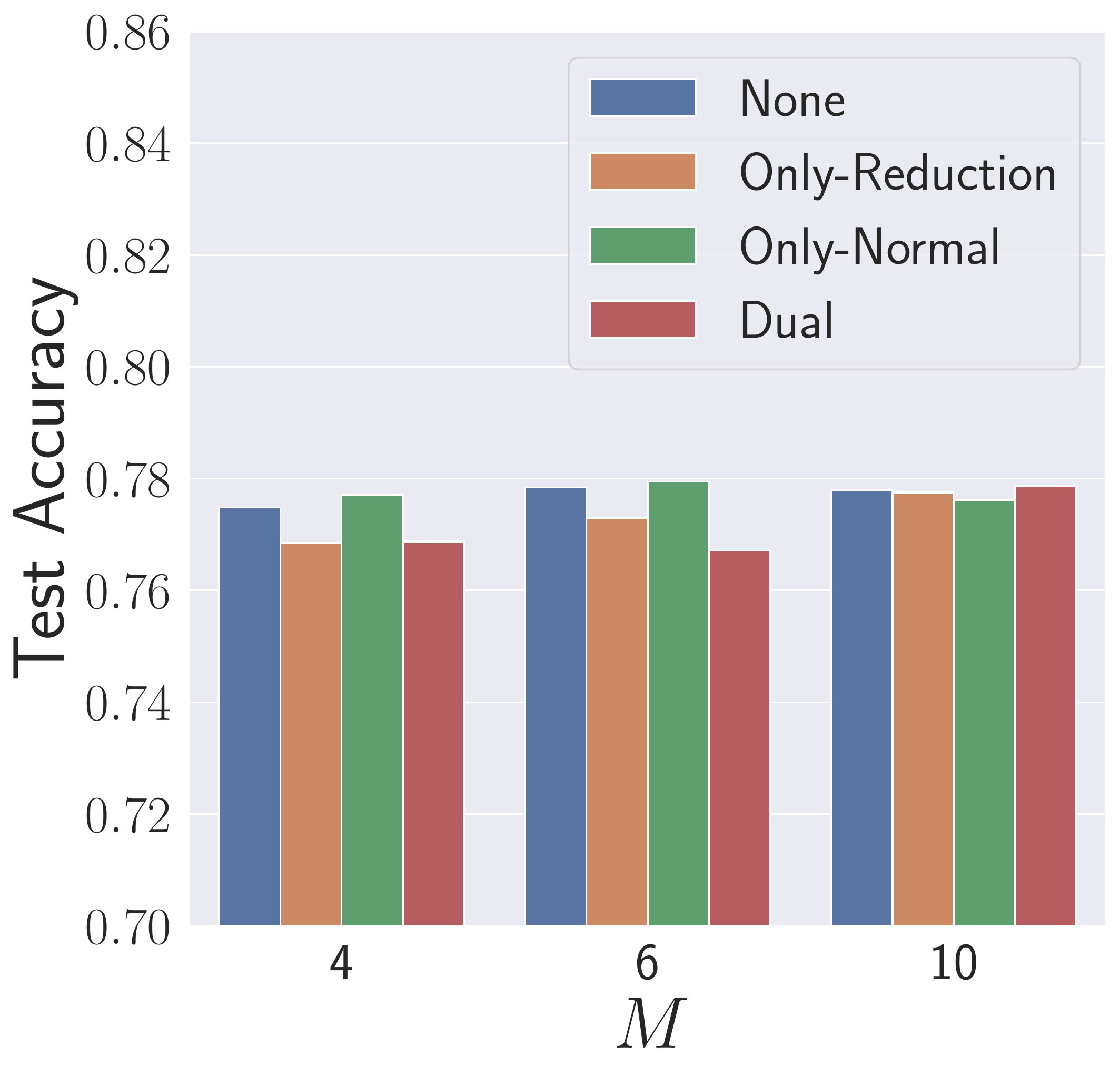}
\caption{Promotion}
\label{figure:test_acc_inter_nodes_promotion}
\end{subfigure}
\caption{Model utility when the cell has different number of intermediate nodes.}
\label{figure:test_acc_intermediate_nodes}
\end{figure}

\mypara{The Impact of the Number of Intermediate Nodes}
The default number of intermediate nodes $M$ for our previous experiments is 4. 
We vary the value of $M$ to estimate its impact on the effectiveness of the cell patterns in this study.
We use the last four cell-based NAS methods (see \autoref{sec:nas_algos}) that use the DARTS search space in our experiments to generate corresponding architectures with different $M$ values. 
The experimental results are shown in \autoref{figure:mia_auc_intermediate_nodes} and \autoref{figure:test_acc_intermediate_nodes}. 
Note that the AUC scores are averaged.
When the number of intermediate nodes $M$ increases, the MIA AUC score of the target model significantly rises even though the test accuracy of the target model has only negligible changes. 
This indicates that the increase in $M$ could degrade model robustness against MIAs without increasing model utility.
The possible reason is that the cell which has more intermediate nodes can offer more internal architectural means for the attacker to aggregate and recover the information from preceding cells.
Besides, we find that our MIA demotion cell patterns do not work well when $M$ is small (e.g., 4), but at the same time, the MIA promotion cell patterns show good performance. 
The reason is that our MIA demotion cell patterns are extracted from the well-performed architectures.
Their MIA AUC scores range between 0.7311 to 0.78, very close to the MIA AUC score of the original target model when $M$ is small (e.g., MIA AUC equals to 0.7722 when $M=4$). 
But the MIA promotion cell patterns are learned from the architectures with relatively high MIA AUC scores.
This also counts for the phenomenon that the performance of MIA demotion cell patterns increases and the performance of MIA promotion cell patterns becomes less relatively significant when $M$ turns out to be larger.
It will be interesting to explore new methods to acquire effective cell patterns not limited to existing cell architectures in the future.

% ----------------------------------------------------
\section{Transferability on the CelebA Dataset}
\label{appendix:transfer_celeba}
% ----------------------------------------------------

\begin{figure}[!t]
\centering
\begin{subfigure}{0.48\columnwidth}
\includegraphics[width=\columnwidth]{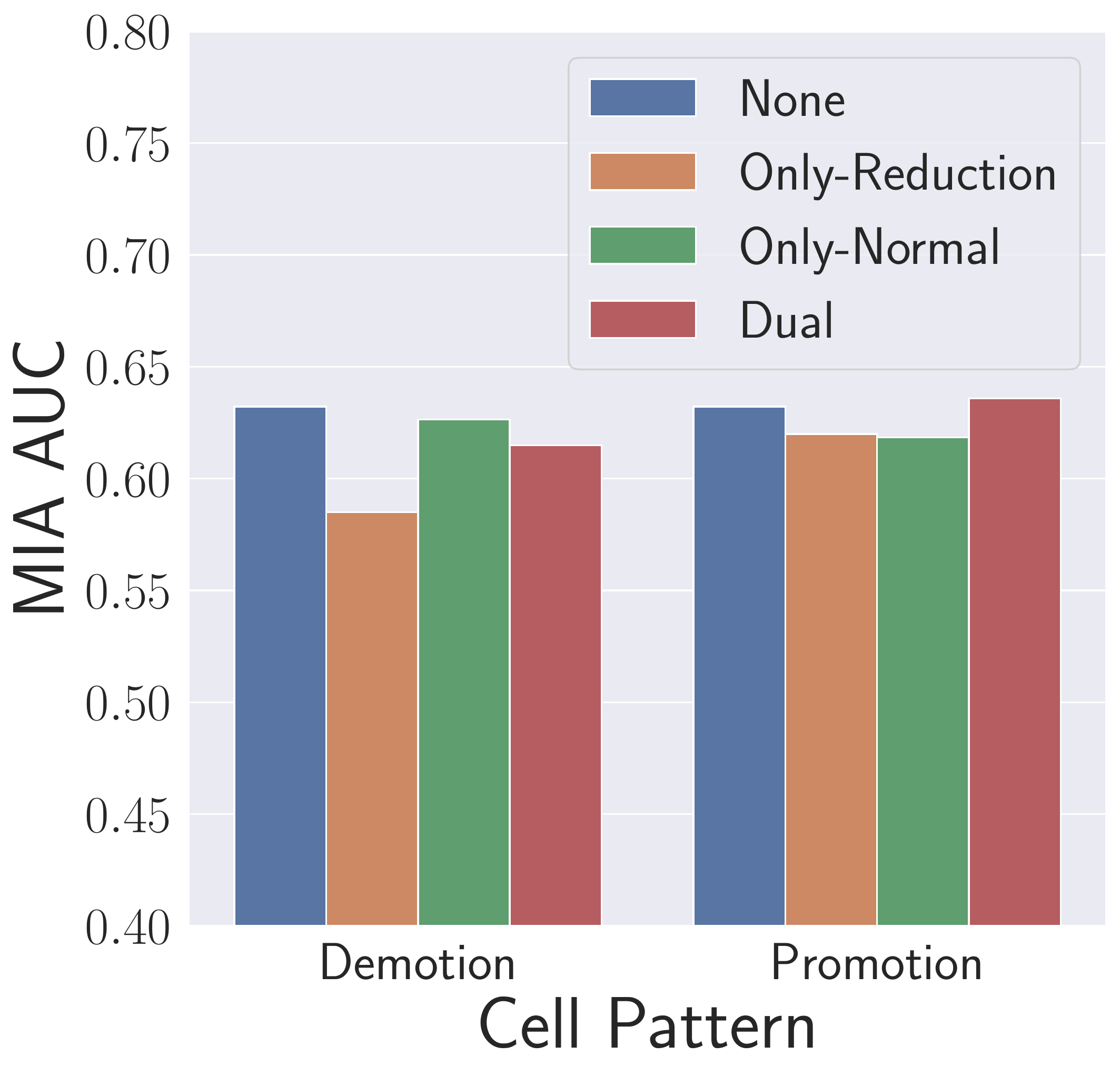}
\caption{MIA Performance}
\label{fig:mia_auc_transfer_celeba}
\end{subfigure}
\begin{subfigure}{0.48\columnwidth}
\includegraphics[width=\columnwidth]{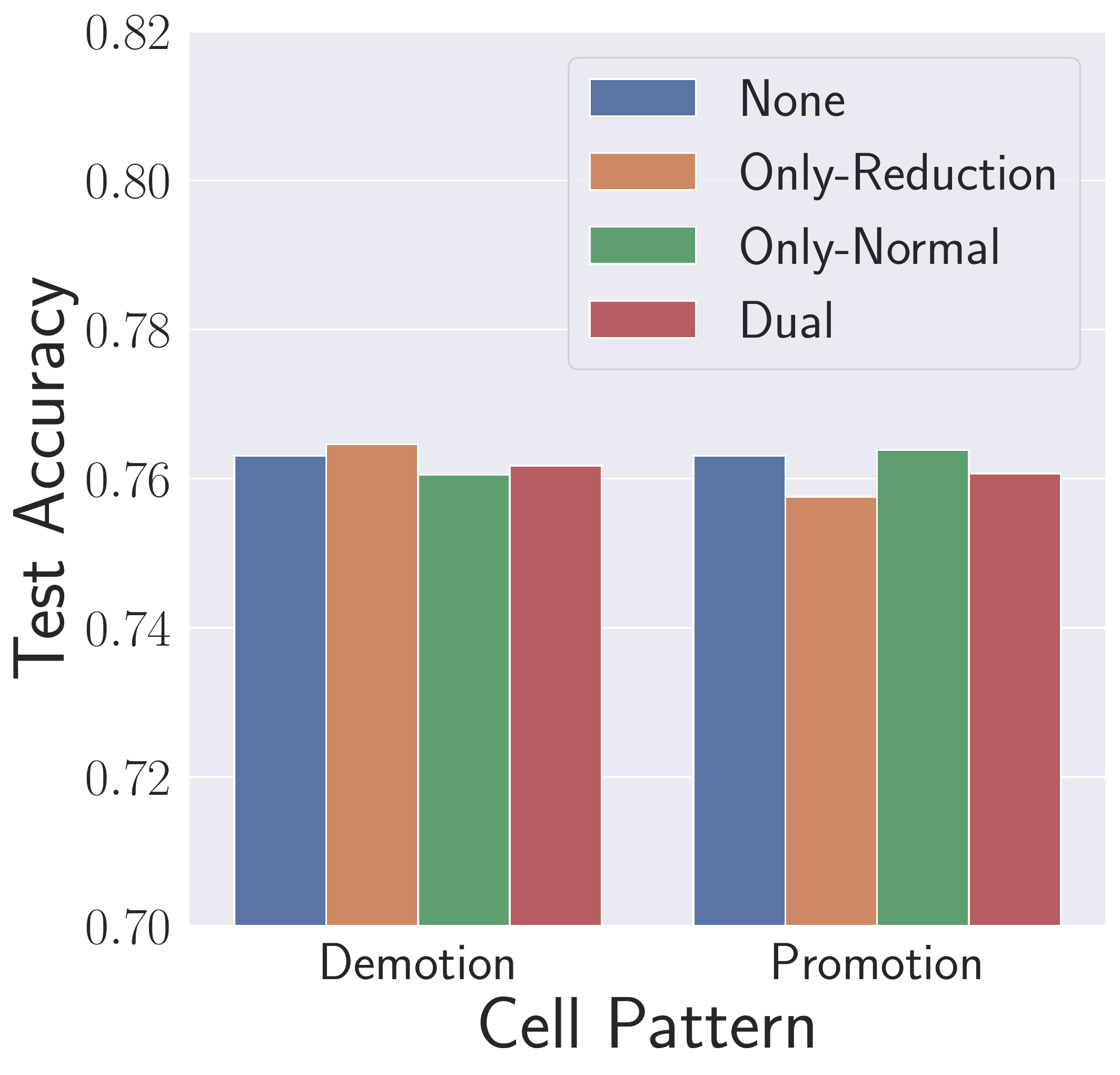}
\caption{Model Utility}
\label{fig:test_acc_transfer_celeba}
\end{subfigure}
\caption{Transferability of the cell patterns on the CelebA dataset.}
\label{figure:transfer_celeba}
\end{figure}

Here we conduct the cell pattern transferability experiments on the CelebA dataset and we keep the same experimental settings as \autoref{subsec:transferability}. 
Our cell patterns are extracted from the architectures searched on the CIFAR10 dataset. 
We aim to test whether they can be successfully transferred to the CelebA dataset. 
The MIA performance and model utility evaluation results for the transferability experiments are presented in \autoref{fig:mia_auc_transfer_celeba} and \autoref{fig:test_acc_transfer_celeba}, respectively.
As we can see from \autoref{fig:mia_auc_transfer_celeba}, our MIA demotion cell patterns, especially the Only-Reduction type, can successfully transfer to the CelebA dataset. 
For instance, the Only-Reduction MIA demotion modifications can decrease the MIA AUC from 0.6322 to 0.5850. 
However, our cell patterns are not always effective for MIA promotion on the CelebA dataset.
Furthermore, as shown in \autoref{fig:test_acc_transfer_celeba},  our cell patterns especially the MIA demotion ones have a slight impact on the model utility of the target model.
Yet, the Only-Reduction cell patterns can still promote the model utility of the target model. 
We speculate that the reason is that the human face dataset CelebA has a very different distribution from the CIFAR10 dataset, resulting in very different NAS-searched architectures on the CelebA dataset. 
In this case, the cell patterns extracted from the CIFAR10 dataset do not all work effectively on the CelebA dataset.
In short, our MIA demotion cell patterns are still effective in this scenario and the Dual modifications work for both MIA demotion and promotion cell patterns. 
They can offer useful insights for designing NAS architectures more robust against MIAs on this human face dataset in the future.

% ----------------------------------------------------
\section{Transferability in Different Search Spaces}
\label{sec:transfer_search_space}
% ----------------------------------------------------

Our cell patterns are extracted from the architectures generated from the DARTS search space, and we want to test whether they can be transferred to some other search space. Here we choose the NAS-Bench-201 search space as a study case.

% ----------------------------------------------------
\subsection{Effectiveness of Transferring to NAS-Bench-201}
% ----------------------------------------------------

Here we choose the representative NAS-Bench-201 search space as the target search space. 
We use the first 6 NAS methods (see \autoref{sec:nas_algos}) to generate 6 architectures in the NAS-Bench-201 search space on the CIFAR10 dataset.
Note that the NAS-Bench-201 search space is a much smaller search space where there is only one type of cell containing 4 identical nodes each. 
We  therefore can not directly apply our cell patterns to this search space.
However, when we do not consider the topological information, our previous findings lead to two interesting observations. 
First, max-pooling operations tend to be favored by the reduction cells for MIA demotion. 
Second, convolution operations are preferred by both normal and reduction cells for MIA promotion in the DARTS search space.
As such, we just use this simplest cell pattern information and modify the architectures in the NAS-Bench-201 search space by assigning specific operations to some edges.
We consider the max pooling operation and convolution operation as the target operations for MIA demotion and promotion respectively in the NAS-bench-201 search space.
Besides, according to previous observations~\cite{CXWT19}, NAS cells tend to have too many skip connections due to the fast error decay of these operations during optimization. 
We replace this operation preferentially and try to avoid modifying the topological edge links between nodes.
We ensure that each node is connected with at least one target operation, and then compare the MIA performance before and after the architectural modifications.

\begin{figure}[!t]
\centering
\begin{subfigure}{0.45\columnwidth}
\includegraphics[width=\columnwidth]{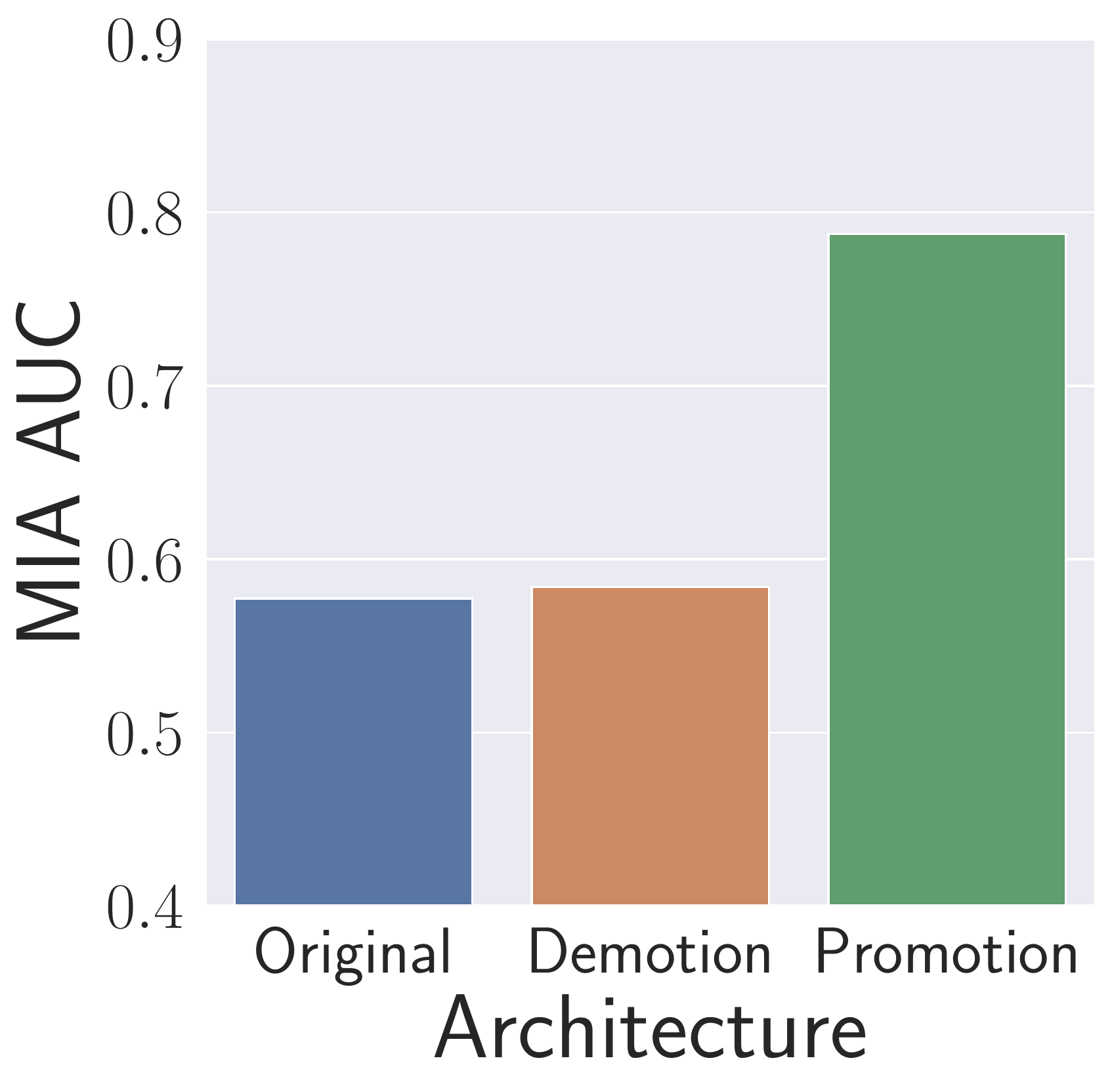}
\caption{MIA Performance}
\label{figure:mia_auc_transfer_nb201}
\end{subfigure}
\begin{subfigure}{0.45\columnwidth}
\includegraphics[width=\columnwidth]{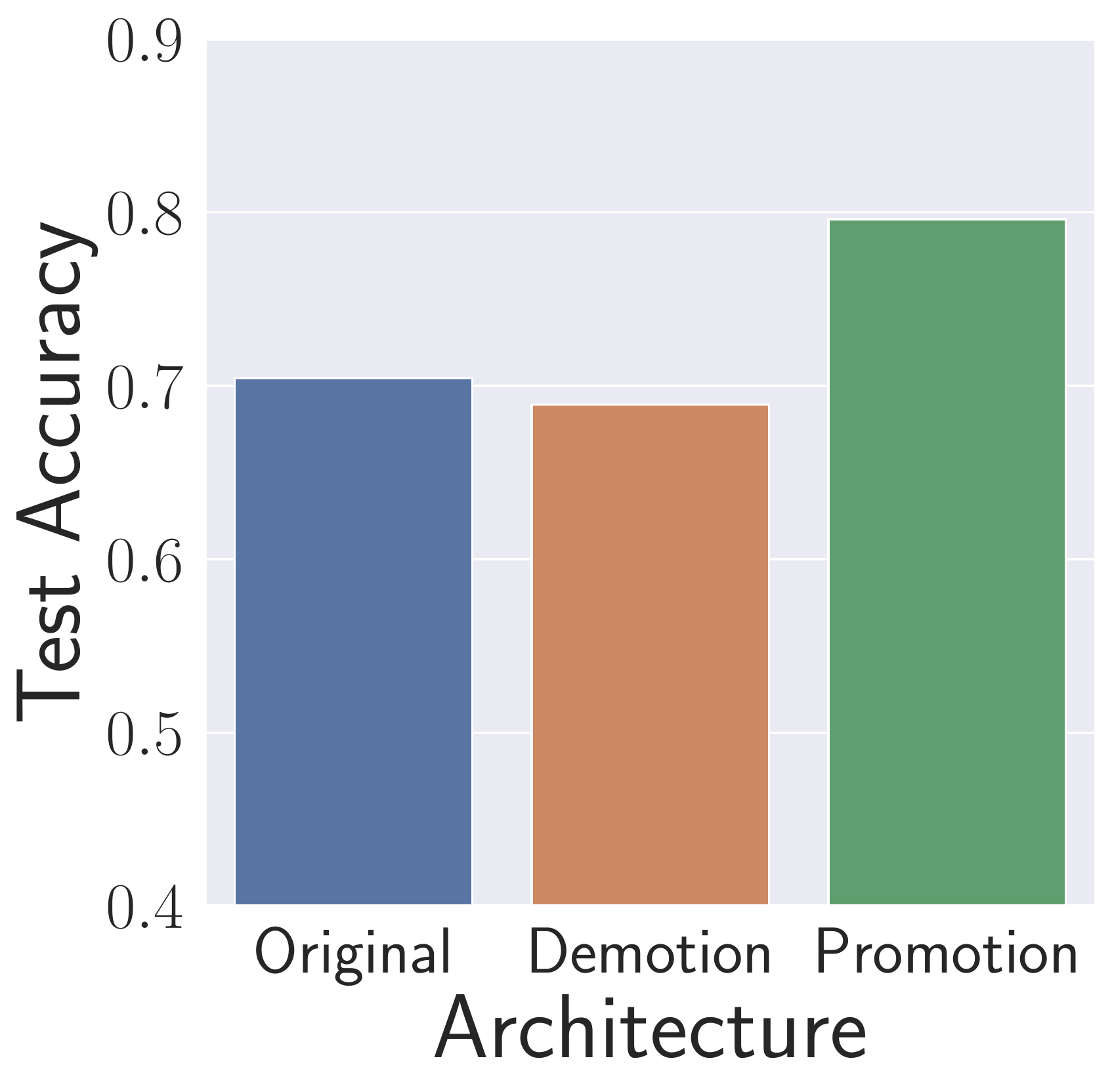}
\caption{Model Utility}
\label{figure:test_acc_tansfer_nb201}
\end{subfigure}
\caption{MIA effectiveness and model utility in the NAS-Bench-201 search space.}
\label{figure:transfer_to_nb201}
\end{figure}

The experimental results are presented in \autoref{figure:transfer_to_nb201}. 
We can see that MIA promotion modifications are  effective while the MIA demotion ones do not work well in this case.
The possible reason is that the MIA AUC score (i.e., 0.5773) of the target architectures is already low, and it would be hard to further degrade MIA performance, but easy to promote it from a small value. 
Interestingly, we find that the MIA demotion modifications make the test accuracy of the target architectures descend while the MIA promotion modifications make that ascend, which is in contrast to our previous observations in \autoref{subsec:effect_cell_pattern}. 
We offer a detailed explanation in ~\autoref{sec:loss_analysis_nb201} for this observation from the viewpoint of loss contour.
Overall, even the transferred cell patterns here with very limited information from our original cell patterns can still ``partially'' transfer to the NAS-Bench-201 search space.
And, it will be interesting to combine more topological information to strengthen the transferred cell patterns, e.g., using our aforementioned general framework to analyze the cell patterns on the architectures sampled from NAS-Bench-201 search space.

% ----------------------------------------------------
\subsection{Loss Contour Analysis for NAS-Bench-201}
\label{sec:loss_analysis_nb201}
% ----------------------------------------------------

\begin{figure}[!t]
\centering
\begin{subfigure}{0.52\columnwidth}
\includegraphics[width=\columnwidth]{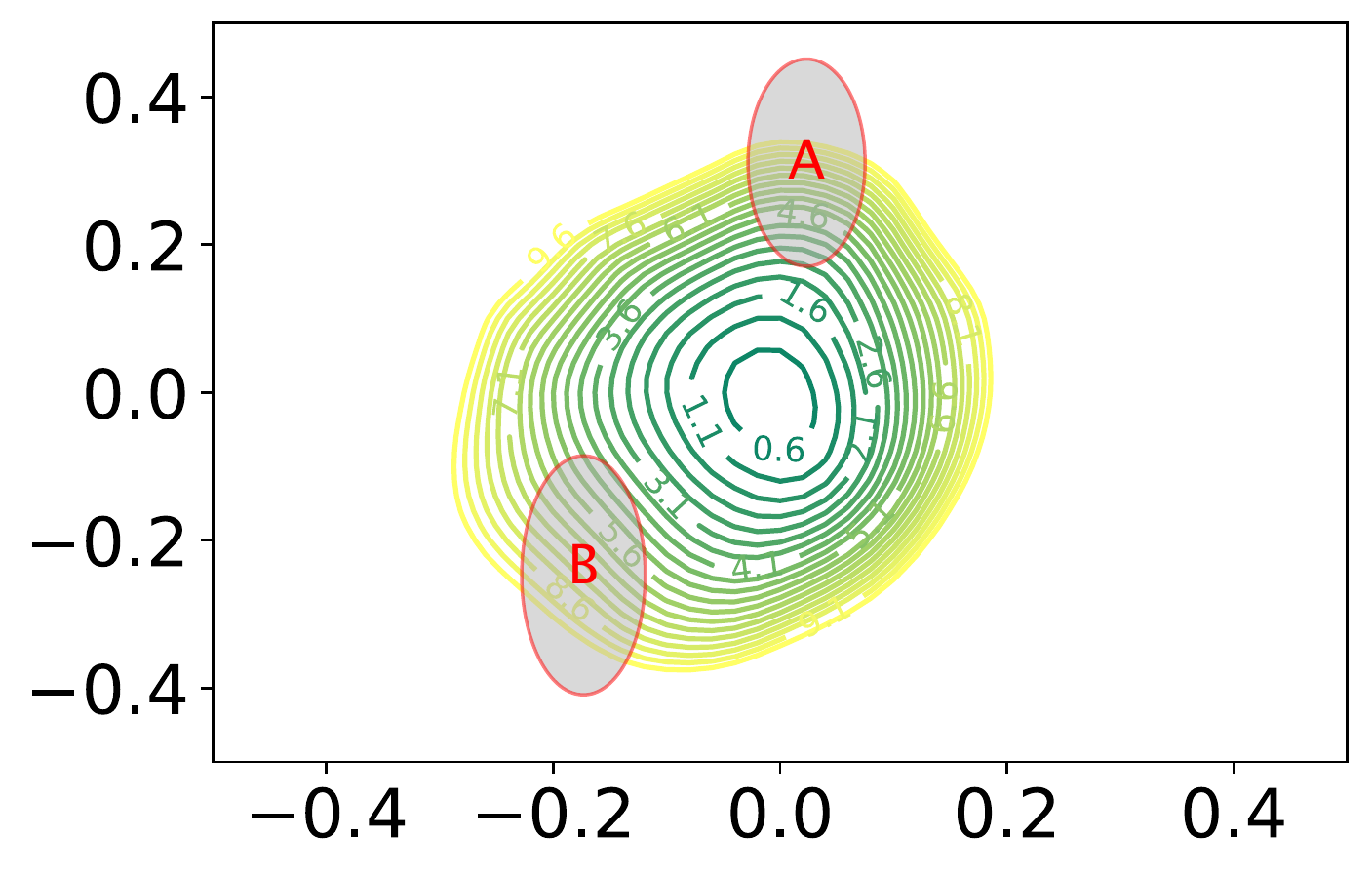}
\caption{Original}
\label{figure:loss_origin_nb201}
\end{subfigure}
\begin{subfigure}{0.49\columnwidth}
\includegraphics[width=\columnwidth]{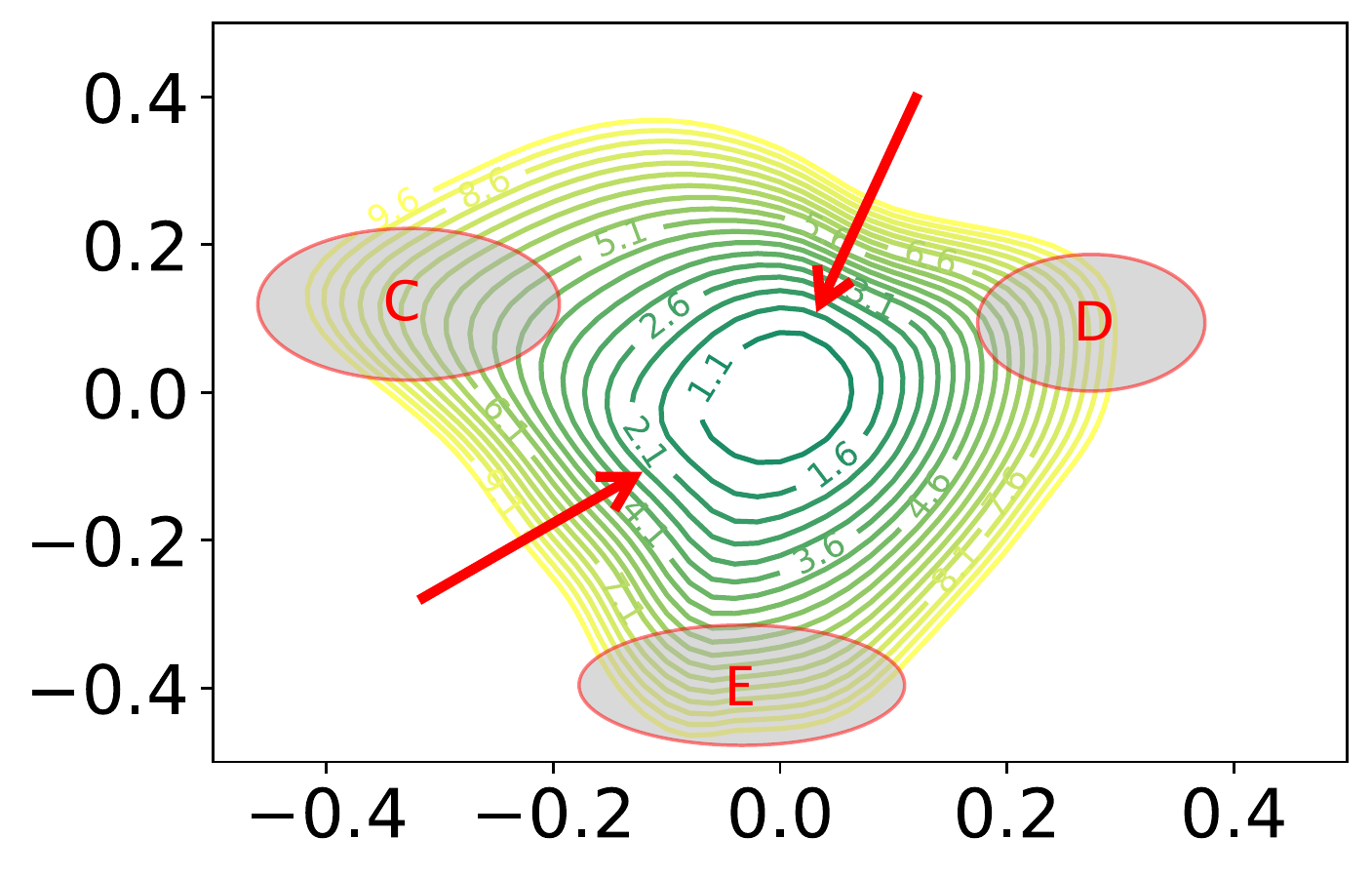}
\caption{After Demotion}
\label{figure:loss_after_demotion_nb201}
\end{subfigure}
\begin{subfigure}{0.49\columnwidth}
\includegraphics[width=\columnwidth]{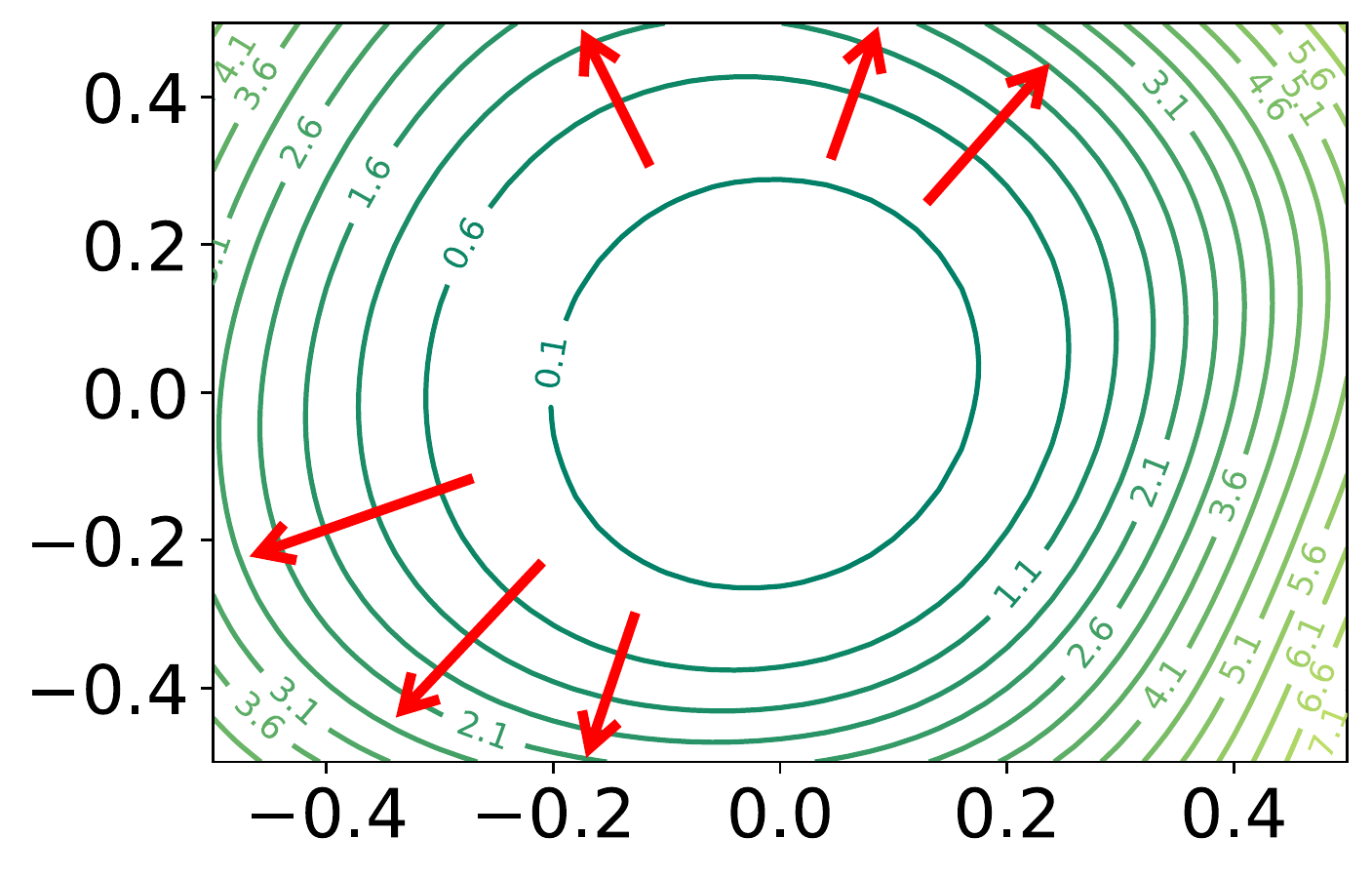}
\caption{After Promotion}
\label{figure:loss_after_promotion_nb201}
\end{subfigure}
\caption{The contours before and after MIA demotion or promotion modifications of the target architecture in the NAS-Bench-201 search space.}
\label{figure:loss_contour_nb201}
\end{figure}

When conducting the transferability experiments on the cell architectures in the NAS-Bench-201 search space, we find that some phenomena are inconsistent with previous results in the DARTS search space.
Here we analyze the possible reasons for the loss contours before and after the architectural modifications in the NAS-Bench-201 search space. 
We select one architecture used in previous transferability experiments in the NAS-Bench-201 search space and plot its loss contour in \autoref{figure:loss_contour_nb201}.
As we can see, the ``trustworthy'' areas in the original loss contour are squeezed as expected after the MIA demotion modifications. 
However, in the loss contour of the modified architecture, some new ``trustworthy'' areas (i.e., $C$, $D$ and $E$ in \autoref{figure:loss_after_demotion_nb201}) appear and are even larger than the original ones. 
In turn, the MIA performance has not been further demoted successfully.
Also, the current loss contour is less smoother than the original one since there are more corners in the current loss contour, which leads to a slight drop in model performance.
As for the MIA promotion, we can see that the loss contour after the modifications is significantly flattened, and almost everywhere in the model weight parameter space is ``trustworthy'' for MIAs. 
Therefore, the MIA performance is improved after the modifications. 
Plus, the loss contour is much smoother than the original one, leading to an ascend in the model performance.

% ----------------------------------------------------
\section{Transferred Defense Effectiveness}
\label{sec:transfer_defense}
% ----------------------------------------------------

\begin{figure}[!t]
\centering
\begin{subfigure}{0.48\columnwidth}
\includegraphics[width=\columnwidth]{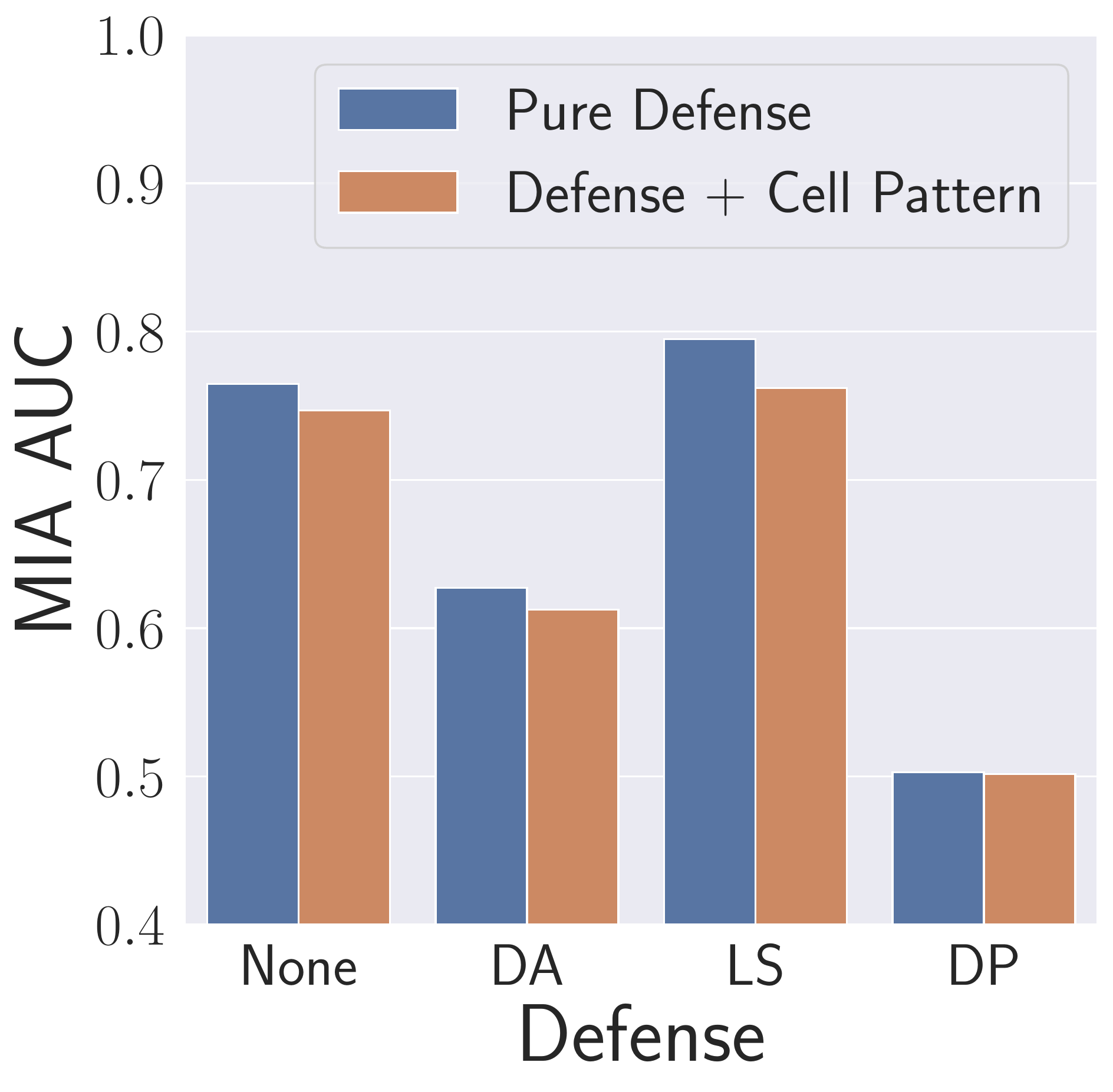}
\caption{\tuple{Black\mbox{-}Box, Shadow}}
\label{figure:mia_auc_defense_meminf0}
\end{subfigure}
\begin{subfigure}{0.48\columnwidth}
\includegraphics[width=\columnwidth]{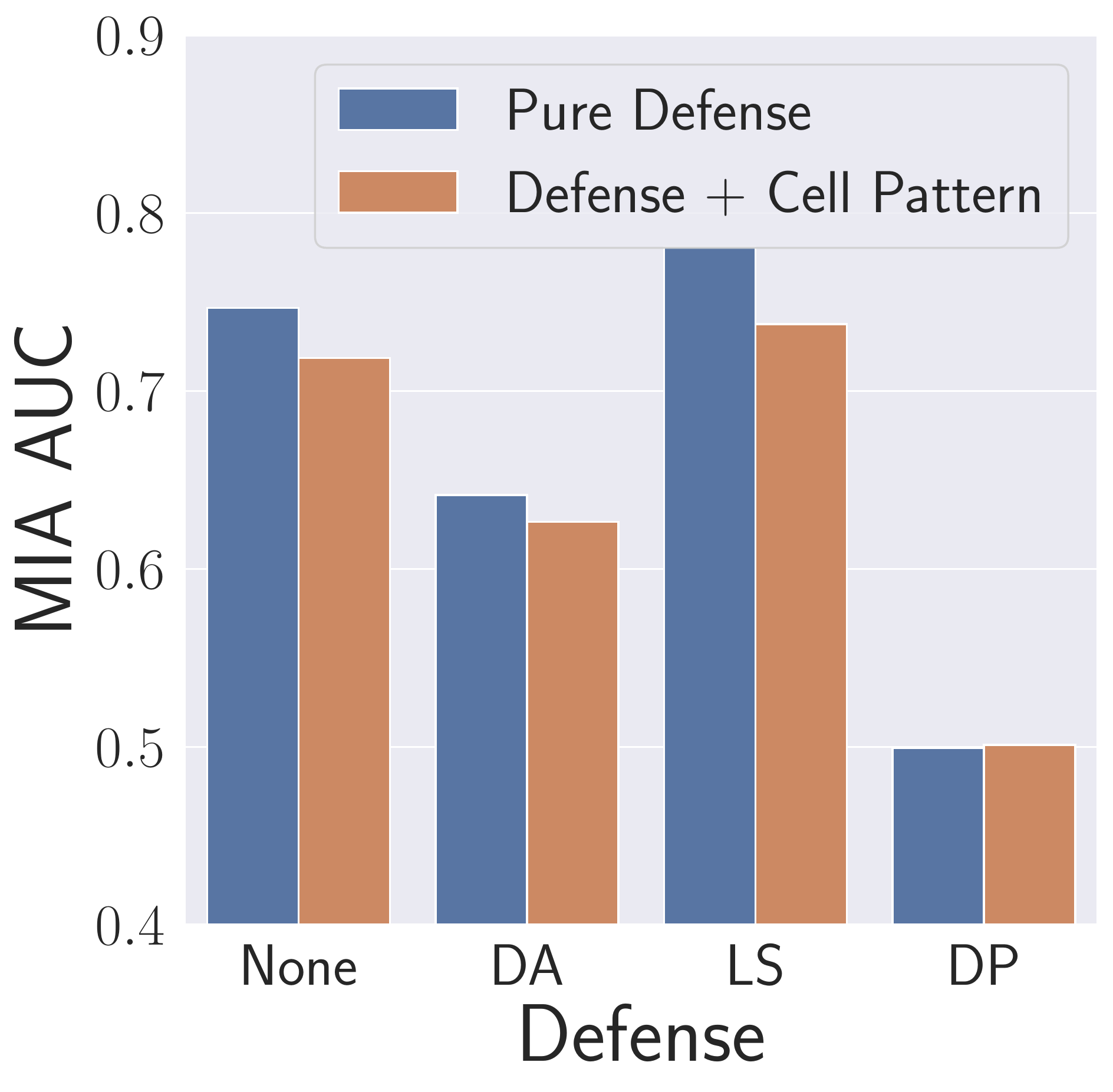}
\caption{\tuple{Black\mbox{-}Box, Partial}}
\label{figure:mia_auc_defense_meminf1}
\end{subfigure}
\begin{subfigure}{0.48\columnwidth}
\includegraphics[width=\columnwidth]{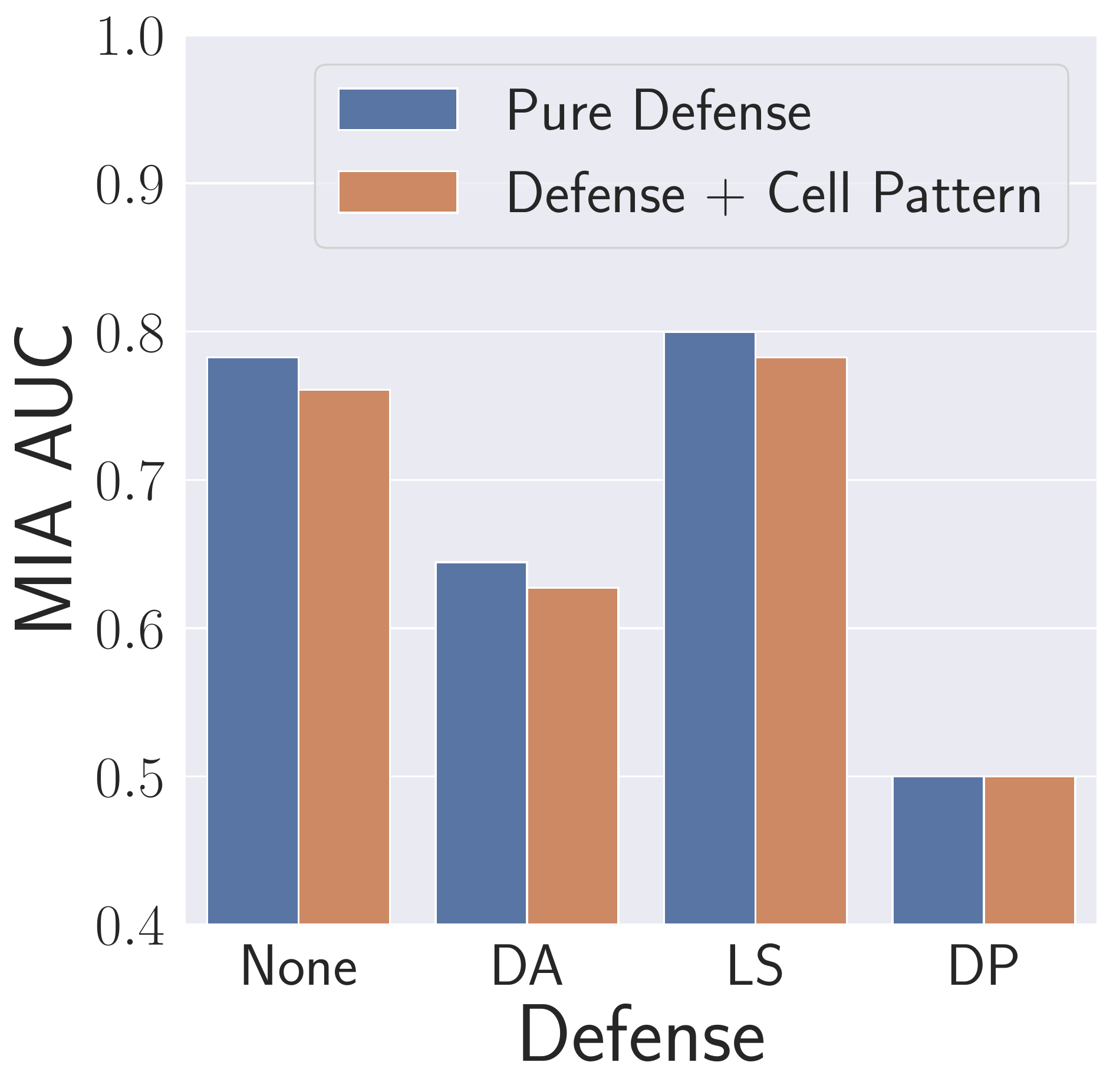}
\caption{\tuple{White\mbox{-}Box, Shadow}}
\label{figure:mia_auc_defense_meminf3}
\end{subfigure}
\begin{subfigure}{0.48\columnwidth}
\includegraphics[width=\columnwidth]{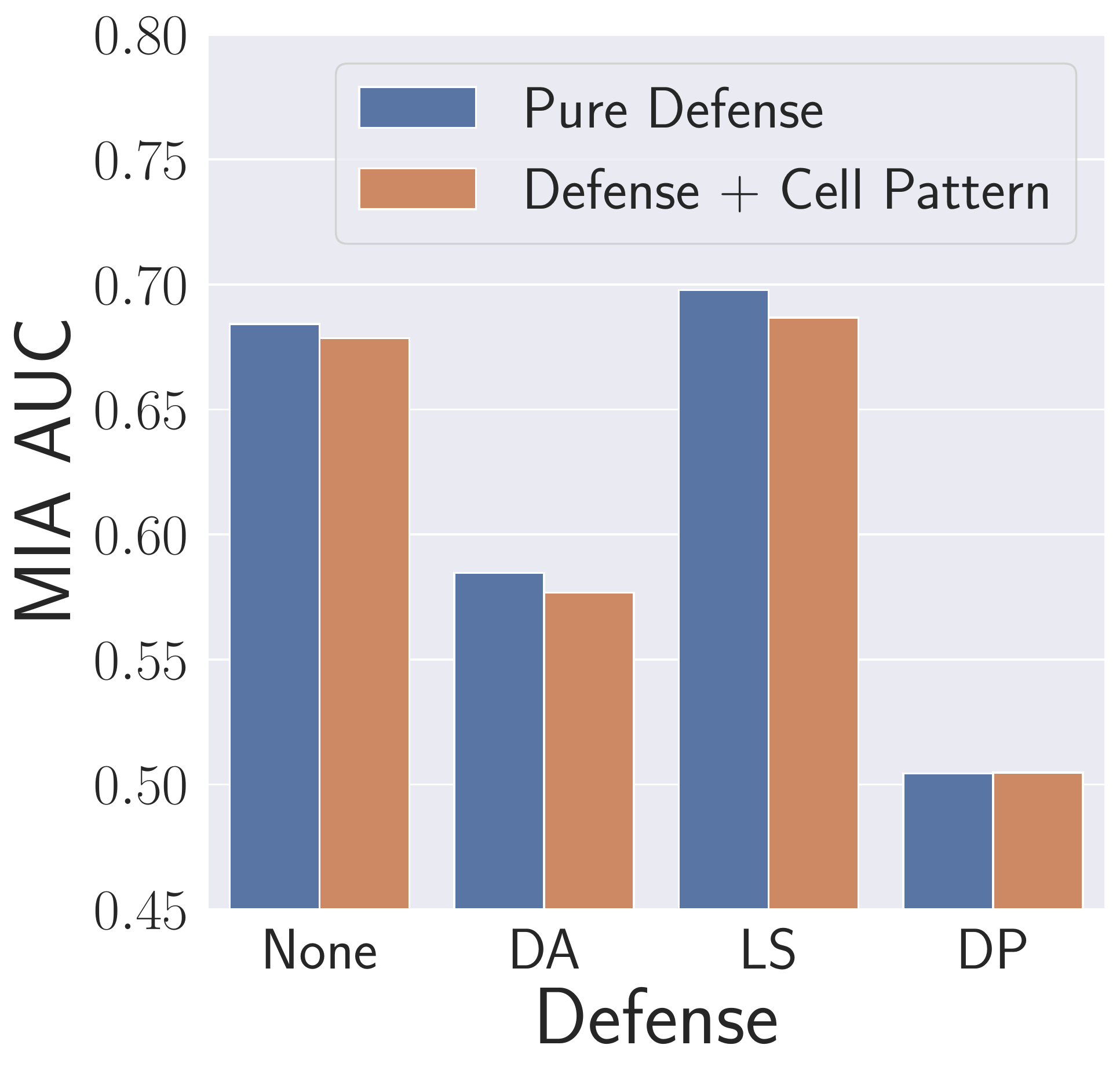}
\caption{\tuple{Label\mbox{-}Only}}
\label{figure:mia_auc_defense_label_only}
\end{subfigure}
\caption{Transferring MIA performance under defense for other attack settings.}
\label{figure:mia_auc_defense_transfer_attacks}
\end{figure}

We also evaluate the effectiveness of defenses when our cell patterns are extracted from the \tuple{White\mbox{-}Box, Partial} attack setting transfer to other attack settings. 
We use the same experimental setup as \autoref{subsec:enhancing_existing_defenses} except the MIA settings. The evaluation results are shown in \autoref{figure:mia_auc_defense_transfer_attacks}.
We can observe that the defense performance is promoted with the transferred MIA demotion cell patterns for almost all cases, which further demonstrates the transferability and effectiveness of our cell patterns.

\end{document}